\documentclass[a4paper,11pt]{emulateapj}
\usepackage{amssymb, latexsym}
\usepackage{graphicx}
\usepackage{longtable}
\usepackage{amsmath} 

\renewcommand\arcsec{\mbox{$^{\prime\prime}$}}%

\renewcommand\ion[2]{#1$\;${\scshape{#2}}}

\def\oii{\ion{O}{ii}}
\def\nii{\ion{N}{ii}}
\def\sii{\ion{S}{ii}}

\def\hii{\ion{H}{ii}}
\def\heii{\ion{He}{ii}}

\def\oiii{\ion{O}{iii}}
\def\neiii{\ion{Ne}{iii}}
\def\rtwo{$R_{23}$}

\def\ha {\mbox{{\rm H}$\alpha$}}
\def\hb {\mbox{{\rm H}$\beta$}}
\usepackage{footmisc}
\usepackage{perpage} 
\MakePerPage{footnote} 

\begin{document}
\title{Physical Properties of Emission-Line Galaxies at $z\sim2$ from
  Near-Infrared Spectroscopy with Magellan FIRE}
\author{Daniel Masters\altaffilmark{1,2}, Patrick McCarthy\altaffilmark{2},
  Brian Siana\altaffilmark{1}, Mathew
  Malkan\altaffilmark{3}, Bahram Mobasher\altaffilmark{1}, Hakim Atek\altaffilmark{4}, Alaina
  Henry\altaffilmark{5}, Crystal L. Martin\altaffilmark{6}, Marc
  Rafelski\altaffilmark{11},
  Nimish P. Hathi\altaffilmark{8}, Claudia Scarlata\altaffilmark{7}, Nathaniel R. Ross\altaffilmark{3},
  Andrew J. Bunker\altaffilmark{9}, Guillermo A. Blanc\altaffilmark{2}, Alejandro G. Bedregal\altaffilmark{10},
  Alberto Dom\'{i}nguez\altaffilmark{1}, James
    Colbert\altaffilmark{11}, 
Harry Teplitz\altaffilmark{12},
  Alan Dressler\altaffilmark{2}}

\altaffiltext{1}{Department of Physics and Astronomy, University of California, Riverside, CA 92521}
\altaffiltext{2}{Carnegie Observatories, Pasadena, CA 91101}
\altaffiltext{3}{Department of Physics and Astronomy, UCLA, Los
  Angeles, 90095}
\altaffiltext{4}{Laboratoire d'Astrophysique Ecole Polytechnique
  F\'{e}d\'{e}rale, Sauverny, Switzerland}
\altaffiltext{5}{Astrophysics Science Division, Goddard Space Flight
  Center, Greenbelt, MD 20771}
\altaffiltext{6}{Department of Physics, Universitey of California,
  Santa Barbara, CA 93106}
\altaffiltext{7}{Department of Physics and Astronomy, University of
  Minnesota}
\altaffiltext{8}{Aix Marseille Universit\'{e}, CNRS, LAM (Laboratoire d'Astrophysique de Marseille) UMR 7326, 13388, Marseille, France}
\altaffiltext{9}{Department of Physics, University of Oxford, UK}
\altaffiltext{10}{Department of Physics and Astronomy, Tufts
  University, Medford, MA 02155}
\altaffiltext{11}{Spitzer Science Center, California Institute of Technology, Pasadena, CA 91125, USA}
\altaffiltext{12}{Infrared Processing and Analysis Center, Caltech, Pasadena, CA 91125}


\begin{abstract}

We present results from near-infrared spectroscopy of 26
emission-line galaxies at $z\sim2.2$ and $z\sim1.5$ obtained with the
Folded-port InfraRed Echellette (FIRE) spectrometer on the 6.5-meter
Magellan Baade telescope. The sample was selected from the WFC3 Infrared
Spectroscopic Parallels (WISP) survey, which uses the
near-infrared grism of the Hubble Space Telescope Wide Field
Camera 3 to detect emission-line galaxies 
over $0.3\lesssim z \lesssim2.3$. Our FIRE
follow-up spectroscopy (R$\sim$5000) 
over 1.0--2.5~$\mu$m permits detailed measurements of
physical properties of the $z\sim2$ emission-line
galaxies. Dust-corrected star formation rates for the sample
range from $\sim$5--100~$\mathrm{M}_{\odot}$~yr$^{-1}$ with a mean of
29~$\mathrm{M}_{\odot}$~yr$^{-1}$. We derive 
a median metallicity for the sample of 12~+~log(O/H)~=~8.34 or $\sim$0.45~Z$_{\odot}$. The estimated stellar masses range from
$\sim$$10^{8.5}-10^{9.5}$~M$_{\odot}$, and a clear positive correlation
between metallicity and stellar mass is observed. The
average ionization parameter measured for the sample, log~$U\approx-2.5$, is significantly higher than 
what is found for most star-forming galaxies in the local universe,
but similar
to the values found for other star-forming
galaxies at high redshift. 
We derive composite spectra from the
FIRE sample, from which we infer typical nebular electron
densities of $\sim$100-400~cm$^{-3}$. Based on
the location of the galaxies and composite spectra on BPT
diagrams, we do not find evidence for significant AGN
activity in the sample.  Most of the galaxies as well as the composites are offset in the BPT diagram toward higher [\oiii]/\hb\
at a given [\nii]/\ha, in agreement with other observations of $z\gtrsim1$
star-forming galaxies, but composite spectra derived from the sample do not show
an appreciable offset from the local star-forming sequence on the
[\oiii]/\hb\ versus  [\sii]/\ha\ diagram. We infer a high 
nitrogen-to-oxygen abundance ratio from the composite spectrum, which
may contribute to the offset of the high-redshift galaxies from the
local star-forming sequence in the  [\oiii]/\hb\ versus [\nii]/\ha\
diagram. We speculate that the elevated nitrogen abundance could result
from substantial numbers of Wolf-Rayet stars in starbursting galaxies at $z\sim2$.

\end{abstract}
\maketitle

\section{Introduction}

The rest-frame optical spectra of star-forming galaxies at all redshifts exhibit emission lines from which
detailed physical properties can be inferred. For galaxies at the peak
of cosmic star formation at $z\sim2$, these
emission lines are shifted into the near-infrared, which, combined
with the intrinsic faintness of the sources, makes them difficult to
observe from the ground. For this reason, relatively few 
near-infrared spectra of galaxies at $z\sim2$ that cover all of the
important rest-frame optical
emission lines have
been published to date (e.g., \citealp{Erb06, Hainline09, Erb10, Rigby11,
 Belli13}).

The available near-infrared spectra of star-forming galaxies at $z\sim2$ 
have revealed differences in comparison with counterparts in the
local universe \citep{Liu08, Newman13}. For example,
star-forming galaxies at $z\sim2$ tend to have higher [\oiii]/\hb\ ratios at a
given [\nii]/\ha\ ratio than local star-forming galaxies. This
observation has been
attributed to more extreme interstellar medium (ISM) conditions, on average, in galaxies at
high redshift, possibly as a result of higher nebular electron
densities, harder ionizing radiation fields, different gas volume
filling factors, or some combination of these  \citep{Shapley05, Brinchmann08, Shirazi13,
Kewley13a,  Kewley13}. The clumpy morphology and relatively high
velocity dispersions observed in many of these sources
\citep{Pettini01, Forster06, Genzel08, Law09} may support the
conjecture that star-formation in the early universe generally occurs
in denser and higher pressure environments than those found in local star-forming galaxies.
Significant contribution of active galactic nuclei (AGN) to emission line
fluxes for $z\sim2$ galaxies has also been suggested 
as a source of the elevated line ratios \citep{Trump11,Trump13}. 
 
The slitless grism spectroscopy provided by the Wide Field Camera 3 (WFC3) on the Hubble Space Telescope
(HST) has enabled the discovery of large numbers of star-forming
galaxies near the peak of cosmic star formation \citep{Atek10, Atek11, Straughn11, Vanderwel11,
Trump11, Brammer12}. Grism surveys such as the WFC3 Infrared
Spectroscopic Parallels (WISP) survey \citep{Atek10} are well-suited to
finding low-mass star-forming galaxies at intermediate redshifts through
their optical emission lines. 
While WFC3 grism spectroscopy detects large numbers
of emission-line galaxies, it is not ideal for extracting the 
physical information encoded in their optical spectra. The spectral resolution
($R\sim130$ in G141 and $R\sim210$ in G102) is insufficient to resolve \ha\ from
[\nii]$\lambda$6548, 6583, or detect line broadening due to AGN
activity. Moreover, despite the broad wavelength coverage
($\sim$0.8--1.7~$\mu$m) of the grism, \ha\ is not detected for galaxies at $z\gtrsim1.6$ and important metallicity diagnostics such as
$R_{23}$~$\equiv$~([\oiii]$\lambda4959,5007$+[\oii]$\lambda$3727)/H$\beta$ are often inaccessible. For these reasons, ground-based spectroscopy with the
new generation of near-infrared spectrometers is required to 
constrain the physical properties of these galaxies. 

Here we present rest-frame optical spectroscopy
of 26 emission-line galaxies from WISP obtained with the Folded-port Infrared Echellette (FIRE, \citealp{Simcoe08, Simcoe10})
on the Magellan Baade 6.5 meter telescope. The sample consists of
13 sources at $z\sim2.2$ and 13 at $z\sim1.5$. Galaxies in the sample were selected
from the WISP survey based on the detection of strong
[\oiii]$\lambda5007$ emission (at $z\sim2.2$) or \ha\ (at $z\sim1.5$)
in the G141 grism data. Follow-up near-IR spectroscopy with
FIRE enables us to: (1) detect \ha\ for sources at
$z>1.6$ and split \ha\ and [\nii] in order to determine star
formation rates (SFRs), (2) get accurate dust reddening estimates using the Balmer
decrement, (3) infer metallicities and ionization parameters from
strong lines, (4) measure
diagnostic line ratios to test for nuclear activity, and (5) resolve emission line velocity dispersions. 
We also construct composite spectra from our sample, which allow us to 
investigate the statistical properties of strongly
star-forming galaxies at $z\sim2.2$ and $z\sim1.5$ in greater detail. 

This paper is structured as follows. In \S2 we provide an overview of
the WISP survey and the emission-line galaxies selected for follow-up
spectroscopy. In \S3 we discuss the Magellan FIRE spectroscopy and data
reduction. In \S4
we present the physical properties measured from the FIRE spectra, 
including dust obscuration, metallicity, ionization parameter, star
formation rate and kinematics. In \S5 we discuss the composite 
spectrum derived from the sample. In \S6 we analyze the
results, exploring the implications of our findings for the nature of
starbursting galaxies at $z\sim2$. In \S7
we conclude with a discussion. We adopt a cosmology with
$\Omega_{M}=0.3$,  $\Omega_{\Lambda}=0.7$ and $H_{0}=70~\mathrm{km~s}^{-1}~\mathrm{Mpc}^{-1}$.

\section{WFC3 Data \& Sample Selection}

\subsection{WISP Survey}

The WISP team (PI Malkan, \citealp{Atek10}) is obtaining
slitless, near-infrared grism spectroscopy over $0.8$--$1.7~\mu$m using
the two infrared grisms installed on the IR channel of Wide Field
Camera 3 (WFC3; \citealp{Kimble08}), with more than 800 $\mathrm{arcmin}^{2}$ of sky
covered at present. WFC3 grism observations are highly sensitive to
emission-line galaxies over $0.3\lesssim z \lesssim2.3$, without suffering from the observational
biases affecting ground-based surveys at these
redshifts.  

WISP is a ``pure parallel'' program, which means 
that the WFC3 pointings are determined by 
primary observing programs using other HST instruments. The WFC3
observations are taken
in a parallel field determined by the offset of the instruments on the
focal plane and
the roll angle of the telescope. WISP consists of two types of observations: a deep survey for 
parallel targets with more than four orbits of visibility, covering a
relatively small number of fields, and a shallower survey 
for parallel targets with one to three orbits of visibility, covering
a much larger number of fields. Parallel observing
targets are selected at galactic 
latitudes $|b|>20^{\circ}$\ with a preference for 
longer visibility times. Typical integration times for a 4--5 orbit
target are $\sim$5000~seconds in G102 and $\sim$2000~seconds in G141. Longer
parallel opportunities are also imaged in F110W and UVIS (the F475X
and/or F600LP filters).
The shallower WISP fields are restricted to the G141 grism and H-band
imaging, with a typical G141 integration time for a 2-orbit target of $\sim$4000~sec.
Because the 2--3 orbit pointings typically achieve deeper G141 integrations and 
cover a larger area than the deep survey, a substantial
fraction of the emission-line galaxies studied here were selected in
these fields.

\begin{figure}
        \centering
	\includegraphics[width=\linewidth]{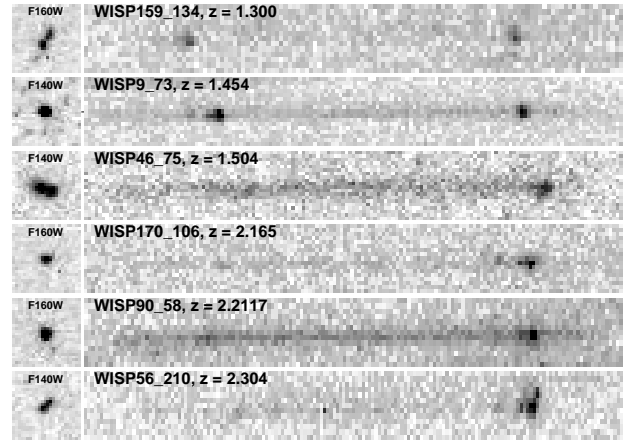}
	\caption{Examples of the WISP sources selected for follow-up
          Magellan FIRE
          spectroscopy. The direct WFC3 H-band images on the left are 3$\arcsec$
          on a side. Prominent rest-frame optical emission lines are visible in the G141
          grism exposures, which span 1.1-1.7~$\mu$m.}
\label{figure:wfc3}
\end{figure}


\begin{center}
\begin{deluxetable*}{lcccccrrr}
\tabletypesize{\scriptsize}
\setlength{\tabcolsep}{0.08in} 
\tablecolumns{9}
\tablecaption{Overview of the emission-line galaxies from the WISP
  survey selected for follow-up near-IR spectroscopy with FIRE.}
\tablehead{   
  \colhead{Object ID} &
  \colhead{R.A.} &
  \colhead{Decl.} &
  \colhead{$z$} &
  \colhead{EW([\oiii])\tablenotemark{a}} &
  \colhead{EW(\ha)\tablenotemark{b}} &
  \colhead{F110W} & 
  \colhead{F140W} &
  \colhead{F160W} \\
  \colhead{} &
  \colhead{(J2000)} &
  \colhead{(J2000)} &
  \colhead{} &
  \colhead{(Rest \AA)} &
  \colhead{(Rest \AA)} &
  \colhead{(AB mag)} &
  \colhead{(AB mag)} &
  \colhead{(AB mag)} 
}
\startdata
WISP159\_134 & 20:56:30.91 & --04:47:56.3 & 1.300 & $\cdots$ & 314 (36) & $\cdots$ & $\cdots$
& 22.82 (0.03) \\
WISP134\_171 & 18:42:33.21 & --68:58:37.4 & 1.354 & 124 (6) & 266 (11) & $\cdots$ & $\cdots$
& 22.55 (0.02) \\
WISP50\_65 & 22:22:15.86 & +09:36:47.1 & 1.437 & $\cdots$ & 202 (15) & $\cdots$ & 22.35
(22.06) & $\cdots$ \\
WISP173\_205 & 01:55:23.64 & --09:03:10.2 & 1.444 & 982 (146) & 603 (42) & $\cdots$ & $\cdots$
& 23.71 (0.03) \\
WISP9\_73 & 12:29:43.35 & +07:48:35.9 & 1.454 & 233 (9) & 221 (12) & 22.80
(0.02) & 22.79 (0.02) & $\cdots$ \\
WISP43\_75 & 21:04:06.18 & --07:22:28.6 & 1.482 & $\cdots$ & 137 (14) & 22.62
(0.16) & 22.37 (0.02) & $\cdots$ \\
WISP25\_53 & 10:08:44.82 & +07:10:20.4 & 1.486 & $\cdots$ & 130 (7) & $\cdots$ & 
22.20 (0.01) & $\cdots$ \\
WISP46\_75 & 22:37:56.48 & --18:42:46.1 & 1.504 & $\cdots$ & 245 (28) & $\cdots$ & 22.77
(0.20) & $\cdots$ \\
WISP126\_90 & 13:41:49.16 & +05:03:06.2 & 1.536 & $\cdots$ & $\cdots$ & $\cdots$ & $\cdots$ &
22.26 (0.02) \\
WISP22\_111 & 08:52:46.09 & +03:09:19.4 & 1.541 & 87 (15) & $\cdots$ & $\cdots$ & 22.60
(0.03) & $\cdots$ \\
WISP22\_216 & 08:52:46.29 & +03:08:45.9 & 1.543 & 99 (8) & $\cdots$ & $\cdots$ & 23.93
(0.06) & $\cdots$ \\
WISP64\_2056 & 14:37:30.20 & --01:50:51.4 & 1.610 & $\cdots$ & $\cdots$ & 27.45
(0.20) & $\cdots$ & 25.12 (0.20) \\
WISP81\_83 & 01:10:06.69 & --02:23:06.5 & 1.677 & 122 (11) & $\cdots$ & 23.11
(0.20) & $\cdots$ & 23.30 (0.06) \\
WISP138\_173 & 15:45:31.03 & +09:33:30.0 & 2.158 & 286 (22) & $\cdots$ & $\cdots$ & $\cdots$
& 23.48 (0.05) \\
WISP170\_106 & 00:12:28.18 & --10:28:33.6 & 2.165 & 154 (7) & $\cdots$ & $\cdots$ & $\cdots$
& 23.31 (0.03) \\
WISP64\_210 & 14:37:28.34 & --01:49:54.4 & 2.177 & 143 (14) & $\cdots$ & 24.55
(0.05) & $\cdots$ & 23.66 (0.06) \\
WISP204\_133 & 11:19:46.37 & +04:10:30.8 & 2.191 & 226 (44) & $\cdots$ & $\cdots$ &
23.93 (0.05) & $\cdots$ \\
WISP27\_95 & 11:33:08.67 & +03:28:27.0 & 2.192 & 150 (14) & $\cdots$ & 23.66
(0.25) & 23.03 (0.05) & $\cdots$ \\
WISP147\_72 & 23:58:22.06 & --10:14:48.7 & 2.196 & 131 (16) & $\cdots$ & 22.99
(0.01) & $\cdots$ & 22.03 (0.02) \\
WISP90\_58 & 01:00:56.20 & +02:25:54.0 & 2.212 & 71 (4) & $\cdots$ & $\cdots$ & $\cdots$ &
22.53 (0.02) \\
WISP70\_253 & 04:02:02.50 & --05:37:19.5 & 2.215 & 257 (23) & $\cdots$ & $\cdots$ &
24.67 (0.48) & $\cdots$ \\
WISP175\_124 & 03:42:19.72 & --20:33:17.1 & 2.216 & 271 (18) & $\cdots$ & $\cdots$ &
23.74 (0.05) & $\cdots$ \\
WISP96\_158 & 02:09:26.37 & --04:43:29.0 & 2.234 & 238 (10) & $\cdots$ & 24.08
(0.31) & $\cdots$ & 23.61 (0.04) \\
WISP138\_160 & 15:45:36.29 & +09:34:26.7 & 2.264 & 116 (4) & $\cdots$ & $\cdots$ & $\cdots$
& 23.30 (0.03) \\
WISP56\_210 & 16:16:50.44 & +06:36:38.0 & 2.304 & 372 (13) & $\cdots$ & $\cdots$ &
23.84 (0.03) & $\cdots$ \\
WISP206\_261 & 10:34:17.56 & --28:30:49.8 & 2.315 & 210 (28) & $\cdots$ & $\cdots$ &
24.26 (0.08) & $\cdots$ 
\enddata
\tablenotetext{a}{Rest-frame equivalent widths derived from the WFC3
  G141 grism data.}
\tablenotetext{b}{1$\sigma$ errors shown in parentheses here and
  throughout the paper.}
\label{table:overview}
\end{deluxetable*}
\end{center}
\normalsize

\vspace{-1.0em}
\vspace{-1.0em}

\subsection{Source Selection}
We select galaxies in WISP fields at $z\sim2.2$ and $z\sim1.5$ with clearly detected emission lines.
These redshifts were chosen because \ha\ and other important
rest-frame optical emission lines are redshifted into
relatively clear atmospheric windows for ground-based near-IR
spectroscopy. To find the candidates, the WFC3 2D G141 grism frames were examined
carefully to identify sources with the prominent 
[\oiii]$\lambda$4959, 5007 emission line
complex and/or \ha\ for sources at $z\sim1.5$. Line fluxes were
measured to ensure that FIRE
spectroscopic follow-up would be feasible; all sources were required
to have [\oiii]$\lambda5007$ or \ha\ fluxes greater than $5\times10^{-17}~\mathrm{erg}~\mathrm{cm^{-2}}~\mathrm{s^{-1}}$.  The WFC3 data for the emission-line galaxy
sample are summarized in Table~\ref{table:overview}, and examples of the WFC3
H-band direct imaging and G141 2D spectra are shown in
Figure~\ref{figure:wfc3}.

The grism selection used here is particularly sensitive to star-forming galaxies with
high equivalent width (EW) emission lines, which tends to favor lower
mass, younger systems, 
with high specific star formation rates (sSFRs). The relation of our emission-line
galaxy selection to galaxies selected using photometric
techniques at similar redshifts, for example the ``BX/BM''
\citep{Steidel04}, ``BzK''  \citep{Daddi04} and UV-dropout \citep{Hathi10} samples, is uncertain. It has been
shown previously that the different
color cut prescriptions non-uniformly sample the
true population of galaxies \citep{Reddy05, Quadri07, Grazian07,
 Ly11}, although there is large overlap between the galaxies selected
with each method. The overlap of our sample
with the UV color-selected
galaxies at similar redshift is likely high, although the
grism-selected emission-line sample can probe lower mass galaxies due to its
relative insensitivity to the galaxy continuum. 

\begin{figure*}[htb]
\centering
  \begin{tabular}{@{}cc@{}}
    \includegraphics[width=.49\textwidth]{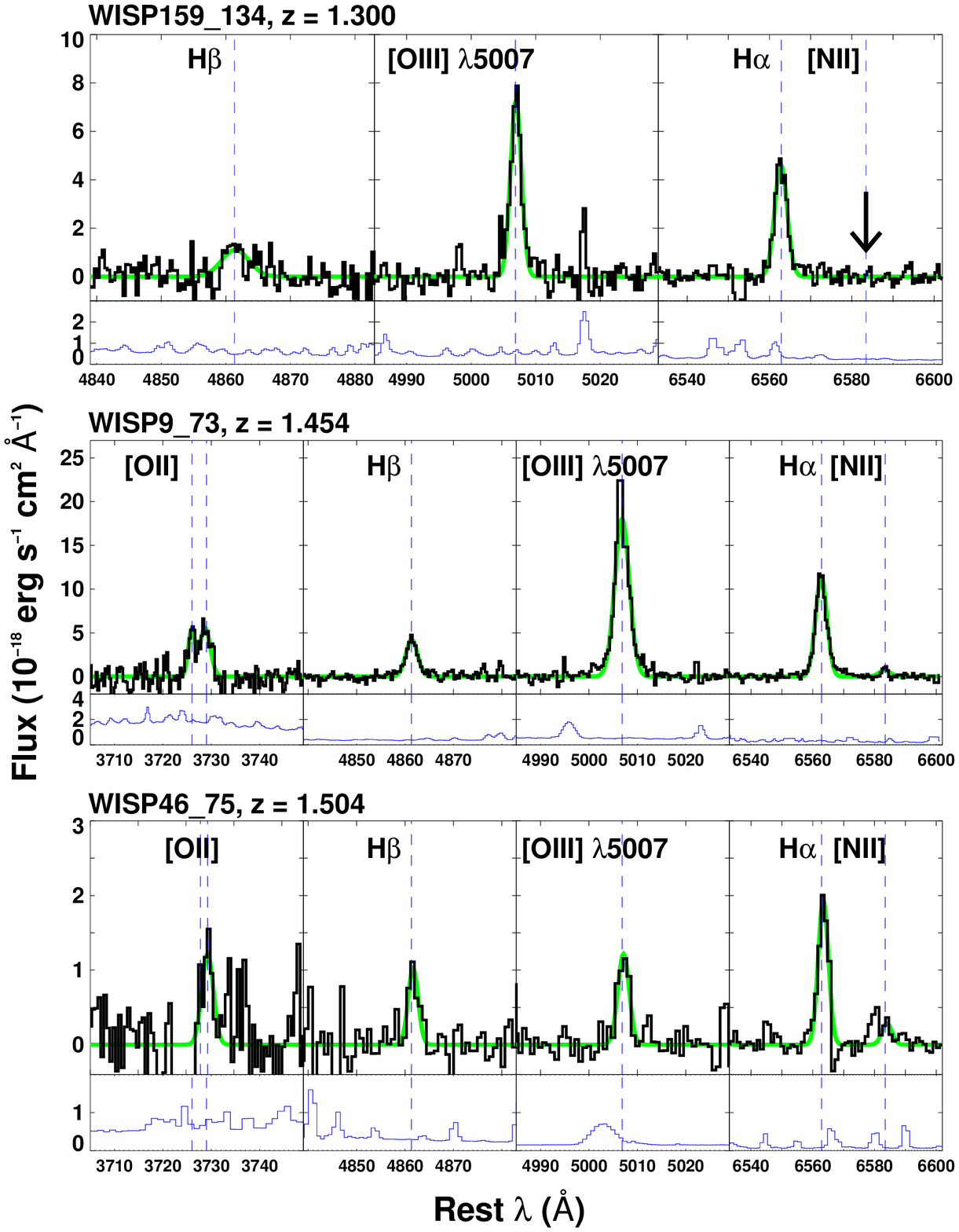} &
    \includegraphics[width=.49\textwidth]{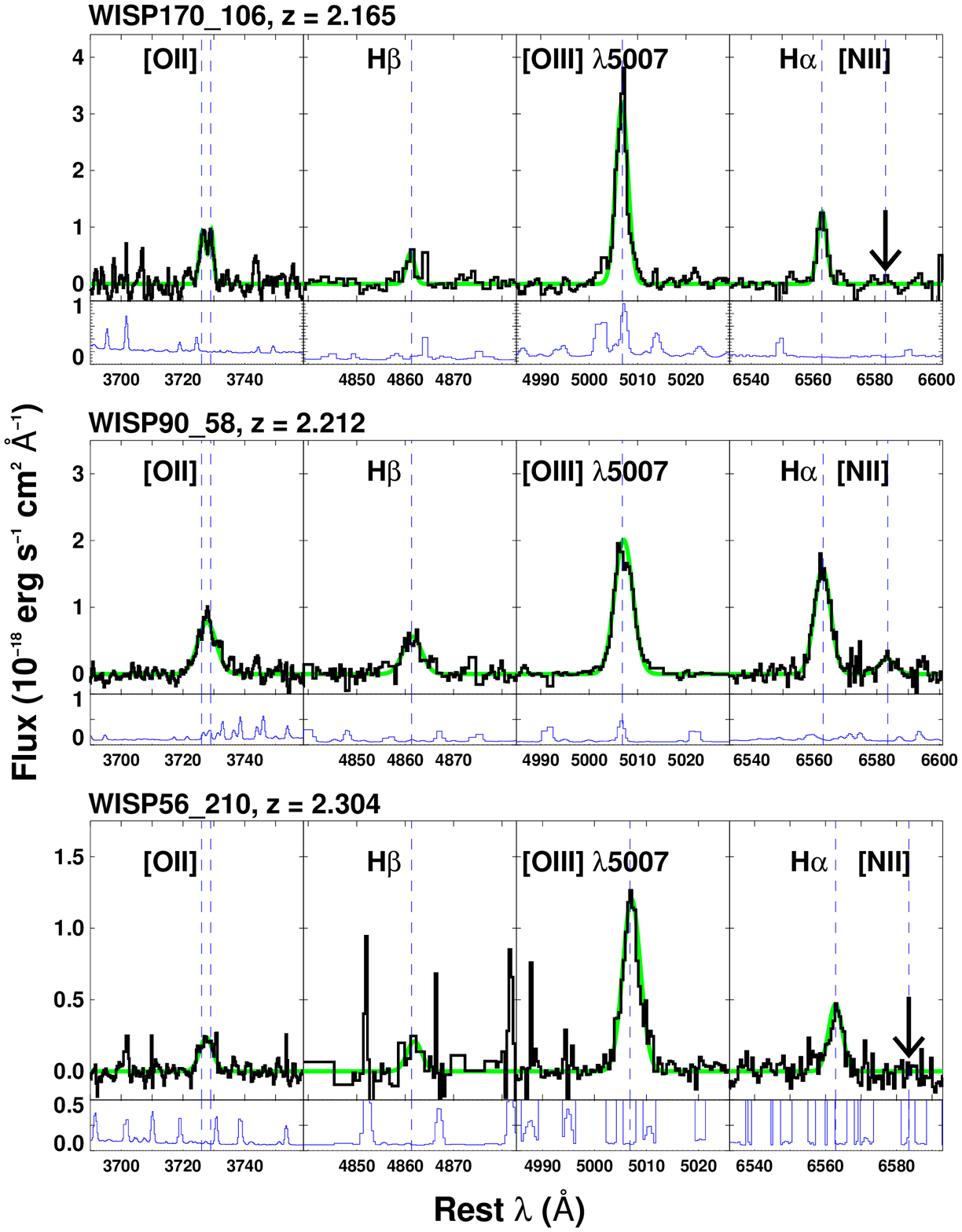} \\
  \end{tabular}
  \label{figure:fire}
  \caption{Example 1D spectra from Magellan FIRE, for the same sources
  as in Figure~\ref{figure:wfc3}. On the left are spectra from
  the $z\sim1.5$ sample and on the right are spectra from the
  $z\sim2.2$ sample. Fits are overlaid in green and the emission lines are labeled. Downward
  arrows indicate non-detections. The lower panels show the error
  spectra. The appendix contains WFC3 and FIRE
  data for all objects in the sample.}
\end{figure*}

\section{Magellan FIRE Near-IR Spectroscopy}

\subsection{Observations \& Data Reduction}
Observations with Magellan FIRE were conducted over six observing
runs from April 2011 to March 2013. FIRE was used in high-resolution
echelle mode to give near-infrared
spectra spanning $\sim$$1.0-2.5~\mu$m. The success rate was nearly
100\% for the candidates, with the only non-detections occurring in
poor observing conditions. Observations were conducted as
follows. The J-band acquisition camera was used to
locate a nearby star from which a blind offset was applied to
position the science target
in the slit. The slits used were either $0.6\arcsec$, $0.75\arcsec$ or
$1.0\arcsec$ in width, depending on the seeing, and were oriented at the parallactic
angle in most cases to minimize differential atmospheric refraction. Exposure times of 900 seconds were used
for sequences of ABBA dither sequences with total integrations
ranging from 1 to 3.5 hours. Readouts were performed with the Sample Up The Ramp (SUTR)
mode in order to minimize
overheads. For each science target, at least one A0V
star was observed at similar airmass for telluric correction. 

Data were reduced using the publicly available
pipeline\footnote[2]{http://web.mit.edu/$\sim$rsimcoe/www/FIRE/ob\_data.htm} developed by
the instrument team. A brief summary of the reduction process
follows. First, the edges of the 2D echelle
orders were traced and the flat field was derived using a combination of 
internal quartz lamp exposures and sky flat exposures. ThAr arc lamp exposures were used to derive the
wavelength solution, which was then refined using the OH sky
lines from the science frames. Sky subtraction was performed using the
method described in \citet{Kelson03}. In most cases optimal extraction was used to
derive the 1D spectrum from the 2D frames,
after fitting a spatial profile to a prominent emission line in a 2D
frame. The 1D spectra were flux
calibrated using observations of an A0V telluric standard star, and
then combined to obtain the final, flux calibrated 1D
spectrum. Examples of reduced FIRE 1D spectra are shown in Figure~2,
and the entire sample is shown in the Appendix.

\subsubsection{Absolute flux calibration} The emission line fluxes measured from the FIRE 1D spectra frequently were 
lower by a factor of 2-4 compared to the same lines measured from the
G141 grism spectra. When multiple lines were measured with
both instruments, for example [\oiii]$\lambda5007$ and \ha\ for
galaxies at $z\sim1.5$, the ratio of the FIRE and WFC3 line fluxes
was found to be consistent for the
two lines to within 20\%, indicating that the relative calibration of FIRE is
reliable. The exact reason for the absolute offset is unclear, but is probably a combination of slit losses in FIRE and
difficulty in deriving the absolute correction for sources with
essentially no detected continuum. 

For the majority of the analysis presented here we do not require absolute line
flux values, but to infer SFRs for the sample we do need to know
the absolute flux of \ha. We use the WFC3 measurements to derive
the absolute \ha\ flux. For galaxies at $z\sim1.5$ we take the WFC3-measured
$F$(\ha) as the reference value, after correcting for blended $F$([\nii]) as measured from the
FIRE spectrum. For galaxies at $z\sim2.2$, \ha\ was not measured with WFC3, so we scale the measured FIRE $F$(\ha)
by the ratio of the WFC3-measured $F$([\oiii]) to the FIRE-measured
$F$([\oiii]), which is justified given the consistent relative
calibrations of the two instruments.

\vspace{1.0em}
\subsection{Line Measurement}

We fit the emission lines with Gaussian profiles using the IDL MPFIT routines
\citep{Markwardt09}. Errors in the derived line fluxes and velocity
dispersions were determined using Monte Carlo simulations in which the
lines were fit repeatedly after applying gaussian noise consistent with the error spectrum. The [\nii]$\lambda$6583 flux was fit simultaneously with 
\ha\ while fixing the width and separation 
of the two lines. For non-detected [\nii]$\lambda$6583, the
95\% confidence upper limit on the flux was determined with a Monte Carlo
simulation in which noise was added to the spectrum
and the [\nii] flux was fit repeatedly as
described. The resulting distribution of [\nii] fluxes determined the
2$\sigma$ upper limit. This limit was verified by adding in and
attempting to recover known fluxes in empty regions of the spectrum
near [\nii]$\lambda6583$.

The instrumental resolution of FIRE was determined by 
measuring the FWHM values of multiple ThAr lines. This was done for slit widths of 0.6\arcsec, 0.75\arcsec, and
1.0\arcsec, giving FWHM resolutions of 59, 65.5, and 72 km~s$^{-1}$,
respectively. These values would be appropriate for galaxies that fill
the slit uniformly; in reality the resolution may be lower
depending on the light profile of the target.
Galaxy line widths are considered to be resolved when the measured FWHM of
the line was at least twice the instrumental resolution. We find that
all of the emission
lines are resolved. Measured line fluxes are presented in
Table~\ref{table:firedata} and the
velocity dispersions for [\oiii]$\lambda$5007 and \ha\ are presented
in Table~5.

\begin{center}
\begin{deluxetable*}{lccccc}
\tabletypesize{\scriptsize}
\tablecolumns{6}
\tablewidth{0pt}
\tablecaption{Line fluxes derived from the
  FIRE near-IR spectroscopy.}
\tablehead{   
  \colhead{Object ID} &
  \colhead{$F$([\oii]$\lambda3727$)\tablenotemark{a}} &
  \colhead{$F$(\hb)} &
  \colhead{$F$([\oiii]$\lambda5007$)} &
  \colhead{$F$(\ha)} &
  \colhead{$F$([\nii]$\lambda6583$)} \\
 \colhead{} &
  \colhead{} &
  \colhead{} &
  \colhead{} &
  \colhead{} &
  \colhead{} 
}
\startdata
WISP159\_134 & $\cdots$ & 13.5 (3.4) & 35.6 (1.3) & 40.3 (2.3) & $<$1.3 \\[0.1cm]    
WISP134\_171 & $\cdots$ & $\cdots$ & 106 (3.6) & 68.3 (1.5) & 7.4 (1.1) \\[0.1cm]	      
WISP50\_65 & 11.7 (2.1) & 2.8 (1.1) & 10.7 (1.1) & 19.3 (0.5) & 6.6 (0.3) \\[0.1cm]	        
WISP173\_205 & 8.9 (2.2) & 7.5 (0.4) & 47.0 (0.7) & 22.3 (0.4) & $\cdots$ \\[0.1cm]	       
WISP9\_73 & 59.4 (7.5) & 31.6 (1.3) & 168 (3.0) & 128 (1.0) & 8.2 (0.8) \\[0.1cm]	      
WISP43\_75 & $\cdots$ & 88.8 (4.8) & $\cdots$ & 296 (37) & 76.8 (20) \\[0.1cm]	      
WISP25\_53 & 65.9 (3.0) & 31.0 (0.9) & 78.9 (0.8) & 86.9 (0.9) & 15.8 (0.9) \\[0.1cm]	      
WISP46\_75 & 18.5 (3.7) & 6.6 (1.1) & 9.0 (0.5) & 20.1 (0.7) & 3.0 (0.3) \\[0.1cm]      
WISP126\_90 & $\cdots$ & 3.8 (0.7) & $\cdots$ & 21.9 (0.4) & 8.7 (1.1) \\[0.1cm]	      
WISP22\_111 & $\cdots$ & 7.5 (1.4) & 26.8 (3.3) & 21.0 (0.5) & 5.3 (1.0) \\[0.1cm]	       
WISP22\_216 & $\cdots$ & 11.7 (0.7) & 53.6 (1.5) & 32.7 (1.6) & $<$ 0.5  \\[0.1cm]	      
WISP64\_2056 & 14.3 (8.0) & $\cdots$ & 30.6 (0.4) & 16.9 (0.4) & $<$ 0.6 \\[0.1cm]      
WISP81\_83 & $\cdots$ & $\cdots$ & 220 (4.1) & 136 (1.6) & $\cdots$ \\[0.1cm]	      
WISP138\_173 & 9.2 (0.5) & 2.7 (0.3) & 14.5 (0.6) & 16.0 (1.3) & $<$1.0 \\[0.1cm]	       
WISP170\_106 & 13.2 (1.6) & 3.7 (0.4) & 32.0 (1.4) & 15.0 (0.5) & $<$ 1.3  \\[0.1cm]	      
WISP64\_210 & 29.9 (1.9) & 20.1 (1.4) & 136 (2.3) & 84.5 (3.5) & 15.1 (3.7) \\[0.1cm]	     
WISP204\_133 & 4.3 (0.8) & 2.6 (0.3) & 5.7 (0.7) & 3.9 (0.4) & $<$ 0.9  \\[0.1cm]	      
WISP27\_95 & 50.0 (1.4) & 18.6 (0.7) & 91.0 (0.7) & 77.5 (1.2) & 9.6 (1.2) \\[0.1cm]	     
WISP147\_72 & 14.4 (7.8) & $\cdots$ & 86.5 (0.5) & 20.2 (0.7) & 1.3 (0.6) \\[0.1cm]	      
WISP90\_58 & 18.7 (0.8) & 8.7 (0.3) & 30.7 (0.5) & 33.0 (0.8) & 4.7 (0.4) \\[0.1cm]	    
WISP70\_253 & 4.1 (0.6) & 7.3 (0.3) & 30.1 (0.2) & 14.0 (0.6) & $<$ 0.6  \\[0.1cm]	      
WISP175\_124 & 13.1 (0.7) & $\cdots$ & 38.1 (0.3) & 25.4 (0.9) & 2.5 (0.4) \\[0.1cm]	    
WISP96\_158 & 19.7 (2.5) & 9.0 (1.6) & 94.4 (2.3) & 44.9 (1.8) & $<$ 2.9  \\[0.1cm]	       
WISP138\_160 & $\cdots$ & 49.8 (5.7) & 256 (1.6) & 170 (1.5) & 14.6 (1.2) \\[0.1cm]	       
WISP56\_210 & 4.2 (0.3) & 2.7 (0.4) & 16.9 (0.7) & 7.5 (1.2) & $\cdots$ \\[0.1cm]	      
WISP206\_261 & 6.9 (1.5) & 7.4 (0.5) & 49.1 (0.4) & 30.3 (0.9) & $<$ 1.1              
\enddata
\tablenotetext{a}{All fluxes are in units of
  $10^{-18}~\mathrm{erg}~\mathrm{cm^{-2}}~\mathrm{s^{-1}}$. Fluxes
  presented here have not been scaled to match WFC3 fluxes, as
  described in \S3.1.1}
\label{table:firedata}
\end{deluxetable*}
\end{center}
\normalsize

\vspace{-1.0em}
\vspace{-1.0em}

\section{Physical Properties of the Sample}

\subsection{Dust Extinction}

Dust preferentially attenuates shorter-wavelength light
and must be taken into account when deriving physical
properties from emission lines. 
We use the ratio \ha/\hb\ (the Balmer decrement) to estimate the
nebular dust extinction of galaxies in the sample. The Balmer decrement is
relatively insensitive to temperature and should equal 2.86 in the
absence of dust, given
reasonable assumptions for the physical conditions of the \hii\
regions producing the line emission \citep{Osterbrock06}.
\begin{figure*}[htb]
\centering
  \begin{tabular}{@{}cc@{}}
    \includegraphics[width=.49\textwidth]{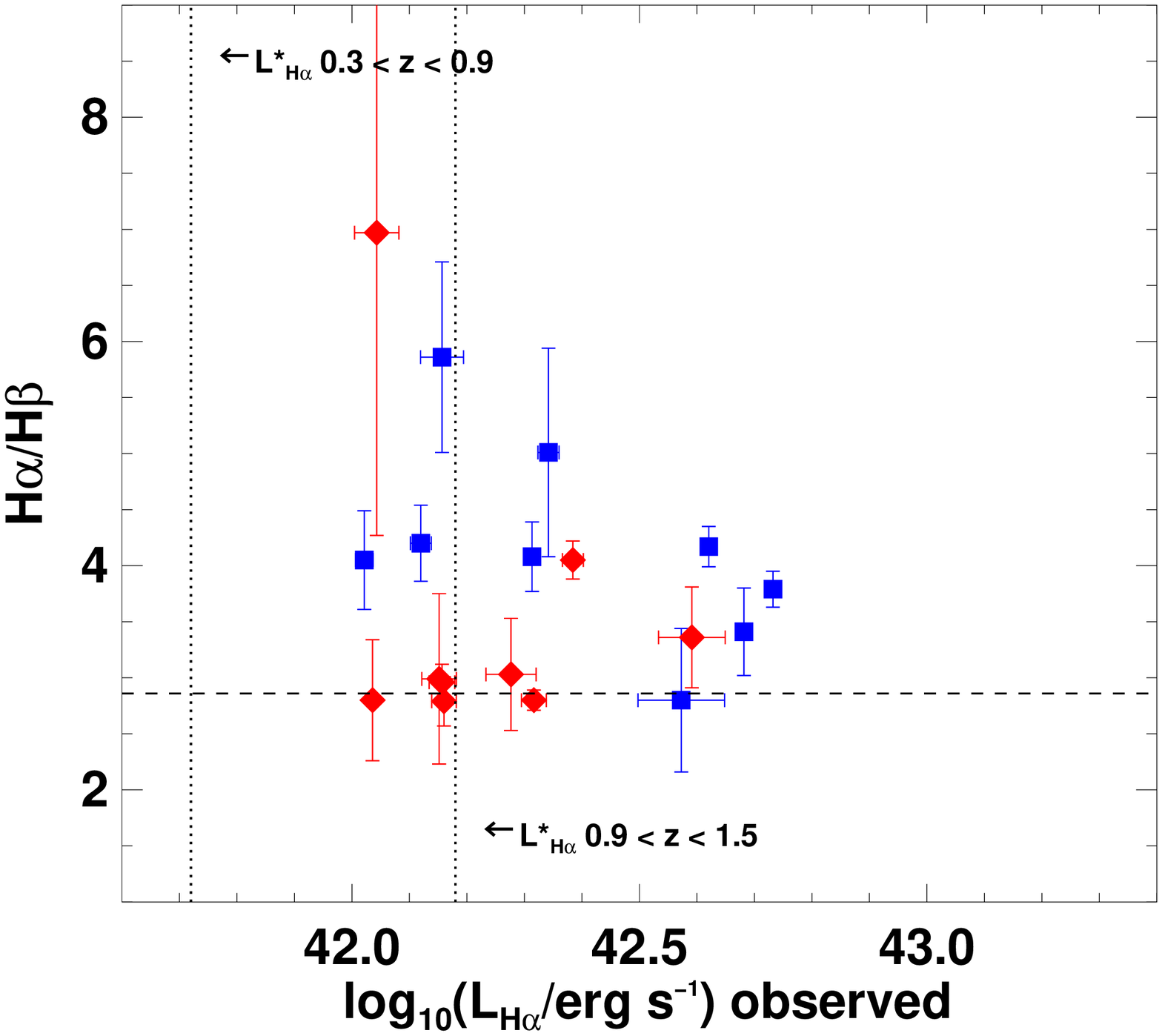} &
    \includegraphics[width=.49\textwidth]{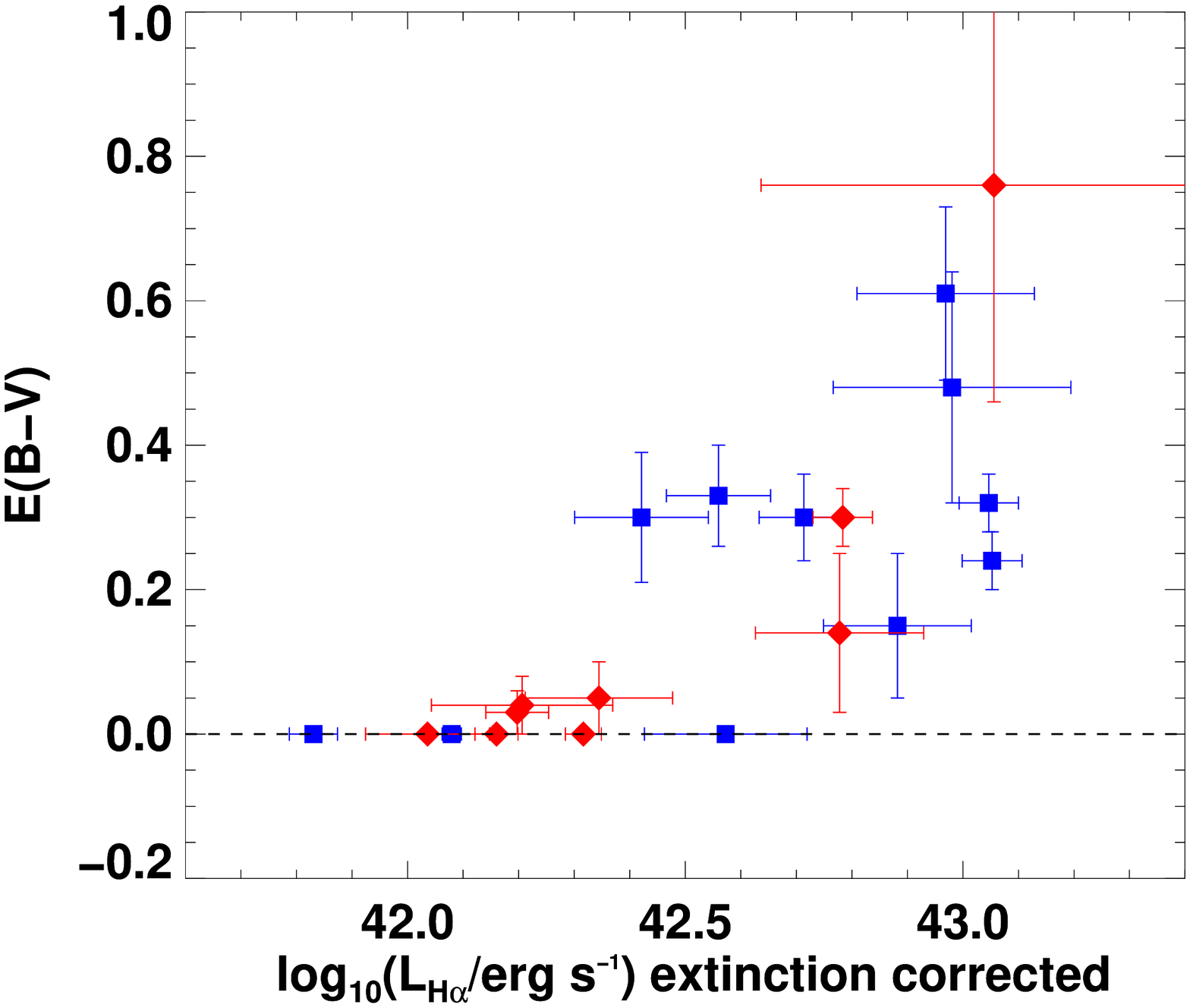} \\
 \end{tabular}
  \label{figure:balmer}
  \caption{\emph{Left:} Balmer decrement as a function of the observed
    \ha\ luminosity. The
          blue squares are from the $z\sim2.2$ sample and the red
          diamonds are from the $z\sim1.5$ sample. There
          is no trend evident here, but the values measured are
          consistent with the results from \citet{Dominguez13} for
          this somewhat narrow range of observed
          $L_{\mathrm{H}\alpha}$. We indicate with dotted lines the values of
          $L^{*}_{\mathrm{H}\alpha}$ for two lower redshift intervals
          recently determined by \citet{Colbert13}. \emph{Right:} The
          $E(B-V)$ values inferred from the Balmer decrements plotted
          against the extinction-corrected \ha\ luminosity. Here we see a clear
          trend, with more luminous \ha\ emitters having
          higher dust extinction. It should be noted that this plot is
          somewhat circular, in that the $E(B-V)$ is used to
          obtain the extinction corrected $L_{\mathrm{H}\alpha}$
          values.}
\end{figure*}
\begin{center}
\begin{deluxetable*}{lccccccc}
\tabletypesize{\scriptsize}
\setlength{\tabcolsep}{0.1in} 
\tablecolumns{8}
\tablecaption{Balmer decrements, extinctions, and SFRs for the sample.}
\tablehead{   
  \colhead{Object ID} &
  \colhead{\ha/\hb} &
  \colhead{\emph{E(B-V)}} &
  \colhead{A$_{\mathrm{H}\alpha}$} &
  \colhead{$F$(\ha) scaled\tablenotemark{a}} &
  \colhead{$F$(\ha) ext. corrected\tablenotemark{b}} &
  \colhead{SFR (uncorr.)} &
  \colhead{SFR (corr.)} \\
 \colhead{} &
  \colhead{} &
  \colhead{(mag)} &
  \colhead{(mag)} &
  \colhead{} &
  \colhead{} &
  \colhead{$M_{\odot}~\mathrm{yr}^{-1}$} &
 \colhead{$M_{\odot}~\mathrm{yr}^{-1}$} 
}
\startdata
WISP159\_134 & 2.99 (0.76) & $0.04_{-0.04}^{+0.20}$ & 0.13 & 14.2 (0.95) & $16.1_{-2.1}^{+13.6}$ & 6.6 (0.5) & $7.5_{-1.0}^{+6.3}$ \\[0.1cm]
WISP134\_171 & $\cdots$ & $\cdots$ & $\cdots$ & 22.6 (0.78) & $\cdots$ & 11.6 (0.4) & $\cdots$ \\[0.1cm]
WISP50\_65 & 6.97 (2.70) & $0.76_{-0.30}^{+0.33}$ & 2.53 & 10.0 (0.9) & $103_{-62}^{+180}$ & 6.0 (0.5) & $62_{-37}^{+108}$ \\[0.1cm]
WISP173\_205 & 2.96 (0.16) & $0.03_{-0.03}^{+0.05}$ & 0.10 & 11.2 (0.6) & $12.3_{-1.2}^{+2.1}$ & 6.8 (0.4) & $7.4_{-0.7}^{+1.3}$ \\[0.1cm]
WISP9\_73 & 4.05 (0.17) & $0.30_{-0.04}^{+0.04}$ & 1.00 & 18.7 (0.8) & $46.9_{-5.4}^{+6.2}$ & 11.5 (0.5) & $29_{-3.3}^{+3.7}$ \\[0.1cm]
WISP43\_75 & 3.36 (0.45) & $0.14_{-0.11}^{+0.11}$ & 0.47 & 29.6 (3.7) & $45.5_{-13.6}^{+18.6}$ & 19.1 (2.4) & $29_{-8.8}^{+12}$ \\[0.1cm]
WISP25\_53 & 2.80 (0.09) & $0.00_{-0.00}^{+0.03}$ & 0.0 & 15.0 (0.7) & $15.0_{-0.7}^{+1.6}$ & 9.8 (0.5) & $9.8_{-0.5}^{+1.0}$ \\[0.1cm]
WISP46\_75 & 3.03 (0.50) & $0.05_{-0.05}^{+0.14}$ & 0.17 & 13.6 (1.3) & $15.9_{-2.6}^{+8.6}$ & 9.1 (0.9) & $10.6_{-1.8}^{+5.8}$ \\[0.1cm]
WISP126\_90 & 5.81 (1.01) & $0.61_{-0.15}^{+0.15}$ & 2.03 & $\cdots$ & $\cdots$ & $\cdots$ & $\cdots$ \\[0.1cm]
WISP22\_111 & 2.80 (0.54) & $0.00_{-0.00}^{+0.16}$ & 0.0 & 7.58 (0.16) & $7.6_{-0.2}^{+4.8}$ & 5.4 (0.1) & $5.4_{-0.1}^{+3.4}$ \\[0.1cm]
WISP22\_216 & 2.79 (0.22) & $0.00_{-0.00}^{+0.04}$ & 0.0 & 9.58 (0.46) & $9.6_{-0.5}^{+1.3}$ & 6.8 (0.4) & $6.8_{-0.4}^{+0.9}$ \\[0.1cm]
WISP64\_2056 & $\cdots$ & $\cdots$ & $\cdots$ & 7.68 (0.17) & $\cdots$ & 6.1 (0.1) & $\cdots$ \\[0.1cm]
WISP81\_83 & $\cdots$ & $\cdots$ & $\cdots$ & 1.14 (0.01) & $\cdots$ & 1.0 (0.01) & $\cdots$ \\[0.1cm]
WISP138\_173 & 5.86 (0.85) & $0.61_{-0.12}^{+0.12}$ & 2.03 & 9.43 (0.78) & $61.2_{-18.8}^{+27.3}$ & 15.3 (1.2) & $99_{-30}^{+44}$ \\[0.1cm]
WISP170\_106 & 4.05 (0.44) & $0.30_{-0.09}^{+0.09}$ & 1.00 & 6.89 (0.24) & $17.3_{-4.2}^{+5.5}$ & 11.3 (0.4) & $28_{-6.9}^{+9.0}$ \\[0.1cm]
WISP64\_210 & 4.20 (0.34) & $0.33_{-0.07}^{+0.07}$ & 1.10 & 7.77 (0.32) & $21.4_{-4.2}^{+5.1}$ & 12.9 (0.5) & $35_{-7.0}^{+8.5}$ \\[0.1cm]
WISP204\_133 & 1.52 (0.25)\tablenotemark{c} & $0.00_{-0.00}^{+0.00}$ & 0.0 & 3.61 (0.36) & $3.6_{-0.4}^{+0.4}$ & 6.1 (0.6) & $6.1_{-0.6}^{+0.7}$ \\[0.1cm]
WISP27\_95 & 4.17 (0.18) & $0.32_{-0.04}^{+0.04}$ & 1.07 & 12.0 (0.2) & $32.0_{-3.7}^{+4.2}$ & 20.2 (0.4) & $54_{-6.2}^{+7.1}$ \\[0.1cm]
WISP147\_72 & $\cdots$ & $\cdots$ & $\cdots$ & 8.27 (0.29) & $\cdots$ & 14.0 (0.5) & $\cdots$ \\[0.1cm]
WISP90\_58 & 3.79 (0.16) & $0.24_{-0.04}^{+0.04}$ & 0.80 & 15.4 (0.4) & $32.2_{-3.8}^{+4.2}$ & 26.5 (0.6) & $56_{-6.6}^{+7.2}$ \\[0.1cm]
WISP70\_253 & 1.92 (0.11)\tablenotemark{c} & $0.00_{-0.00}^{+0.00}$ & 0.0 & 3.38 (0.13) & $3.4_{-0.1}^{+0.1}$ & 5.8 (0.2) & $5.8_{-0.2}^{+0.3}$ \\[0.1cm]
WISP175\_124 & $\cdots$ & $\cdots$ & $\cdots$ & 9.53 (0.33) & $\cdots$ & 16.5 (0.6) & $\cdots$ \\[0.1cm]
WISP96\_158 & 5.01 (0.93) & $0.48_{-0.16}^{+0.16}$ & 1.60 & 6.09 (0.25) & $26.5_{-10.3}^{+16.9}$ & 10.8 (0.5) & $47_{-18}^{+30}$ \\[0.1cm]
WISP138\_160 & 3.41 (0.39) & $0.15_{-0.10}^{+0.10}$ & 0.50 & 13.3 (0.1) & $21.1_{-5.6}^{+7.5}$ & 24.2 (0.3) & $38_{-10}^{+14}$ \\[0.1cm]
WISP56\_210 & 2.80 (0.64) & $0.00_{-0.00}^{+0.16}$ & 0.0 & 10.3 (1.6) & $10.3_{-1.6}^{+6.7}$ & 19.6 (3.1) & $20_{-3.2}^{+12.7}$ \\[0.1cm]
WISP206\_261 & 4.08 (0.31) & $0.30_{-0.06}^{+0.06}$ & 1.00 & 5.57 (0.17) & $14.0_{-2.4}^{+2.8}$ & 10.7 (0.4) & $27_{-4.5}^{+5.4}$ 
\enddata
\tablenotetext{a}{This is the FIRE flux scaled to WFC3 as described in
\S3.1.1, accounting for the uncertain absolute calibration of the FIRE
data. The units are $10^{-17}~\mathrm{erg}~\mathrm{cm^{-2}}~\mathrm{s^{-1}}$.}
\tablenotetext{b}{Scaled $F$(\ha) extinction corrected using the $E(B-V)_{gas}$
  derived from the Balmer decrement and assuming a Calzetti extinction
  curve \citep{Calzetti00}.}
\tablenotetext{c}{Low value likely due to measurement error in \hb.}
\label{table:sfrs}
\end{deluxetable*}
\end{center}
\normalsize
\vspace{-1.0em}
\vspace{-1.0em}

In computing the nebular extinction from the Balmer decrement we follow the
method outlined in \citet{Momcheva13} and \citet{Dominguez13}, which
uses the extinction curve of \citet{Calzetti00} to convert the observed
Balmer decrement into $E(B-V)$ and total extinction values. The Balmer
decrements we measure for 
the sample and the derived
$E(B-V)$ and A$_{\mathrm{H}\alpha}$ values are given in
Table~\ref{table:sfrs}. The nebular extinctions for the sample are generally
low, ranging from no evidence for extinction up to
A$_{\mathrm{H}\alpha}$ = 2.53~mags, with an average extinction of
A$_{\mathrm{H}\alpha}=0.7$~mags. In this analysis we do not
correct for the possible effect of stellar absorption lines on the
measured line fluxes due to the lack of reliable continuum measurements
for most objects. This may affect the measured Balmer decrements and
lead to slightly overestimated dust extinctions, but the effect will
be small for high-EW galaxies. For example, the typical
equivalent width of the stellar absorption in \hb\ is 3-4~\AA\ (see,
e.g., \citealp{Dominguez13}), whereas the equivalent widths of the
\hb\ lines in this sample are likely to be $\gtrsim30$~\AA\ judging
from the measured EW([\oiii]) values and [\oiii]/\hb\
ratios. We conclude that the effect will not have a substantial impact on the results.

On the left panel of Figure~3 we
show the Balmer decrement as a function of the observed \ha\
luminosity (which was scaled to account for the uncertain absolute FIRE
calibration, as described in \S3.1.1).
There is no obvious trend between the Balmer decrement and the
observed $L_{\mathrm{H}\alpha}$, which is unsurprising given the
relatively narrow range of observed \ha\ luminosities in our
sample. Nevertheless, the Balmer decrements we find for
$\mathrm{log}_{10}(L_{\mathrm{H}\alpha})\sim42-42.5$~erg~s$^{-1}$ are in good
agreement with those found by \citet{Dominguez13} using stacked WFC3
spectra from WISP. 

On the right panel of Figure~3 we show the inferred
$E(B-V)$ values plotted against the extinction-corrected \ha\
luminosities. There is circularity in this plot, in that the
reddening on the y-axis was used to estimate the 
correction for the
\ha\ luminosity on the x-axis. If taken at face value, a clear trend is evident, with
the galaxies that are more luminous in \ha\ having more dust
extinction. This result is consistent with the trend seen at both low and
high redshift that more luminous galaxies tend to have more dust
obscuration \citep{Dominguez13}, as well as with
previous studies (e.g., \citealp{Ly11}) that have found that the star
formation rate density at $z\sim2$ is dominated by a less numerous
population of highly dust-extinguished galaxies. A word of caution is
in order, however, as this might be a selection effect due to the
grism selection missing obscured sources with lower intrinsic \ha\ luminosities.

\subsection{Metallicity}

A number of different metallicity calibrations based on emission line
strengths have been proposed in the literature 
(see \citealp{Kewley08} for an overview). One of the most
commonly used indicators, \rtwo, is based on the ratio
([\oiii]$\lambda$4959,5007+[\oii]$\lambda$3727)/\hb\ \citep{Pagel79, McGaugh91,
  Kewley02, Kobulnicky04}.  A well-known difficulty with \rtwo\ is
that it is double-valued with 12+log(O/H), making it necessary to determine which ``branch'' of
the \rtwo\ curve a galaxy is on (\citealp{Dopita13} and references therein). The \rtwo\ calibration also has a 
strong dependence on ionization parameter, although this quantity can
be measured from the ratio [\oiii]/[\oii] when R23 is observed. Another well-known
metallicity indicator is $N2$ $\equiv$
log([\nii]$\lambda6583$/\ha) \citep{Pettini04}. This estimator has the
advantage of increasing monotonically with metallicity below $\mathrm{12+log(O/H)}\sim9.2$, but is also
sensitive to ionization parameter \citep{Kewley02}. 

\begin{figure}[htb]
	\includegraphics[scale=0.5]{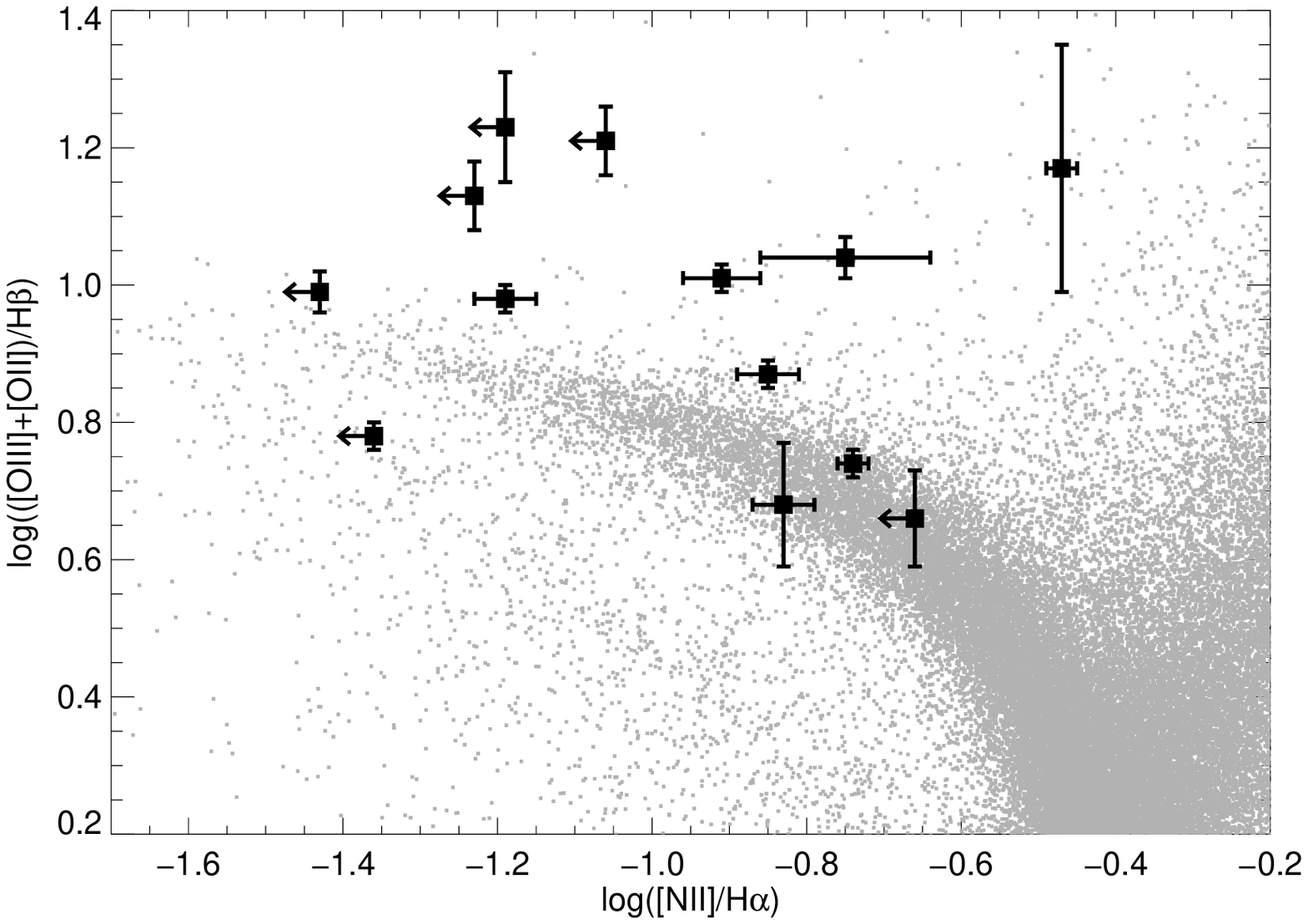}
	\caption{$R23$ plotted against $N2$. Both of
          these quantities are metallicity indicators, and in the
          FIRE data a hint of a turnover in R23 at low metallicities (low $N2$
          values) is apparent in the two leftmost FIRE points. 60,000 data points from SDSS galaxies
          are shown for comparison; no turnover in R23 is
          apparent in this data, but this is because of the combination of
          the rarity of lower branch galaxies in the local universe
          and the lack of sensitivity of SDSS to such low-mass
          sources. Close to half of the galaxies in our sample show
          higher \rtwo\ values than found in 
          local star-forming galaxies.}
\label{figure:r23vn2}
\end{figure}

We find unusually large \rtwo\
values compared with local samples for about half of the sources for which
we were able to measure \rtwo. We illustrate this in Figure~\ref{figure:r23vn2}, in which the dust-corrected \rtwo\ values from
the FIRE sample are plotted against the ratio log([\nii]/\ha). This plot is similar to
the BPT diagram, but in this case both axes are commonly-used metallicity
indicators. Noteworthy is the offset of many of the FIRE-observed points to
higher \rtwo\ values at a given [\nii]/\ha\ ratio compared with the
Sloan Digital Sky Survey (SDSS; \citealp{Abazajian09}) data points shown for comparision.
Calibrations in the literature generally do not account for such high
values. We are relatively confident that the majority of the high \rtwo\ values
are not due to AGN activity, a point that is discussed further in \S6.1.

We used the code Inferring Metallicities (Z) and Ionization parameters (IZI, Blanc et
al. in prep) to further analyze the sample. This software uses
Bayesian inference to compute the joint and marginalized posterior
probability distribution functions (PDF) for the metallicity and
ionization parameter by comparing the dust-corrected line fluxes to
the predictions of a photoionization model grid. We used the grids of
\citet{Levesque10}, which have been recently extended to higher
ionization parameter values by \citet{Richardson13}. Again, we
find that the model grids are often unable to reproduce the observed
lines for those sources with high [\oiii]/\ha\ and
[\oiii]/\hb\ line ratios. 

To estimate metallicities, we use
the $N2$ calibration of \citet{Pettini04}. We then use these 
estimates to determine which branch of the \rtwo\ curve to
adopt when computing the \rtwo\ metallicity, with the lower branch
being used for $N2$ metallicities less than 8.4. The metallicity estimates from
\rtwo\ and $N2$ are given in
Table~\ref{table:metallicity}. In cases for which \rtwo\ was
higher than normal, we approximate the \rtwo\ metallicity as
$\mathrm{12+log(O/H)}\sim8.4$, the value at the nominal turnover of the \rtwo\ curve. 

We find a median ratio [\nii]/\ha\ of
$\sim$10\%, in close agreement with the value of $\sim$12\%
derived from the overall composite spectrum presented in \S5. The $N2$ metallicities of the sample range
from $<$13\% solar to roughly solar, with a median of
0.45~Z$_{\odot}$ (Z$_{\odot}$=8.69 in units of 12+log(O/H),
\citealp{Asplund09}). The \rtwo\ metallicities are similar, with a
median metallicity of 0.50~Z$_{\odot}$. As pointed out by
\citet{Dopita13}, \rtwo-based metallicities should be treated with
caution, particularly as the ratio is only weakly sensitive to
abundance for $8.3\lesssim\mathrm{12+log(O/H)}\lesssim9.0$. Therefore,
we use the $N2$ metallicities in the analysis that follows, although
these can also be problematic.

\begin{figure}[htb]
        \centering
	\includegraphics[scale=0.35]{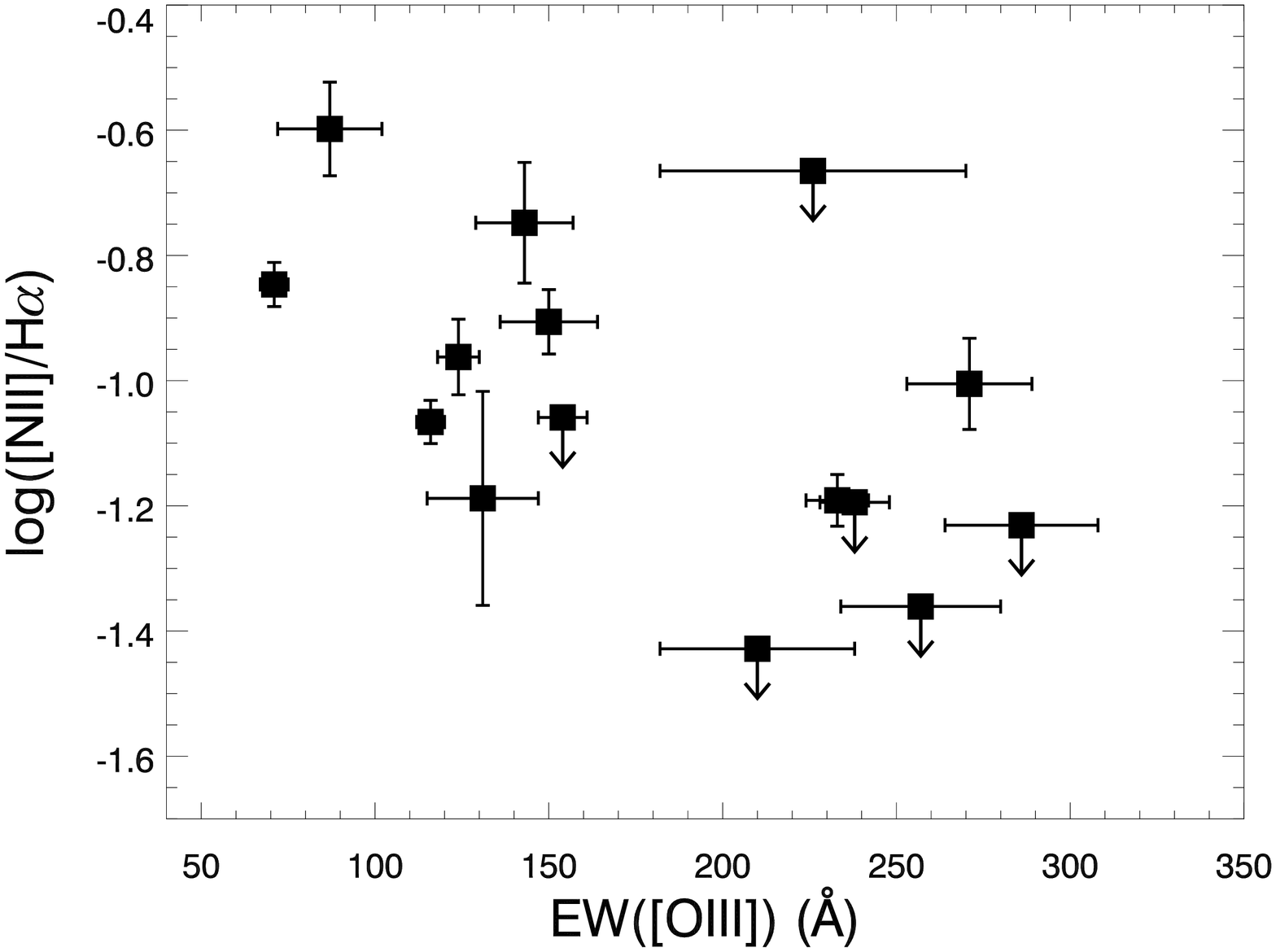}
	\caption{The ratio [\nii]/\ha\
          vs. EW([\oiii]). A fairly tight negative correlation between
          [\nii]/\ha\ and EW(\ha) was shown by  
          \citealp{Cowie11} (Figure 9f).
          A similar trend is apparent
          here, with EW([\oiii]) as a proxy for EW(\ha).}
\label{figure:nii_ha_ratio_v_ewoiii}
\end{figure}

Both \citet{Villar08} and \citet{Cowie11} have shown a strong relationship
between the [\nii]/\ha\ ratio and EW(\ha) for star-forming galaxies at
lower redshifts. For most of our sample we are unable to determine
EW(\ha), but the  [\nii]/\ha\ ratios we find are consistent with EW(\ha) values of $\sim$100--300 as
determined by those authors. Based on the few sources for which we do
have EW(\ha), this is very reasonable. In
Figure~\ref{figure:nii_ha_ratio_v_ewoiii} we plot the [\nii]/\ha\
ratio against EW([\oiii]$\lambda$5007) as a proxy for
EW(\ha). A negative correlation of EW([\oiii]) and
log([\nii]/\ha) is evident. Assuming EW([\oiii]) is a rough
tracer of age, as is the case for EW(\ha), the
increasing [\nii]/\ha\ ratio with decreasing EW([\oiii]) may reflect a
smooth metal buildup with age.

\subsubsection{Mass-metallicity relation}

The relationship between the gas-phase metallicity and the stellar mass of galaxies has
been explored extensively in the literature because it directly constrains
models of the interplay between star formation, pristine gas inflows, chemical
enrichment of the ISM, and feedback
in the form of supernova-driven galactic-scale winds
(e.g., \citealp{Lequeux79, Skillman89, Zaritsky94, Tremonti04, Erb06,
  Henry13b}). It is now well-established that a strong positive correlation exists
between metallicity and stellar mass in both the local
\citep{Tremonti04} and high redshift \citep{Erb06} universe. This
relationship probably reflects the higher gas fractions found in lower mass
systems and/or the relative inability of low-mass galaxies to retain metals
accelerated by supernova-driven winds.  The
mass-metallicity relation at $z\sim2$ was recently explored at lower masses
than previously possible using stacked data from the WISP survey
\citep{Henry13b}. We explore this regime further using our FIRE data. 

On the left panel of Figure~5 we show the 
$N2$ metallicities plotted against the rest-frame V band absolute magnitudes derived from WFC3 H band
photometry (either the F140W or F160W filter), corrected for emission line contamination. There is a clear positive
correlation between metallicity and luminosity. On the right panel of
Figure~5 we show the $N2$-based metallicity estimates
against rough stellar mass estimates from the WFC3
photometry. The masses were estimated by using a M/L ratio in V band
from the \citealp{Bruzual03} (BC03) models assuming constant
star-formation and an age of 100~Myr. The age was chosen to roughly
match the values of EW(\ha) for the sample (the relation between age
and EW(\ha) is well-known; see, e.g., \citealp{Shim11}). The large error
bars primarily reflect the uncertainty in the M/L ratio, which depends
on the uncertainty in the age estimates. This was in turn estimated
from the likely spread in EW(\ha) values for the sample, as only a
handful are directly measured from the WFC3 grism data. 

The strong
mass-metallicity relation we find is in relatively good agreement with
the recent determination by \cite{Henry13b} using stacked WISP data at
similar redshift, although that result shows a somewhat higher
normalization. We also compare to the determination of the
mass-metallicity relation at $z\sim2$ from \citet{Erb06} and find that our points
show a similar trend but tend to be offset upwards from this
relation. This offset
may be due to a combination of the somewhat lower average redshift of points in our
sample and/or a nitrogen enhancement in the emission-line sample,
discussed further in \S6.3. Such an enhancement, if real, could bias
our $N2$ metallicity estimates toward higher values.  

\begin{figure*}[htb]
\centering
  \begin{tabular}{@{}cc@{}}
    \includegraphics[width=.49\textwidth]{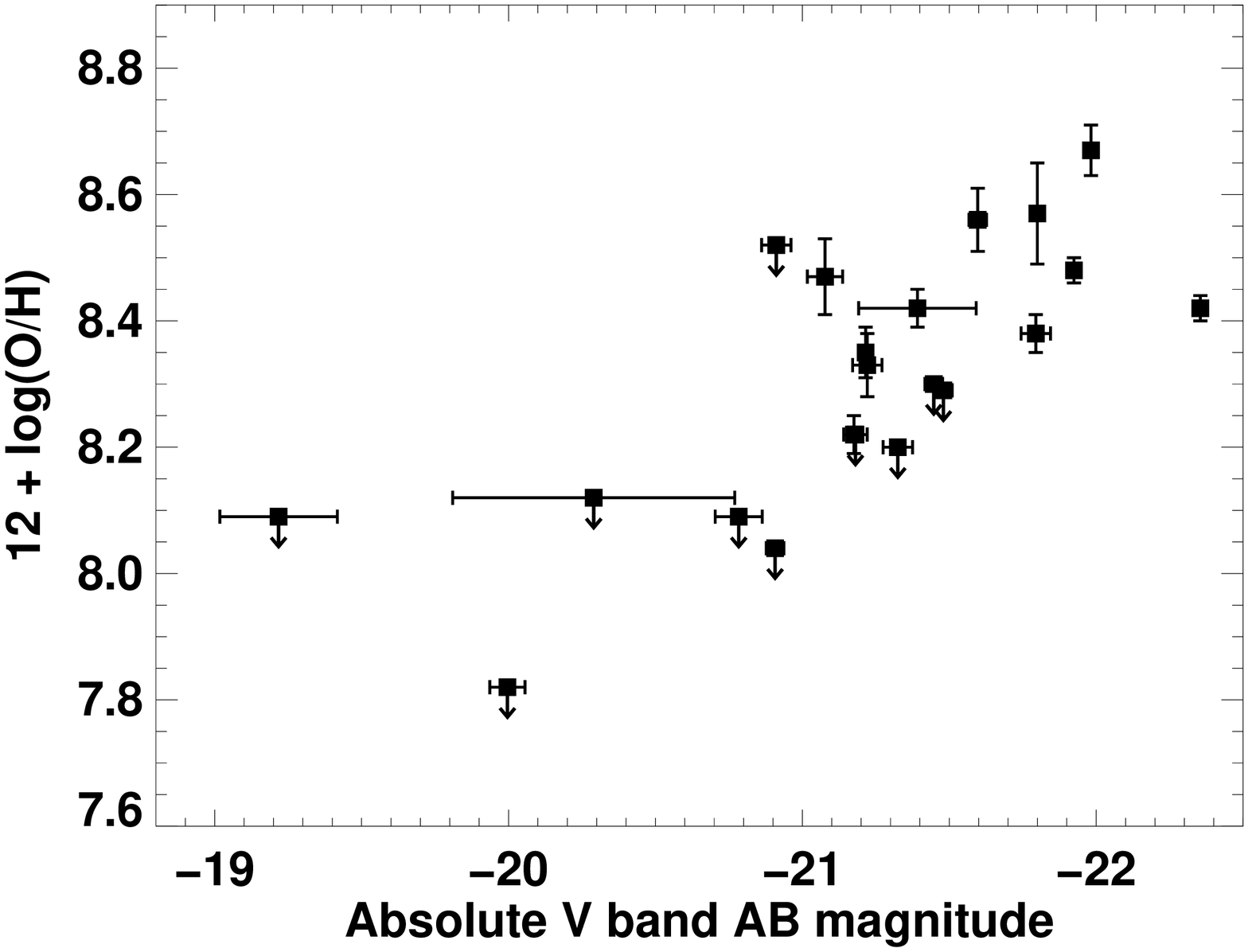} &
    \includegraphics[width=.49\textwidth]{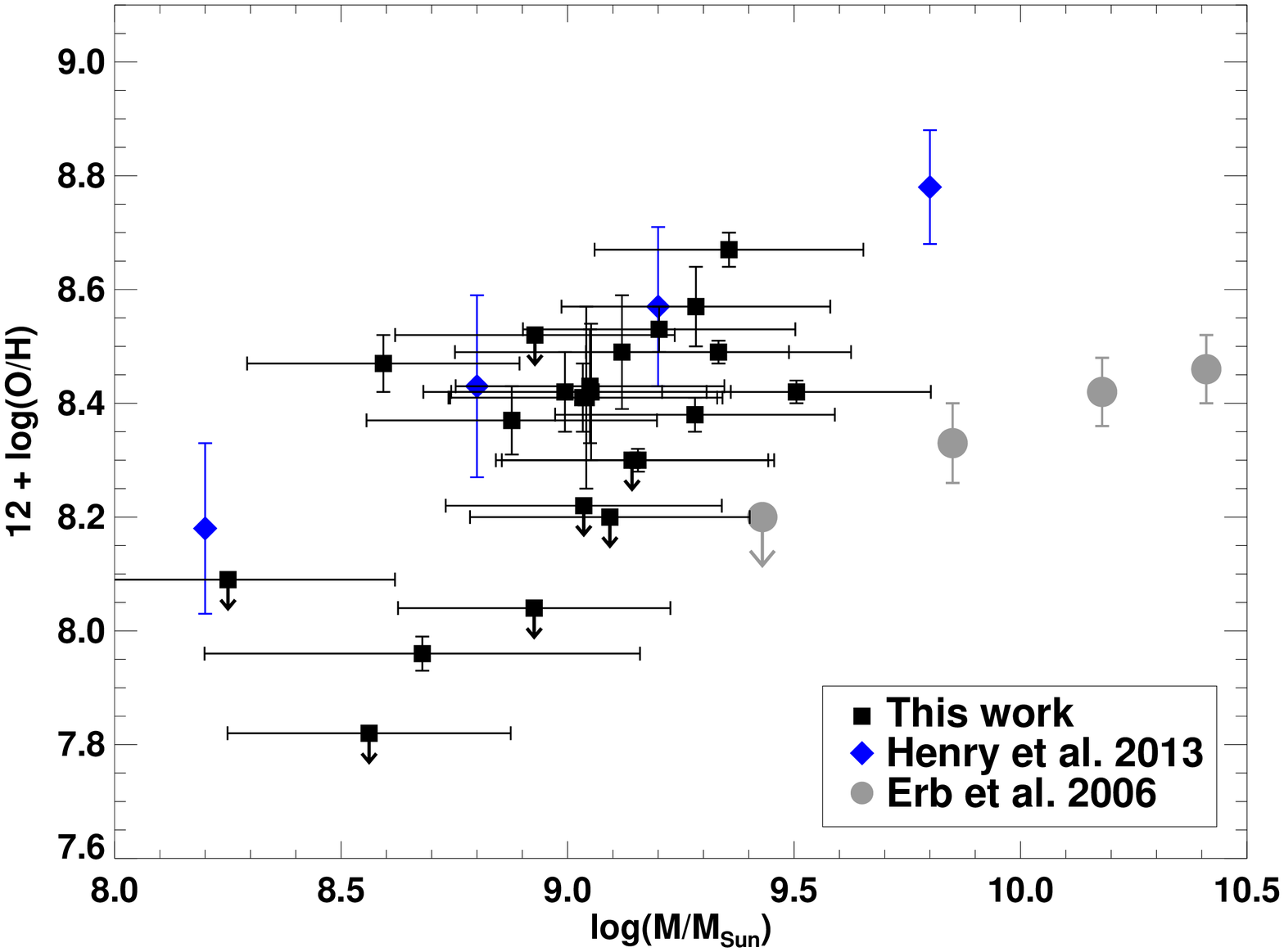} \\
  \end{tabular}
  \label{figure:fire_metallicity}
  \caption{\emph{Left:} The $N2$ metallicity vs. absolute magnitude from WFC3
          H-band photometry, corresponding to rest-frame
          optical. The photometry has been corrected for emission line contamination. \emph{Right:} Rough mass estimates from rest-frame
          optical photometry corrected for emission line flux against
          the $N2$ metallicity estimates. The blue diamonds are from the
         recent determination of the MZ relation by \citet{Henry13b}
         using stacked data from WISP. These values were found using
         \rtwo\, so a systematic offset between the two estimates might
         exist \citep{Kewley08}. Nevertheless, we see relatively good
         agreement. The gray circles are points determined by
         \citet{Erb06}. Our points are offset upward from
         these, which may be due to a combination of a slightly lower
         average redshift for our sample and possible nitrogen
         enhancement in the emission-line selected sample, as discussed in \S6.3. Such an enhancement could
         bias our $N2$ metallicities.}
\end{figure*}

\begin{center}
\begin{deluxetable}{lcccc}
\tabletypesize{\scriptsize}
\tablewidth{0.46\textwidth}
\tablecolumns{5}
\tablewidth{0pt}
\tablecaption{$N2$ and \rtwo\ metallicity estimates and ionization parameters for the sample.}
\tablehead{   
  \colhead{Object ID} &
  \colhead{$Z$\tablenotemark{a}} &
  \colhead{$Z$\tablenotemark{a}} &
  \colhead{$O32$} &
  \colhead{$\mathrm{log}~U$} \\[0.06cm]
  \colhead{} &
  \colhead{($N2$)} &
  \colhead{(\rtwo)} &
  \colhead{} &
  \colhead{} 
}
\startdata

WISP159\_134 & $<8.04$ & $\cdots$ & $\cdots$ & $\cdots$ \\[0.06cm]
WISP134\_171 & 8.35 (0.04) & $\cdots$ &   $\cdots$ & $\cdots$ \\[0.06cm]
WISP50\_65 & $\cdots$ & $\sim8.4$ &  -0.46 (0.19) & -2.98$_{-0.2}^{+0.07}$ \\[0.06cm]
WISP173\_205 & $\cdots$ & $\cdots$ &   0.71 (0.15) & -2.20$_{-0.35}^{+0.63}$ \\[0.06cm]
WISP9\_73  & 8.22 (0.03) & 8.44 (0.04) &   0.28 (0.19)  & -2.77$_{-0.07}^{+0.28}$ \\[0.06cm]
WISP43\_75  & 8.57 (0.08) & $\cdots$ & $\cdots$ & $\cdots$ \\[0.06cm]
WISP25\_53 & 8.48 (0.02) & 8.81 (0.02) &  0.08 (0.25) & -2.91$_{-0.07}^{+0.21}$ \\[0.06cm]
WISP46\_75 & 8.42 (0.03) & 8.87 (0.09) &  -0.34 (0.26)  & -3.26$_{-0.07}^{+0.21}$ \\[0.06cm]
WISP126\_90 & 8.67 (0.04) & $\cdots$ &  $\cdots$ & $\cdots$ \\[0.06cm]
WISP22\_111 & 8.56 (0.05) & $\cdots$ & $\cdots$ & $\cdots$ \\[0.06cm]
WISP22\_216 & $<7.82$ & $\cdots$ & $\cdots$ & $\cdots$ \\[0.06cm]
WISP64\_2056 & $<8.09$ & $\cdots$ &  $\cdots$ & $\cdots$ \\[0.06cm]
WISP81\_83 & $\cdots$ & $\cdots$ &  $\cdots$ & $\cdots$ \\[0.06cm]
WISP138\_173 & $<8.20$ & $\sim8.4$ &   -0.14 (0.21)  & -2.76$_{-0.14}^{+0.14}$ \\[0.06cm]
WISP170\_106 & $<8.30$ & $\sim8.4$ &    0.22 (0.25) & -2.76$_{-0.14}^{+0.21}$ \\[0.06cm]
WISP64\_210 & 8.47 (0.06) & $\sim8.4$ & 0.47 (0.06) & -2.41$_{-0.28}^{+0.14}$ \\[0.06cm]
WISP204\_133 & $<8.52$ & 8.00 (0.11) &  0.12 (0.15) & -2.91$_{-0.07}^{+0.28}$ \\[0.06cm]
WISP27\_95 & 8.38 (0.03) & $\sim8.4$ &  0.08 (0.15) & -2.93$_{-0.07}^{+0.21}$ \\[0.06cm]
WISP147\_72 & $\cdots$ & $\cdots$ &  0.78 (0.30) & -2.06$_{-0.49}^{+0.80}$ \\[0.06cm]
WISP90\_58 & 8.42 (0.02) & 8.62 (0.03) &  0.08 (0.25) & -2.93$_{-0.09}^{+0.28}$ \\[0.06cm]
WISP70\_253 &  $<8.12$ & 7.96 (0.03) & 0.86 (0.07) & -1.88$_{-0.53}^{+0.40}$ \\[0.06cm]
WISP175\_124 & 8.33 (0.05) & $\cdots$ &  0.46 (0.05) & -2.62$_{-0.07}^{+0.35}$ \\[0.06cm]
WISP96\_158 & $<8.22$ &  $\sim8.4$ &  0.41 (0.13) & -2.55$_{-0.21}^{+0.28}$ \\[0.06cm]
WISP138\_160  & 8.29 (0.02) & $\cdots$ &  $\cdots$ & $\cdots$ \\[0.06cm]
WISP56\_210  & $\cdots$ & $\sim8.4$ & 0.61 (0.06) & -1.78$_{-0.60}^{+0.73}$ \\[0.06cm]
WISP206\_261 & $<8.09$ & 8.37 (0.06) &  0.69 (0.13)  & -2.34$_{-0.21}^{+0.56}$

\enddata
\tablenotetext{a}{In units of 12+log(O/H).}
\label{table:metallicity}
\end{deluxetable}
\end{center}
\normalsize

\vspace{-1.0em}
\vspace{-1.0em}

\subsection{Ionization Parameter}

The ionization parameter $q$, defined as the ratio of the flux
of ionizing photons through a unit area to the local
number density of hydrogen nuclei, is frequently used to characterize the
radiation field in \hii\ regions. The effective ionization parameter
is defined in terms of the Str\"{o}mgrem radius $R_{S}\colon$
\begin{equation} q = \frac{Q_{H^{0}}}{4\pi R^{2}_{S} n} \end{equation} 
where $Q_{H^{0}}$ is the ionizing photon flux produced
by the radiation source(s) and $n$ is the local number density of
hydrogen atoms. It is often expressed in terms of the logarithm of the
dimensionless ionization parameter $U$ where \begin{equation}U \equiv \frac{q}{c}\end{equation} The
ionization parameter depends on the nature of the
ionizing radiation source(s) as well as
the density and geometric distribution of the
gas and stars in \hii\ regions. It is useful because it can be measured easily using
the ratio of fluxes from different ionization stages of the same
element, for example $O32\equiv$[\oiii]$\lambda5007$/[\oii]$\lambda$3727.

We used the $O32$ values from the sample
together with the IZI code and priors on the metallicity to derive
ionization parameter estimates. The extinction-corrected values
of $O32$ and inferred log~$U$ values are given in
Table~\ref{table:metallicity}. The ionization parameters for our
sample range from $-3.26<
\mathrm{log}~U<-1.78$, with a mean log~$U$ of
$-2.59$. These values are signficantly elevated with respect to local
samples, for which ionization parameters in the range $-4<
\mathrm{log}~U<-3$ are typical. The ionization parameters we measure are, however, in close agreement with those
measured in other high redshift samples (e.g., \citet{Pettini01}, \citet{Hainline09},
\citet{Richard11}, \citet{Wuyts12}, \citet{Nakajima13}). 

\citet{Erb10} found a very high
ionization parameter of log~$U\sim-1$ for the low mass, low
metallicity, star-forming
galaxy Q2343-BX418 at z=2.3. This galaxy was selected for detailed
study based on its low mass and blue UV-continuum slope. While our sample includes galaxies analogous to
BX418 in many respects, including stellar mass, metallicity, SFR, half-light
radius, and reddening, we do not find any sources with
a comparably high ionization parameter. BX418 may be
somewhat exceptional in this regard. 

\vspace{1.0em}
\subsection{Star Formation Rates}

The star formation rates for the sample were determined using the
extinction-corrected values of $F$(\ha) together with the calibration of
\citet{Kennicutt98} converted to a Chabrier IMF.  In this analysis we have assumed that AGN
contribution is negligible (this issue is discussed further in \S6.1). The results are summarized in
Table~\ref{table:sfrs}. The estimated star
formation rates range from
$\sim5-100~\mathrm{M}_{\odot}~\mathrm{yr}^{-1}$ with an average of
29~$\mathrm{M}_{\odot}~\mathrm{yr}^{-1}$ and a median of 27~$\mathrm{M}_{\odot}~\mathrm{yr}^{-1}$. As noted previously, the
sources with the highest extinction-corrected SFRs tend to be those
sources with more dust extinction overall, but this could be a selection
effect.

\subsection{Line Widths \& Dynamical Mass Estimates}

\begin{figure}[htb]
       \centering
	\includegraphics[scale=0.43]{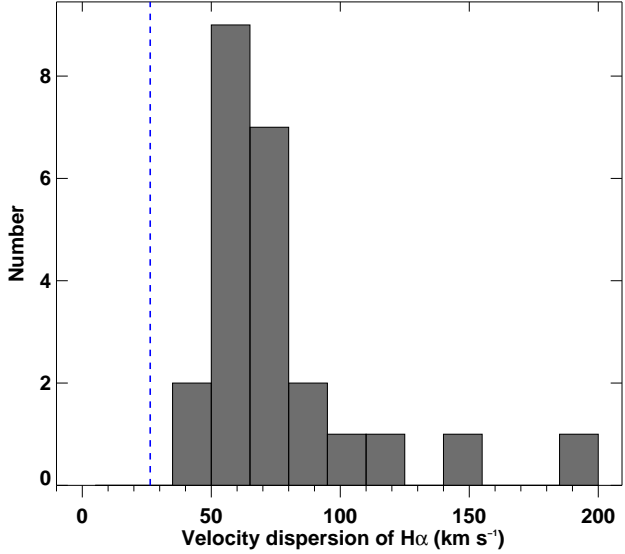}
	\caption{The velocity dispersions of the \ha\ lines. The blue
          dashed line shows the average resolution of FIRE
          (62~km~s$^{-1}$ FWHM = 26.4~km~s$^{-1}$ in sigma). The line velocity dispersions have
          been deconvolved with the instrumental resolution.}
\label{figure:vdisp}
\end{figure}

The emission line velocity dispersions are resolved in the
FIRE spectra and are summarized in
Table~\ref{table:dispersions}.  Figure~\ref{figure:vdisp} shows 
a histogram of the measured dispersions (deconvolved with the
instrumental resolution), which range from 50 to
200~km~s$^{-1}$, but show a peak around
$\sim$70~km~s$^{-1}$.  These dispersions are comparable to those found
for the sample of Lyman-break galaxies at $z\sim3$ analyzed in \citet{Pettini01}.

Comparable (though
generally more massive)
star-forming galaxies at $z\sim2$ have been shown through integral
field studies (e.g. \citealp{Law09}, \citealp{Genzel11})  to have gas kinematics that are
often dominated by turbulent motions with large intrinsic velocity
dispersions. For any particular galaxy it is
difficult to know if the line width results from
rotation-dominated galaxy dynamics, turbulent motions of gas 
within regions of high star formation surface density, or some combination
of these. Two galaxies in our sample (WISP43\_75 and WISP25\_53) show evidence for rotational motion in
the FIRE 2D spectra, but the majority show no clear
spatial gradient in the 2D emission lines. One source, WISP81\_83, shows
double-peaked emission lines separated by 290~km~s$^{-1}$,
likely due to either a small merger in progress or an unresolved disk. WISP43\_75 also shows what
appears to be two-component emission in \ha, though less pronounced
than WISP81\_83. 

\begin{figure}[htb]
       \centering
	\includegraphics[scale=0.35]{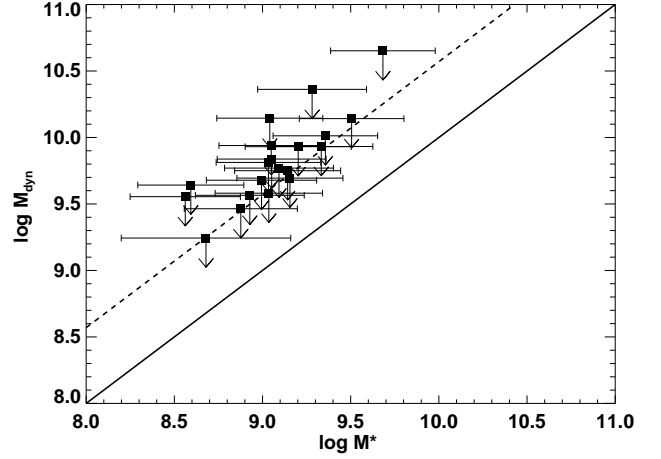}
	\caption{The dynamical mass estimated from
          Equation~3 (in units of M$_\odot$) against the stellar mass
          estimates from the WFC3 photometry. The dynamical mass
          estimates (plotted as upper limits) are systematically higher by $\sim$0.6 dex, which
          may reflect a low ($\sim30\%$) contribution of stars to the
          dynamical masses of the systems. However, the dynamical
          mass estimates are also likely to be elevated by random/turbulent
          motions of the gas affecting the measured line dispersions.}
\label{figure:mdyn}
\end{figure}

Following the analysis of \citet{Maseda13} and references therein, we
can estimate the dynamical mass of galaxies in the sample using the
emission line velocity dispersions and half-light radii from the
formula \begin{equation} M_{dyn} =
  C\frac{r_{\mathrm{eff}}\sigma^{2}}{G} \end{equation} The constant
$C$ is a geometric correction factor that can vary depending on the
assumed shape and orientation of the galaxies. We adopt $C=3$ as in
Maseda et al., with a quoted uncertainty of 33\%. The half-light
radii are measured from the direct H-band images from WFC3, and are
summarized in Table~\ref{table:dispersions}. In
Figure~\ref{figure:mdyn} we plot the dynamical masses obtained in this
way against the rough stellar mass estimates obtained as described in
\S4.2.

The dynamical mass estimates from Equation~3 are plotted as upper limits,
because the line dispersions are a result of both
rotational and turbulent/thermal motions of the gas and the relative
contribution of these is uncertain. Our dynamical and stellar
mass estimates show a strong correlation and are in relatively good
agreement with the 0.57 dex offset found by Maseda et al. between the
two estimates. This offset indicates that stars contribute 
$\sim$27\% to the mass, assuming all of the dispersion is dynamical in origin.
However, random motions of the gas may contribute substantially to the
line dispersions, which would in part explain the offset
between the dynamical and stellar mass estimates.

\begin{center}
\begin{deluxetable}{lccc}
\tabletypesize{\scriptsize}
\tablecolumns{4}
\tablewidth{0.48\textwidth}
\tablecaption{Velocity dispersions and half-light radii for the sample.}
\tablehead{   
  \colhead{Object ID} &
  \colhead{$\sigma$([\oiii])} & 
  \colhead{$\sigma$(\ha)} &
  \colhead{$R_{e}$} \\
  \colhead{} &
  \colhead{(km~$\mathrm{s^{-1}}$)} &
  \colhead{(km~$\mathrm{s^{-1}}$)} &
  \colhead{(kpc)} 
}

\startdata
WISP159\_134 & 45.7 (3.0) & 64.9 (4.3)   &  $\cdots$  \\[0.05cm]    
WISP134\_171 & 111.8 (5.2) & 71.5 (2.1) &  2.4 \\[0.05cm]	      
WISP50\_65 & 152.4 (20.0) & 145.5 (3.8) & 2.7 \\[0.05cm]	        
WISP173\_205 & 65.4 (1.3) & 61.3 (1.4) & 1.5 \\[0.05cm]	       
WISP9\_73 & 85.9 (2.6) & 79.0 (1.0) &  1.5 \\[0.05cm]	      
WISP43\_75 & $\cdots$ & $\cdots$    &      3.2 \\[0.05cm]	      
WISP25\_53 & 74.5 (0.8) & 75.9 (1.2) & 2.2 \\[0.05cm]	      
WISP46\_75 & 66.1 (10.4) & 69.4 (3.9) &  $\cdots$ \\[0.05cm]      
WISP126\_90 & $\cdots$ & 78.1 (1.5)   &   2.4 \\[0.05cm]	      
WISP22\_111 & 80.2 (12.5) & 68.3 (1.9) &  2.6 \\[0.05cm]	       
WISP22\_216 & 40.9 (1.7) & 48.2 (1.7) &  2.2 \\[0.05cm]	      
WISP64\_2056 & 52.8 (1.0) & 58.5 (1.9)  &  $\cdots$ \\[0.05cm]      
WISP81\_83 & $\cdots$ & $\cdots$       &     2.8 \\[0.05cm]	      
WISP138\_173 & 62.3 (2.7) & 60.5 (6.4) & 2.2 \\[0.05cm]	       
WISP170\_106 & 71.4 (4.0) & 55.2 (3.2) & 1.6 \\[0.05cm]	      
WISP64\_210 & 64.8 (1.5) & 74.5 (4.7) &  1.6 \\[0.05cm]	     
WISP204\_133 & 55.3 (8.4) & 54.9 (11.2)  & 1.7 \\[0.05cm]	      
WISP27\_95 & 128.4 (1.3) & 109.6 (3.0)  &  2.0 \\[0.05cm]	     
WISP147\_72 & 202.4 (1.3) & 186.7 (7.2) &  1.6 \\[0.05cm]	      
WISP90\_58 & 110.4 (1.9) & 111.9 (2.9) & 1.6 \\[0.05cm]	    
WISP70\_253 & 39.6 (0.7) & 51.2 (3.2) &  1.6 \\[0.05cm]	      
WISP175\_124 & 71.1 (1.0) & 85.1 (4.5) & 2.0 \\[0.05cm]	    
WISP96\_158 & 47.5 (1.5) & 50.7 (3.2) & 2.4 \\[0.05cm]	       
WISP138\_160 & 66.0 (0.4) & 56.5 (0.8) & 1.6 \\[0.05cm]	       
WISP56\_210 & 97.7 (5.7) & 84.0 (23.3) &  2.1 \\[0.05cm]	      
WISP206\_261 & 46.5 (0.8) & 39.4 (1.7) &   1.9              
\enddata
\label{table:dispersions}
\end{deluxetable}
\end{center}
\normalsize

\vspace{-1.0em}
\vspace{-1.0em}
\subsection{Morphologies}

\begin{figure}[htb]
        \centering
	\includegraphics[scale=0.35]{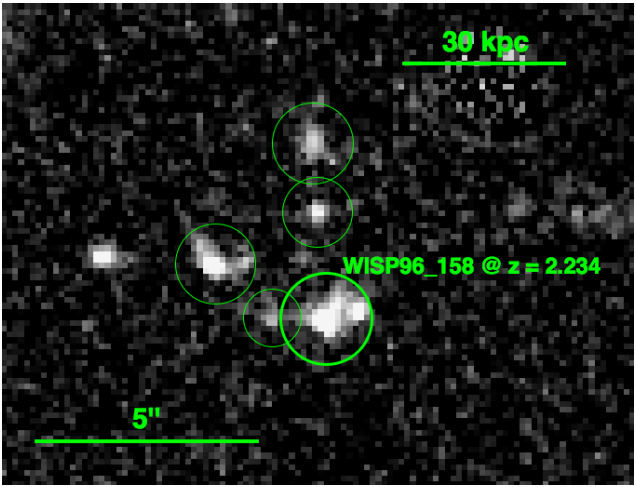}
	\caption{The emission-line
          galaxy WISP96\_158 lies in one of the deepest WISP fields
          (28081 sec and 11430 sec exposures in the G102 and G141
          grisms, respectively,
          and a 4295 sec direct image in the F110W filter, shown). A serendipitous detection of
          [\oiii]$\lambda$5007 in the FIRE 2D spectrum from a
          neighboring galaxy led to the identification
          of the small group of galaxies (circled) at the same
          redshift as WISP96\_158. A significant fraction of strong emission-line
          galaxies at $z\sim2$ are plausibly triggered by small gas-rich mergers in
          systems such as this one.}
\label{figure:smallmerger}
\end{figure}

While detailed morphological modeling of the galaxies is beyond the
scope of this paper, we may make some observations based on
visual inspection of the WFC3 imaging data. The H-band imaging
(corresponding to rest-frame optical) for the galaxies in the sample
reveals that a large fraction ($\gtrsim$40\%) show 
evidence for disruption or asymmetry, probably indicative of merging
or the existence of multiple star-forming clumps. Additionally, most sources are
quite compact, with measured half-light radii typically $\lesssim2.5$~kpc.

An interesting example of a probable merger-driven starbursting galaxy
is WISP96\_158, which lies in one of the deepest WISP pointings. A
serendipitous detection of [\oiii] emission in the FIRE 2D spectrum from a
neighboring galaxy led to the identification of a small group of
objects at the same redshift as the main emission-line
galaxy. In Figure~\ref{figure:smallmerger} we show the unusually deep WISP
F110W exposure, in which we can clearly see the system of small galaxies surrounding
WISP96\_158, which itself appears clumpy and disturbed. It seems
likely that this starburst is
triggered by the interaction/merger of small systems. 

\section{Composite Optical Spectrum}

\begin{figure*}[htb]
        \centering
	\includegraphics[width=0.9\linewidth]{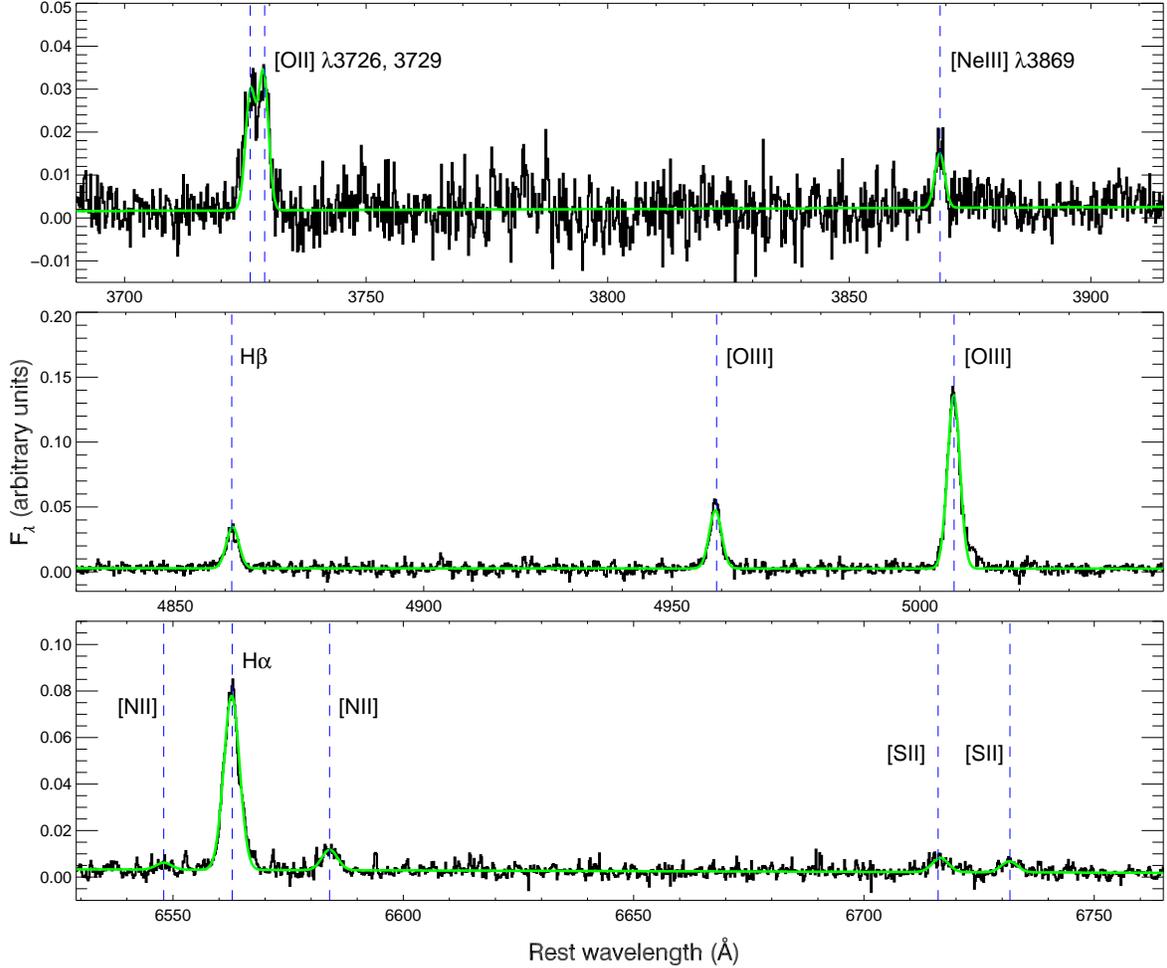}
	\caption{The composite FIRE spectrum of 24 emission-line galaxies from the WISP survey, with an average redshift
          of $\langle z \rangle = 1.85\pm0.39$. The fit is overlaid in
         green.}
\label{figure:composite}
\end{figure*}

We constructed a composite 1D spectrum using the FIRE
spectra, in order to probe the average properties of emission-line
galaxies at $z\sim2$ in greater detail. The composite contains 24
sources from FIRE, with two sources, WISP81\_83 and WISP43\_75, excluded due to
the presence of
multiple velocity components in
their emission lines. To create the composite we shifted the individual 1D spectra to rest
frame, interpolated to a common dispersion of 0.2~\AA, and performed
an inverse-variance weighted average of the flux at each wavelength
after scaling the spectra to a common $F$(\ha). The weighting was done to minimize the contribution of OH
line residuals from individual spectra. To verify that the result 
is not driven by a handful of individual spectra, we also
performed a median combine and an unweighted average, both of which gave nearly
identical results. 

The composite 1D spectrum is
shown in Figure~\ref{figure:composite}. In
Table~\ref{table:composite_fluxes} we give the emission line fluxes
for the composite, normalized to $F$(\hb)=1. 

\subsection{Average Physical Properties from the Composite Spectrum}

The physical properties of the
composite are summarized in Table~\ref{table:compositeproperties}. 
The composite spectrum has a low extinction of $E(B-V)=0.08\pm0.01$ as derived from its Balmer decrement of
\ha/\hb\ = 3.17. We
computed the metallicity of the composite using the $N2$ and \rtwo\ (upper
and lower branch) calibrations, as well as
with the interpolation code IZI. The metallicity estimates are in
relatively good agreement and yield a similar metallicity
($\sim$0.50~Z$_{\odot}$) as found on average from the individual spectra. The ionization parameter for the
composite is log~$U\sim-2.49$, which is also consistent with the estimates
from the individual FIRE spectra. 

The velocity dispersion in \ha\ for the composite spectrum is
78~km~s$^{-1}$. This value is 
in agreement with the $\langle \sigma_{mean} \rangle
= 78\pm17$~km~s$^{-1}$ found for the similar sample analyzed by \citet{Law09}
at $z\sim2$, which comprised 12 star-forming galaxies selected to have well-detected \ha\
or [\oiii] emission.

\subsubsection{Limit on $T_{e}$ from the Composite}
The ratio of [\oiii]$\lambda4959,5007$ to [\oiii]$\lambda4363$ is a
direct temperature indicator as it probes the relative abundance of
electrons with the different excitation energies needed to produce
these lines. [\oiii]$\lambda$4363 is not detected in the composite, but we are able
to put a limit on its value: $F$([\oiii]$\lambda4363$)~$<$~0.11
(2$\sigma$) on a scale in which
$F$(\hb)=1. This gives
an upper limit of $T_{e}<16,800$~K on the average gas temperature
\citep{Osterbrock06}. 

\subsubsection{Electron Density from [\sii] and [\oii]}
Electron densities can be estimated from the collisionally excited forbidden lines [\sii]$\lambda6716, 6731$
and [\oii]$\lambda3726, 3729$, both of which are cleanly detected in
the FIRE composite spectrum.  We measure [\sii]$\lambda6716/\lambda6731 =
1.29\pm0.07$ and [\oii]$\lambda3729/\lambda3726 = 1.16\pm0.13$. These
values are both consistent with
$n_{\mathrm{e}}\simeq100-400$~cm$^{-3}$. This density range is
elevated in comparison with typical \hii\ regions
 in the local universe, which generally have 
$n_{\mathrm{e}}\sim$50--100~cm$^{-3}$ \citep{Brinchmann08,
  Shirazi13}. 

\citet{Shirazi13} argued that the elevated ionization parameters
observed in high redshift star-forming galaxies are a result of
higher electron densities, on the order of $\sim$10$^{3}$~cm$^{-3}$. The
higher average densities we find are consistent with this being a factor driving the higher ionization
parameters at high redshift. However, ionization parameter
scales weakly with electron density: $U \propto
n_{\mathrm{e}}^{1/3}$ \citep{Charlot01}, and the $\sim$4$\times$ higher electron density we find
with respect to local samples is, therefore, probably not the only cause of the $\sim$0.5-1.0~dex
higher ionization parameters in the high redshift emission-line
galaxies. The volume-averaged rate of production of ionizing photons
in these galaxies, which have concentrated, low metallicity star
formation, may be
another major factor producing the high ionization parameters.

\begin{deluxetable}{ll}
\tablewidth{0.38\textwidth}
\tablecolumns{2}
\tablecaption{Line flux measurements from composite spectrum,
  normalized such that $F$(\hb)=1.}
\tablehead{   
  \colhead{Line} &
  \colhead{Value} 
}
\startdata
$F([\mathrm{OII}]\lambda3726)$ & 0.702 (0.064) \\[0.15cm]
$F([\mathrm{OII}]\lambda3729)$ & 0.815 (0.057)  \\[0.15cm]
$F([\mathrm{NeIII}]\lambda3869)$ & 0.324 (0.017) \\[0.15cm]
$F([\mathrm{H}\beta])$ & 1.000 (0.008)  \\[0.1cm]
$F([\mathrm{OIII}]\lambda5007)$ & 4.116 (0.007) \\[0.15cm]
$F([\mathrm{H}\alpha])$ & 3.170 (0.013) \\[0.15cm]
$F([\mathrm{NII}]\lambda6583)$ & 0.372 (0.010) \\[0.15cm]
$F([\mathrm{SII}]\lambda6716)$ & 0.264 (0.010) \\[0.15cm]
$F([\mathrm{SII}]\lambda6731)$ & 0.205 (0.009) 

\enddata
\label{table:composite_fluxes}
\end{deluxetable}
\normalsize

\begin{center}
\begin{deluxetable}{ll}
\tablewidth{0.38\textwidth}
\tablecolumns{2}
\tablecaption{Physical parameters derived from the composite spectrum.}
\tablehead{   
  \colhead{Parameter} &
  \colhead{Value} 
}
\startdata
$\sigma_{(\mathrm{H}\alpha)}$ &  78 km s$^{-1}$ \\[0.15cm]
$\sigma_{(\mathrm{[OIII]})}$ &  75 km s$^{-1}$ \\[0.15cm] 
H$\alpha$/H$\beta$ & 3.17 (0.03) \\[0.15cm] 
$E(B-V)_{\mathrm{nebular}}$ & 0.08 (0.01) \\[0.15cm]
12+log(O/H) ($N2$) & 8.37 (0.01) \\[0.15cm]
12+log(O/H) ($R23$, lower) & 8.20 (0.01) \\[0.15cm]
12+log(O/H) ($R23$, upper) & 8.68 (0.01) \\[0.15cm]
12+log(O/H) (IZI) & 8.46 (0.08) \\[0.15cm]
$O32$\tablenotemark{a} & 0.39 (0.05)
\\[0.15cm]
log~$U$ & -2.49$^{+0.07}_{-0.14}$
\enddata
\tablenotetext{a}{Corrected for dust extinction using Balmer decrement.}
\label{table:compositeproperties}
\end{deluxetable}
\end{center}
\normalsize


\begin{figure*}[htb]
\centering
  \begin{tabular}{@{}cc@{}}
    \includegraphics[width=.50\textwidth]{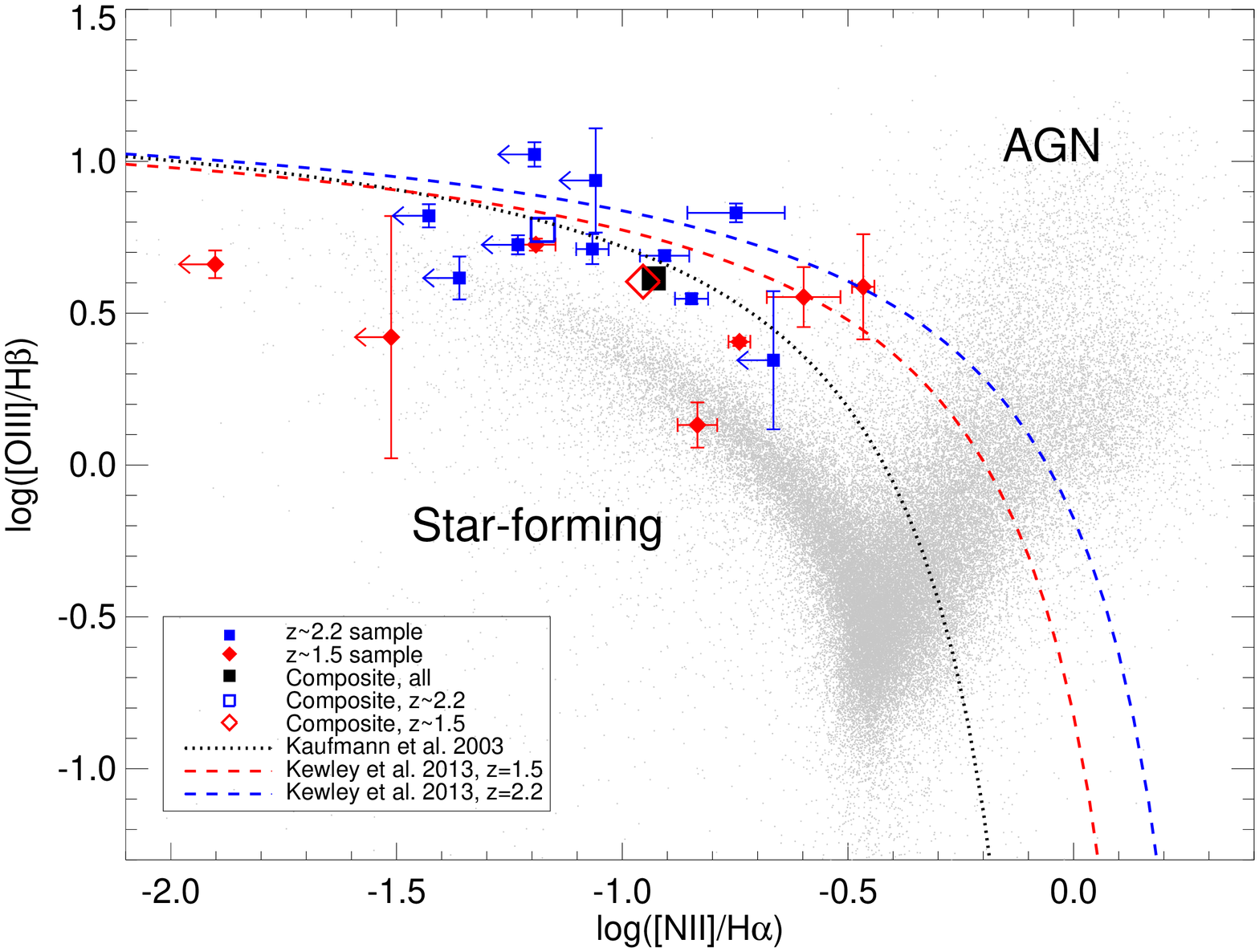} &
    \includegraphics[width=.50\textwidth]{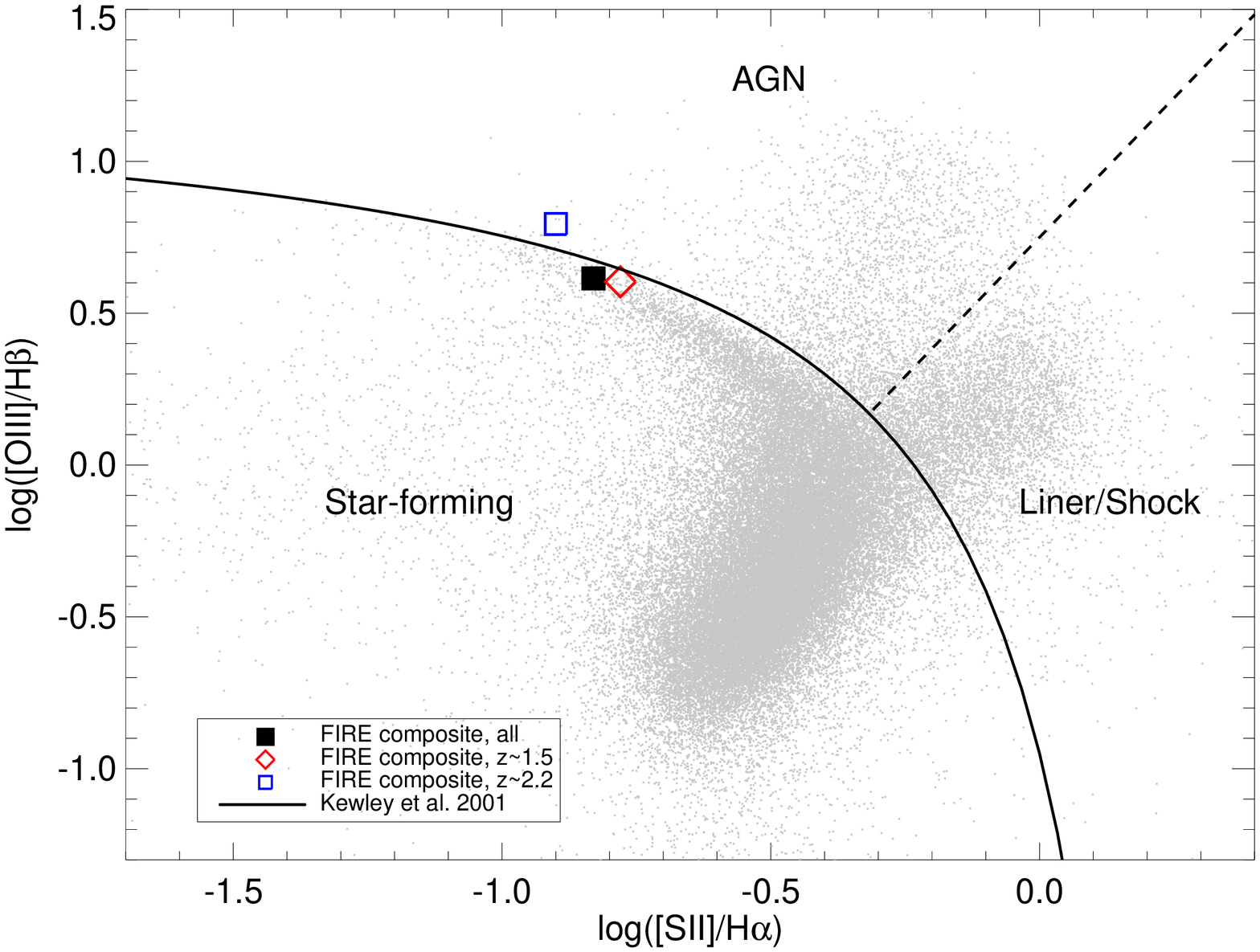} \\
  \end{tabular}
  \label{figure:fire_bpt_diagrams}
  \caption{\emph{Left:} The BPT diagram for the galaxies in the
          sample for which we could measure the required
          lines.  The small gray points are 60,000 sources from SDSS
          for comparison. Upper
          limits on log([\nii]/\ha) at the 95\% confidence level are given for those
          sources for which [\nii]$\lambda6583$ was not detected. The
          black square and open symbols are the result for the composite
          FIRE 1D spectra (note that the error bars for these are
          smaller than the symbols).
          The [\oiii]/\hb\ ratio for our
          sources is systematically
          higher than for star-forming galaxies in the local universe,
          a trend that has been noted previously. We show the empirical maximum starburst line
          from \citet{Kauffmann03}. Recently,
          \citet{Kewley13} proposed a redshift-dependent formulation
          of the starburst/AGN dividing line, shown here
          for $z=1.5$ and $z=2.2$. The majority of our points are
          consistent with being highly star-forming galaxies according to
          these demarcations. \emph{Right:} The BPT diagnostic diagram using the ratio [SII]/\ha\
          in place of [NII]/\ha\ \citep{Veilleux87}. Our composite
          spectrum from FIRE falls along the metal-poor tail of the
          local star-forming sequence from SDSS. No
          evidence for significant AGN or LINER/shock contribution to
          the emission lines is evident from this diagnostic.}
\end{figure*}

\vspace{-1.0em}
\vspace{-1.0em}

\section{Analysis}

\subsection{BPT Diagram and AGN Contribution}

Baldwin, Phillips \& Terlevich (BPT) diagrams \citep{Baldwin81} use
emission line ratios to distinguish star-forming galaxies from
AGN. We were able to measure the four lines required for the
[\oiii]/\hb\ vs. [\nii]/\ha\ BPT diagnostic 
for nine galaxies in our sample, and eight more for which a
meaningful limit on $F$([\nii]) could be placed.  The resulting BPT plot
is shown on the left panel of Figure~11. Galaxies that are not shown
had one or more of the relevant emission lines contaminated by
an OH sky line. 

The position occupied by the
overall composite spectrum is shown as a black filled square. Additionally, we
generated composite spectra at $z\sim1.5$ and $z\sim2.2$ using the
procedure described in the previous section and measured
the diagnostic ratios from them in order to search for possible evolution
over this redshift interval. It should be noted that all of these composites
contain additional objects that we were not able to include
individually in the plot. 

A clear feature of this plot is the average upward offset of [\oiii]/\hb\ for a given [\nii]/\ha\ of the $z\sim2$ sample relative to local
star-forming galaxies. This trend has been noted in previous studies
\citep{Shapley05, Erb06, Liu08, Hainline09, Newman13}, and has been
linked to higher ionization parameters, on average, in star-forming
galaxies at high redshift (e.g. \citealp{Brinchmann08,
  Kewley13}). The one galaxy in the sample that does fall on the local star-forming
sequence in this diagram, WISP46\_75, also has the lowest measured
ionization parameter.

Recently \citet{Kewley13} presented a
redshift-dependent parameterization of the AGN/starburst boundary line,
motivated by the observational evidence for systematic offsets in the
diagram for galaxies at high redshift. These lines are plotted on the
left panel of
Figure~11 for $z=2.2$ and $z=1.5$. It can be
seen both from the individual galaxies and the composites 
 that the sources at $z\sim2.2$ tend to have higher
[\oiii]/\hb\ ratios in comparison with those at $z\sim1.5$. The composite spectra are in fact consistent with being ``maximal
starbursts'' according to the empirical demarcation of
\citet{Kauffmann03}, while falling well above the locus of star-forming
galaxies in the local universe. The $z\sim2.2$ composite is offset to the
upper left of the $z\sim1.5$ composite, hinting at an
evolutionary trend over this redshift interval. However, this may be (at
least in part) due to a selection effect resulting from the fact that
the $z\sim2.2$ sample was selected on the basis of [\oiii] emission, while the
$z\sim1.5$ sample was often selected on the basis of \ha\ emission
without regard to the strength of [\oiii]. 

We plot the positions of the FIRE composite
spectra on the log([\oiii]/\hb) versus log([\sii]/\ha)
diagnostic \citep{Veilleux87} on the right panel of
Figure~11. The $z\sim2.2$ composite is again
offset to the upper left from the $z\sim1.5$ composite. Notably,
there is no obvious offset of the overall composite spectrum in this
diagram with respect to the local, metal-poor
star-forming sequence from SDSS. 

The elevated [\oiii]/\hb\ ratios in the classical BPT diagram could be
at least partially due to AGN activity in such galaxies
\citep{Trump11, Trump13, Kewley13a}. While a few objects in our sample fall
close to the composite region of the BPT diagram, most are consistent
with a purely star-forming classification. The positions of the
composite spectra in the BPT diagrams, in particular,
argue against signficant AGN contribution. Nevertheless, caution is required in
interpreting these diagrams at high redshift. For example, \citet{Kewley13a} showed
that the effect of a low-metallicity AGN on the [\oiii]/\hb\ versus
[\nii]/\ha\ diagram can be indistinguishable from the offset caused by more extreme
star-forming conditions. However, the position of the composites on the [\oiii]/\hb\ versus
[\sii]/\ha\ diagram, as well as the relatively low
($\sim$75~km~s$^{-1}$) average velocity dispersion of the emission
lines, also disfavor significant AGN contribution. We conclude that AGN
activity in high-EW emission-line galaxies at $z\sim2$ is likely to be
minimal.

\subsection{High [\oiii]/\ha\ and [\oiii]/\hb\  Ratios}

A significant fraction of galaxies in the our sample show quite high
[\oiii]$\lambda$5007 emission relative to the recombination lines \ha\
and \hb. Both \citet{Dominguez13} and \citet{Colbert13}
also noted the relatively large number of sources in the WISP survey with high
[\oiii]/\ha\ ratios and found a negative
correlation between the ratio [\oiii]/\ha\ and \ha\ luminosity. High
[\oiii]/\hb\ ratios were also found to be characteristic of
Lyman-break selected galaxies at $z\sim3.5$ by \citet{Holden14}.

The [\oiii]/(\ha,\hb) line ratios we find for a number of galaxies in our sample are higher than
the maximum values predicted by photoionization model grids for low-metallicity galaxies (e.g. 
\citealp{Levesque10}). The $O32$ ratios for these sources indicate
high ionization parameters, but not high enough to account for the
[\oiii]/(\ha,\hb) ratios. To investigate this further we made composite FIRE spectra of subsamples
with high and low [\oiii]/\ha\ ratios, with the dividing ratio being
[\oiii]/\ha=1.55 to give a roughly equal number of galaxies in each composite. The physical parameters derived from these composite
spectra are summarized in Table~\ref{table:separate_composite_properties}.
A few things emerge from these composites:

\begin{enumerate}
\item{The cut on the ratio [\oiii]/\ha\ largely splits in a similar
    way as the cut on redshift, with high [\oiii]/\ha\ ratios more
    common among the $z\sim2.2$ sample. As mentioned previously, this
    could be a selection effect given that the $z\sim2.2$ sample was
    selected based on [\oiii] emission.}
\item{The [\nii]/\ha\ ratio is lower for
the composite of galaxies with high [\oiii]/\ha, most likely
indicative of lower
metallicity for those sources, although a higher ionization
parameter can also lower this ratio. The correlation of [\oiii]/\ha\ and
metallicity has been noted previously \citep{Liang06,
 Colbert13}, and is related to the shape of the star-forming sequence
in the BPT diagram.}
\item{The sources with higher [\oiii]/\ha\ have significantly
stronger [\neiii]/[\oii] and [\oiii]/[\oii] ratios than those with
lower [\oiii]/\ha. Both of these ratios are sensitive to ionization
parameter \citep{Levesque13}, suggesting (as expected) that the higher ionization parameter is
a key factor in elevating the [\oiii]/\ha\ ratio.}
\item{Roughly the same electron density
 is inferred for both composites from the density-sensitive doublets
 [\oii]$\lambda$3726,3729 and [\sii]$\lambda$6716,6731, although there
 is considerable uncertainty in the measurements.}
\end{enumerate}

\begin{center}
\begin{deluxetable}{lcc}
\tabletypesize{\scriptsize}
\tablewidth{0pt}
\tablecolumns{3}
\tablecaption{Comparison of the properties of composite spectra binned
  by the ratio [\oiii]/\ha.}
\tablehead{   
  \colhead{Parameter} &
  \colhead{[\oiii]/\ha\ $>$ 1.55} &
  \colhead{[\oiii]/\ha\ $<$ 1.55}
}
\startdata
12+log(O/H) ($N2$) & 8.22 (0.04) & 8.34 (0.03)  \\[0.15cm]
12+log(O/H) (Model grid) & 8.30 (0.03) & 8.46 (0.10) \\[0.15cm]
O3O2 & 0.59 (0.05) & 0.26 (0.09) \\[0.15cm]
Ne3O2 & -0.52 (0.09) & -0.74 (0.08) \\[0.15cm]
log~$U$ & -2.1 (0.2)   & -2.6 (0.2) \\[0.15cm]
[\sii]$\lambda$6716/[\sii]$\lambda$6731 &  1.28 (0.32)  & 1.34 (0.18)
\\[0.15cm]
[\oii]$\lambda$3729/[\oii]$\lambda$3726 &  1.09 (0.33)  & 0.95 (0.35)
\enddata
\label{table:separate_composite_properties}
\end{deluxetable}
\end{center}
\normalsize


Overall, what we infer from this comparison is that the sources with elevated [\oiii]/\ha\ and [\oiii]/\hb\ ratios
have higher ionization
parameters, as expected, but that the difference seems more linked to their 
lower metallicity than a difference in electron density. The relationship between ionization parameter and metallicity is also
known to exist for local
star-forming galaxies \citep{Dopita00, Kewley02}. Metallicity influences ionization parameter through
 its effect on stellar radiation: at low metallicity the UV flux from
 metal-poor stars is more intense due
to decreased metal-line blanketing in stellar atmospheres. Moreover,
rotational mixing in massive, rotating stars can lead to more
efficient mass loss and facilitate the onset of the Wolf-Rayet phase
and a corresponding hardening of the spectrum; 
this effect is predicted be more pronounced in low-metallicity environments
\citep{Leitherer08, Levesque12, Kewley13a}. While electron density does not
differentiate the two composites, both have higher electron
densities than local samples. 

\subsection{What drives the BPT diagram offset?}

The upward offset of many high redshift star-forming galaxies in the
[\oiii]/\hb\ versus [\nii]/\ha\ diagram relative to the local
star-forming sequence has been
interpreted as resulting from more extreme star forming
conditions in the early universe. \citet{Brinchmann08} showed that the
offset from the ridgeline in the BPT diagram correlates closely with
EW(H$\alpha$) in SDSS, and from this inferred that higher ionization
parameter (which likely is related to EW(H$\alpha$)) is behind the effect.

The galaxies in our sample do have elevated
ionization parameters relative to local
samples, and show the offset in the [\oiii]/\hb\ versus
[\nii]/\ha\ diagram seen in other high
redshift star-forming galaxies. However, the composite spectrum from
the FIRE sample does not show
an analogous offset from the local metal-poor star-forming sequence
in the [\oiii]/\hb\ versus [\sii]/\ha\ diagram, which is 
somewhat puzzling if more extreme star-forming
conditions are the root of the offset in the [\oiii]/\hb\ versus
[\nii]/\ha\ diagram. Moreover, the ionization parameters we measure
for the sample, corresonding to log~$q$ values of $\sim$7.5-8.5, are
simulated in up-to-date photoionization models
(e.g., \citealp{Dopita13}), and 
generally do not produce as large of an offset in the [\oiii]/\hb\ versus
[\nii]/\ha\ diagram as what is observed. 

Thus it is worth considering alternative
explanations for the offset of high redshift star-forming galaxies
from the local star-forming sequence in the [\oiii]/\hb\ versus
[\nii]/\ha\ diagram. One possible cause of such an offset is a higher average
nitrogen-to-oxygen abundance ratio \citep{Perez09}. The
N/O ratio for galaxies is roughly flat (though with significant scatter) at low
metallicities, with a plateau value of $\mathrm{log(N/O)}\sim-1.5$, and
increases at higher metallicities due to the
secondary production of nitrogen through the CNO cycle \citep{Henry00}. Cases of
enhanced nitrogen abundance have been found in starburst galaxies in
the local universe. For example, the
blue compact starburst galaxy NGC~5253 is known to have regions of 3$\times$
higher nitrogen abundance than normally found in metal-poor galaxies
\citep{Welch70, Walsh87, Kobulnicky97}, which seem to
be associated with super star clusters
\citep{Westmoquette13}. \citet{Andrews13}, using stacked spectra from
SDSS, found that the log(N/O) ratio tends to be higher at a given
metallicity for galaxies with higher SFRs.  
Additionally, \citet{Amorin10} presented evidence that the so-called ``green pea''
galaxies found in SDSS (which, with strong [\oiii]
emission lines and compact sizes, may be low-redshift analogues of the galaxies in our
$z\sim2$ sample) show systematically high
log(N/O) values for their metallicities. 

\begin{figure}[htb]
        \centering
	\includegraphics[scale=0.43, trim=0.4in 0.in 0.in 0.in]{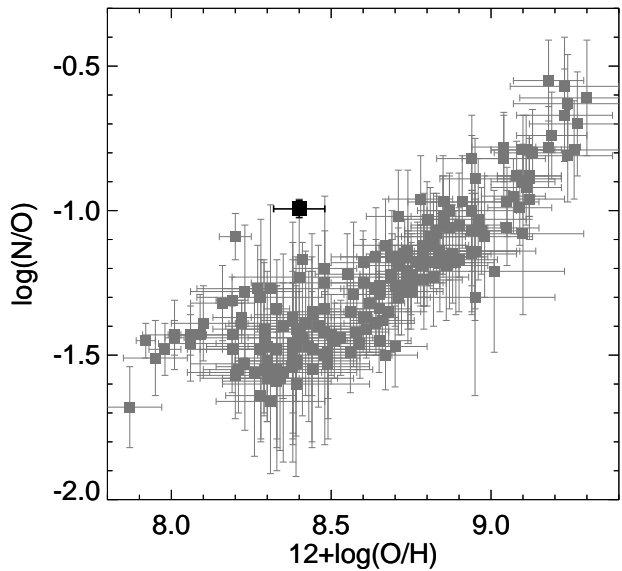}
	\caption{The nitrogen-to-oxygen abundance ratio vs. metallicity computed from the FIRE
          composite spectrum (black square) compared with values from local \hii\
          regions \citep{Vanzee98}.}
\label{figure:nitrogen}
\end{figure}

To investigate this possibility, we used the calibration of
\citet{Thurston96}, based on the dust-corrected ratio log(N$^{+}$/O$^{+}$), to compute the value of
log(N/O) for the composite FIRE spectrum. We adopted a nebular temperature of
10,000~K, yielding a value of log(N/O)~=~$-0.99\pm0.03$. For higher
nebular temperatures log(N/O) increases, so we consider this to be a
relatively conservative lower limit\footnote[8]{If, rather than use the
  calibration of \citet{Thurston96}, we assume that
log(N/O)$\simeq$log(N$^{+}$/O$^{+}$) (as is often done in the literature), we obtain a higher value: log(N/O)=$-0.60\pm0.06$.}. We also
computed log(N/O) using the ratio
N2S2~=~log([\nii]/[\sii]) as in \citet{Amorin10},
yielding log(N/O)~=~$-0.94\pm0.05$. Comparing these results to the measured
log(N/O) values as a function of 12+log(O/H) in the literature
\citep{Vanzee98, Liang06, Vanzee06, Berg12, Andrews13}, we find that the
log(N/O) of the FIRE composite 
is higher by approximately 0.4 dex than the
average value for galaxies at similar
metallicity locally. The offset is illustrated in
Figure~\ref{figure:nitrogen}, in which we show the FIRE
composite spectrum in comparison with the data from local \hii\
regions from \citet{Vanzee98}. 

From this analysis, it seems plausible that the
offset of star-forming galaxies
at high redshift in the  [\oiii]/\hb\ versus
[\nii]/\ha\ diagram is, to some extent, an offset in [\nii]/\ha\ caused by
systematically high N/O ratios. We now consider
why this might be the case.

One mechanism for producing a nitrogen enhancement is through the 
inflow of metal-poor gas. Inflows can lower the metallicity (O/H) of
a galaxy with a high N/O abundance ratio (from secondary nitrogen production) while preserving the high N/O ratio
\citep{Vanzee98, Koppen05}. We consider this unlikely to be the
explanation of 
enhanced N/O in emission-line galaxies at $z\sim2$, which are
relatively young and metal-poor, and thus would not have had time to
enter the secondary nitrogen production regime. 

A more likely cause of the nitrogen enhancement, in our view, is a
significant population of
Wolf-Rayet (W-R) stars in the $z\sim2$ emission-line galaxies. The connection between nitrogen-enriched galaxies and
the influence of W-R stars has been known for some time
\citep{Pagel86, Henry00}, and recent evidence has been found that low-redshift
galaxies with W-R features in their optical spectra
often display enhanced nitrogen abundances \citep{Brinchmann08a,
 Lopez10, Berg11}. However, the effect is generally smaller than the
enhancement we find in the $z\sim2$ sample. 

\citet{Kobulnicky97} suggested that the
significantly nitrogen-enhanced regions of NGC~5253 (comparable to the enhancement we see
in the $z\sim2$ sample) result from the winds of W-R and/or
related luminous blue variable (LBV)
stars. \citet{Westmoquette13} argued that the hot winds of these stars
are able to mix with the
cooler surrounding ISM because of the unusually high pressure
environment the stars inhabit. This explanation may hold for 
the high redshift emission-line galaxies as well, which have higher
nebular gas densities than local galaxies and probably harbor very dense star
clusters to produce the observed emission
lines. 

A potential problem with this explanation is that the fraction of massive stars entering the W-R phase
should decline with metallicity, so one would expect that the ratio of
W-R to O stars would be lower in galaxies at high redshift. However,
the metallicity of NGC~5253, $\mathrm{12+log(O/H)}\sim8.3$, is comparable to that of the galaxies in
our emission-line sample at $z\sim2$, and \citet{Brinchmann08a} also
find a large number of low-metallicity galaxies with W-R
features. Moreover, the ratio of 
nitrogen- (WN) to carbon (WC)-sequence W-R stars increases strongly
at lower
metallicities \citep{Crowther07}, and WN-sequence stars would be most
responsible for ejecting nitrogen into the ISM, leading to the
observed enhancement.

It should be noted that evidence already exists for significant numbers of W-R
stars in star-forming galaxies at high redshift. The composite
rest-frame UV spectrum of $z\sim3$ Lyman-break galaxies
presented in \citet{Shapley03} shows relatively strong and 
broad \heii$~\lambda1640$ emission, as does 
the rest-frame UV spectrum obtained by \citet{Erb10} for the
low-metallicity galaxy BX418 at $z=2.3$. This feature is attributed to
the fast, dense winds of W-R stars. As discussed by those authors,
reproducing the \heii$~\lambda1640$ emission is difficult with existing stellar population synthesis
models. In order to do so, very young starburst ages, a  
top-heavy initial mass function (IMF), or physical effects due to binary stellar
evolution must be invoked. Clearly, there is still much that is not
fully understood about
the star formation in these environments.

\vspace{2.0em}
\section{Summary \& Discussion}
We have presented rest-frame optical spectra taken with Magellan FIRE of a sample
of  26 emission-line galaxies at $z\sim1.5$ and $z\sim2.2$
selected from the HST-WFC3 grism spectroscopy of the WISP survey. The
FIRE spectra provide significant additional information about these
galaxies, allowing us to measure metal abundances, dust reddening,
kinematics, star formation rates and important diagnostic line
ratios. We also created composite spectra that reveal
further details about the average properties of the sample.

Direct WFC3 imaging shows the sources to be compact,
with many ($\gtrsim$40\%) exhibiting clumpiness or asymmetry. The galaxies are low mass
($\sim$10$^{8.5}$--10$^{9.5}$~M$_{\odot}$), with a median metallicity
of 0.45~Z$_{\odot}$. A clear mass-metallicity relation is found for
the sample. As seen in other high-redshift samples, the ionization parameters of the galaxies in our sample
are significantly higher, on average, than those found for typical
star-forming galaxies in the local universe. The velocity dispersions
of the emission lines for the
sample range from $\sim$50--200~km~s$^{-1}$ but usually are around 75~km~s$^{-1}$. Assuming that the
dispersions are almost entirely due to gravitational motions (which is
probably not the case), this implies that
stars make up roughly 30\% of the dynamical mass of the galaxies. 

The composite spectra we generated from the FIRE sample reveal more
detail about the average properties of emission-line galaxies at
$z\sim2.2$ and $z\sim1.5$.  From the overall composite we infer an
average electron density for
the star-forming regions of 100-400~cm$^{-3}$ and put a limit on the average
temperature in these regions of $T_{e}<16,800$~K from the
non-detection of [\oiii]$\lambda4363$. The locations of galaxies and composites on the BPT
diagnostic diagrams favor a starburst classification,
and we find little evidence for substantial AGN contribution.

Our sample includes galaxies with unusually high [\oiii]/\ha\ and [\oiii]/\hb\
line ratios that are not reproduced with existing
photoionization models. Composite spectra sorted on the ratio [\oiii]/\ha\ indicate that those sources with high
[\oiii]/\ha\ ratios show stronger [\oiii]$\lambda$5007 and
[\neiii]$\lambda$3869 emission relative [\oii]$\lambda$3727,
indicating higher ionization parameters. The high
[\oiii]/(\ha,\hb) sources tend to be lower metallicity, which seems to be the main
factor driving their higher ionization parameters. Reproducing the
line ratios observed in such objects may require modifications to
photoionization models to better account for the star-forming
conditions in high redshift galaxies. 

The well-known offset of high redshift star-forming galaxies from the local star-forming sequence in the [\oiii]/\hb\ versus [\nii]/\ha\ diagnostic diagram is
observed in our sample, but we do not find a similar offset in the [\oiii]/\hb\ versus [\sii]/\ha\
diagram. We therefore investigated the possibility that the offset of
high-redshift galaxies in the
[\oiii]/\hb\ versus [\nii]/\ha\ diagram may result, at least in part,
from elevated nitrogen abundances. The composite FIRE spectrum was used to infer the N/O
abundance ratio, which we find to be $\sim$0.4 dex higher in
comparison with local galaxies of similar metallicity. We speculate
that an elevated
nitrogen abundance in high redshift star-forming galaxies may be
common, and could be explained by the presence of a substantial population
of Wolf-Rayet stars embedded in super star clusters in such galaxies.

\acknowledgments

We thank the anonymous referee for thoughtful comments and suggestions
that significantly improved this paper, and Evan Skillman for a
careful reading and helpful comments. We are also grateful to Rob Simcoe, Lisa Kewley, Ryan Quadri, Daniel Kelson and
Louis Abramson for discussions regarding the FIRE
observations and science. DCM gratefully
acknowledges the support provided by the Carnegie
Observatories graduate research fellowship, 
as well as the excellent support provided by the staff of the Las
Campanas Observatory. CLM acknowledges support from NSF AAG 1109288.

\bibliographystyle{apj}
\bibliography{biblio}

\appendix
\section{WFC3 data and FIRE spectra}
Here we show the WFC3 grism spectra, direct H-band images, and
measured lines from FIRE spectroscopy for the sample.

\begin{figure}[htb]
\centering
  \begin{tabular}{@{}cc@{}}
    \includegraphics[width=.47\textwidth]{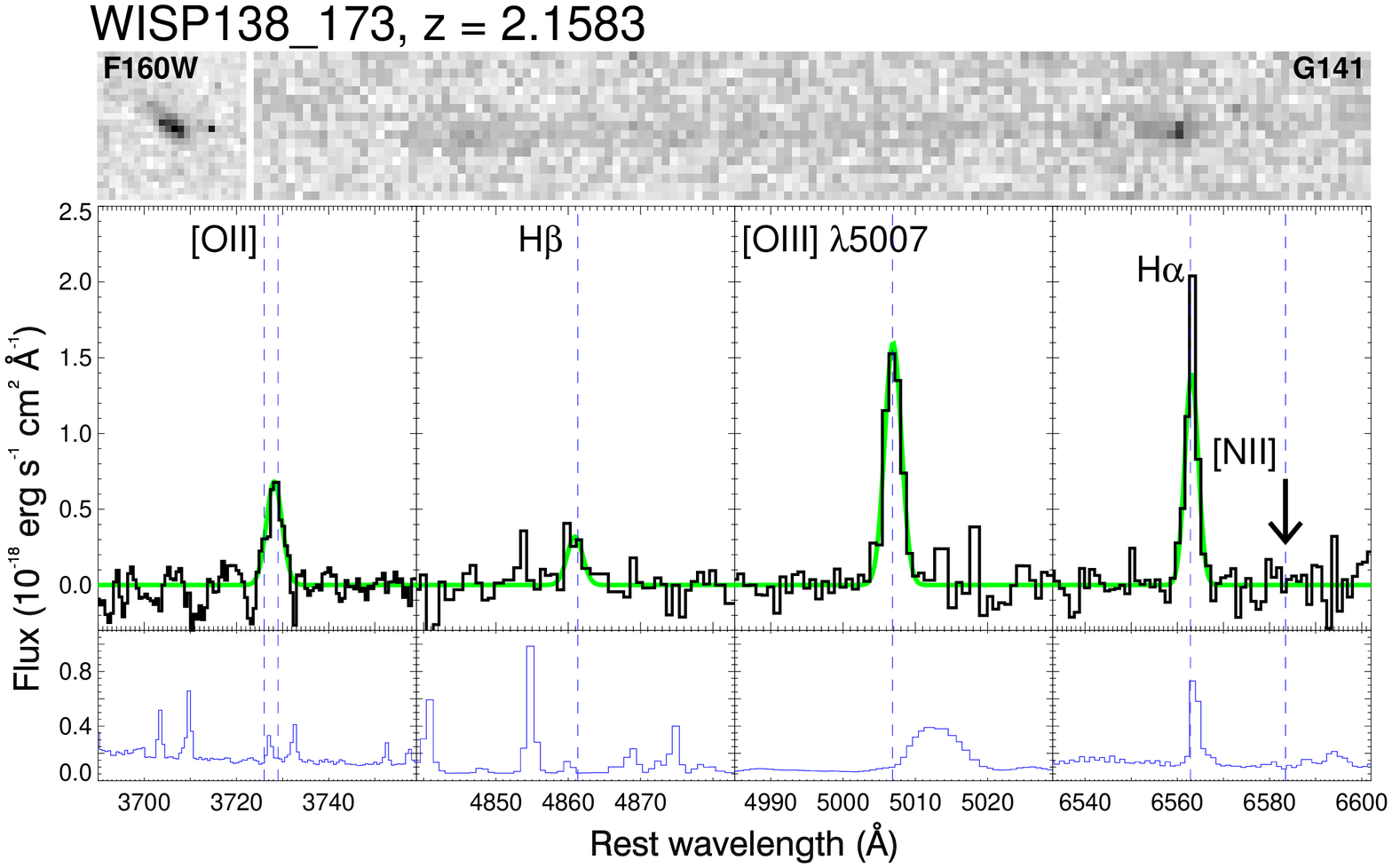} &
    \includegraphics[width=.47\textwidth]{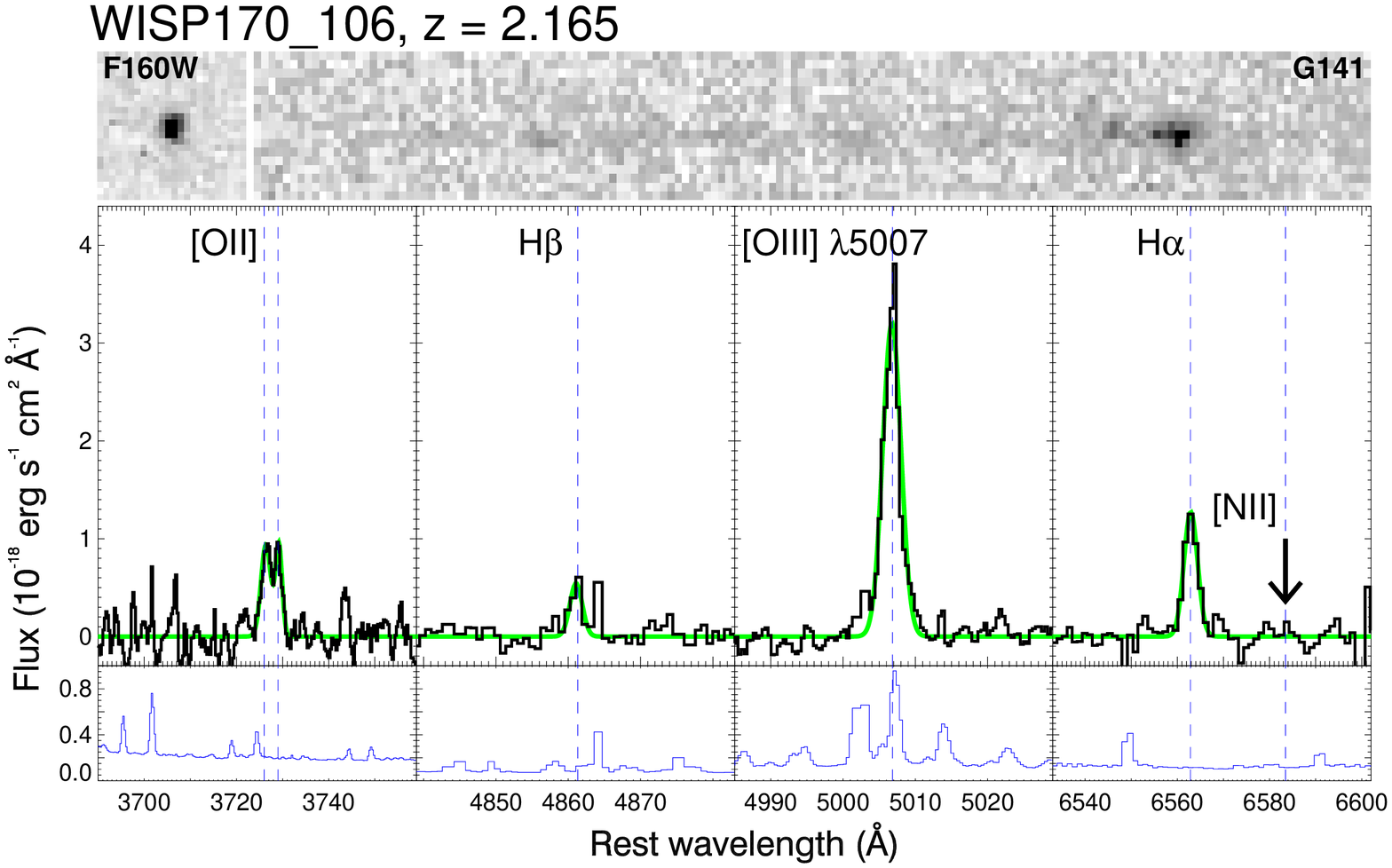} \\
    \includegraphics[width=.47\textwidth]{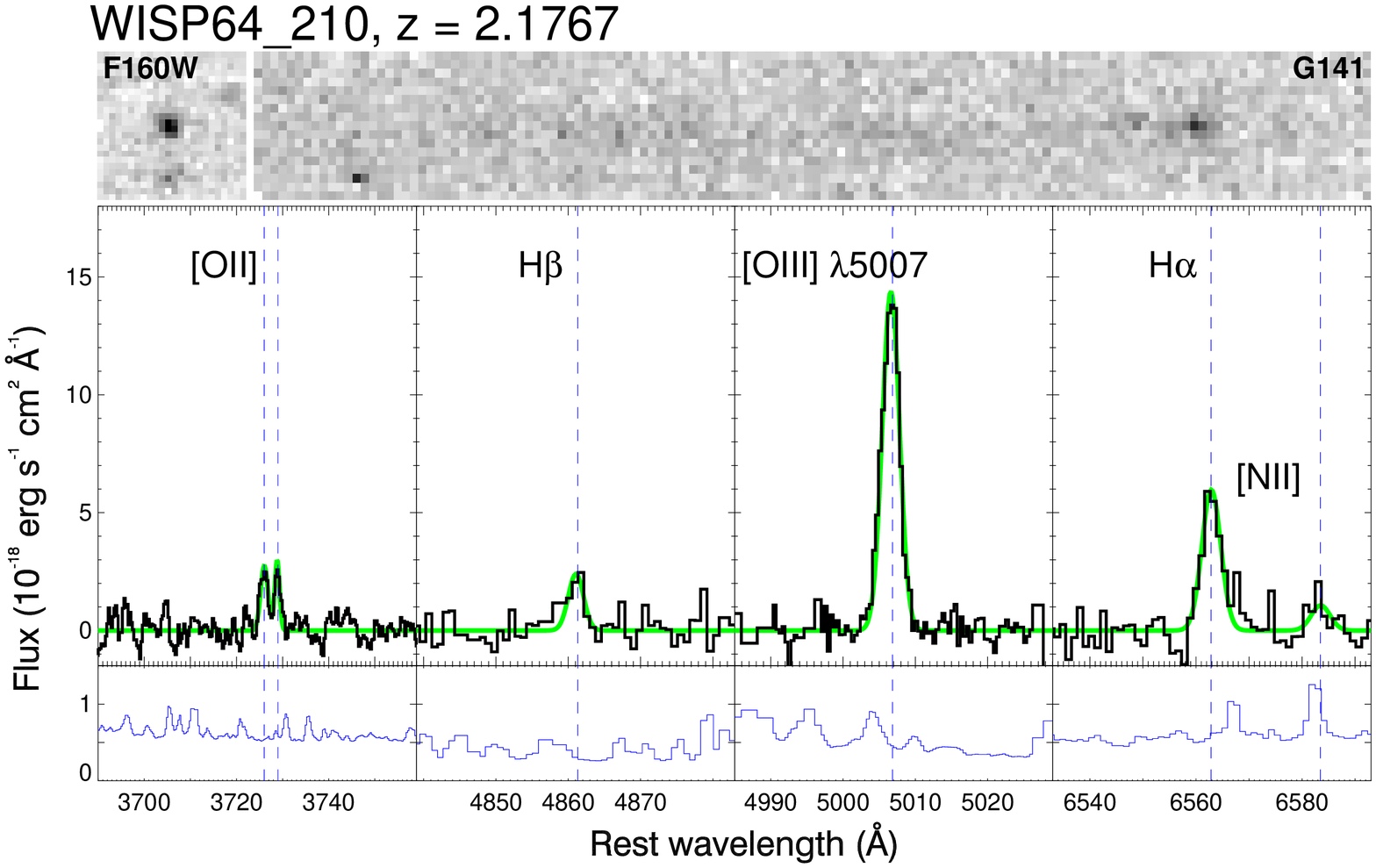} &
    \includegraphics[width=.47\textwidth]{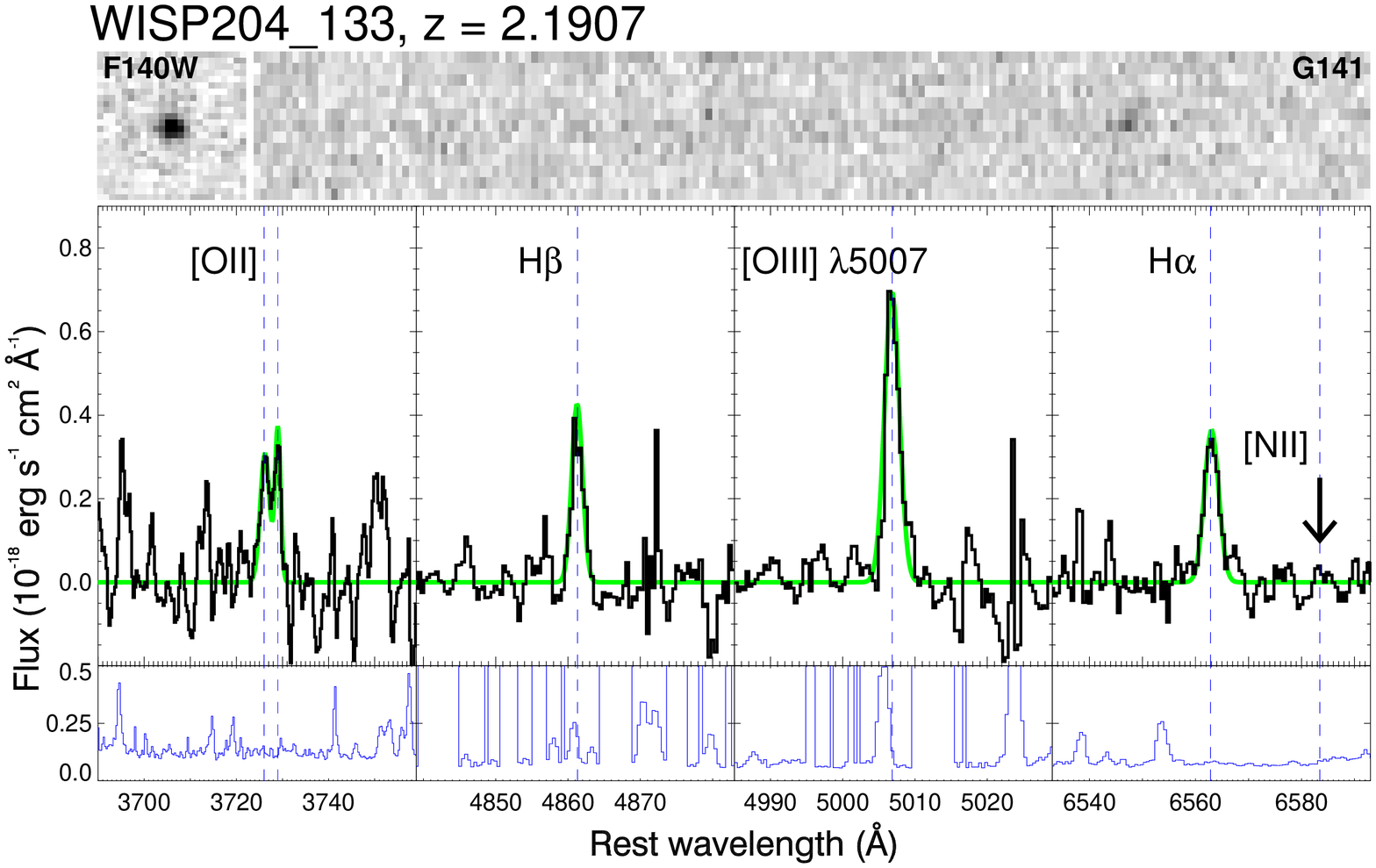} \\
    \includegraphics[width=.47\textwidth]{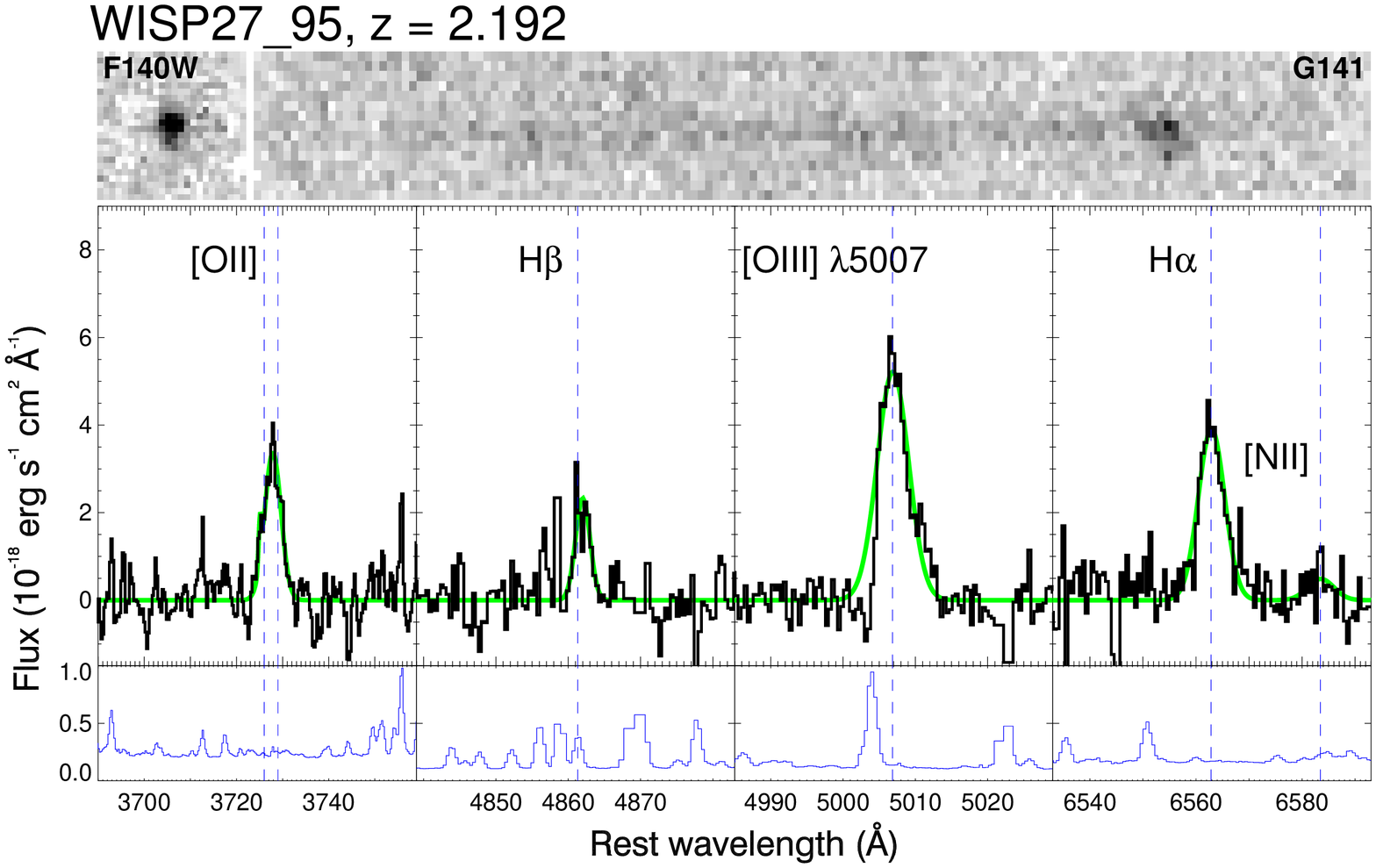} &
    \includegraphics[width=.47\textwidth]{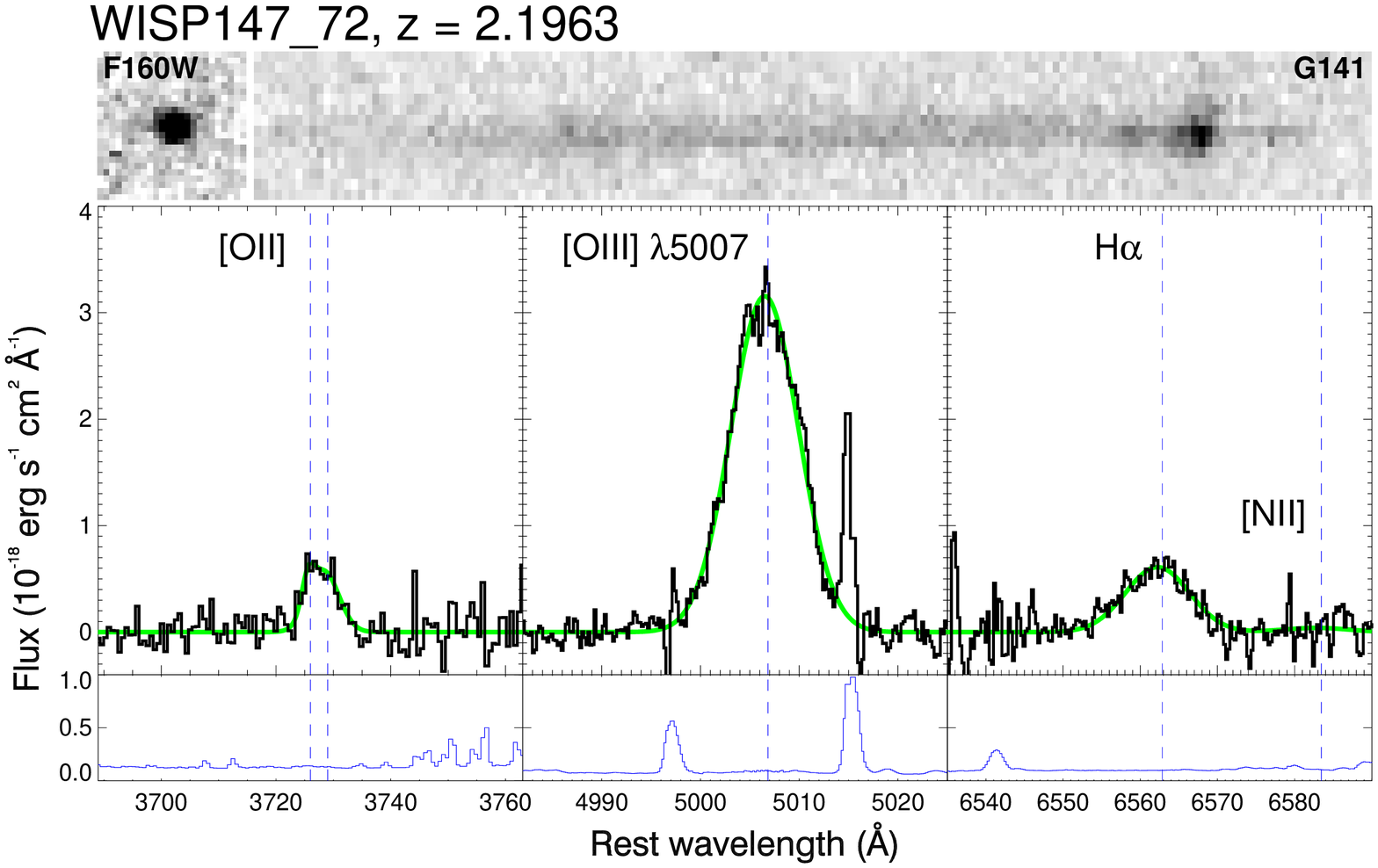} \\
    \includegraphics[width=.47\textwidth]{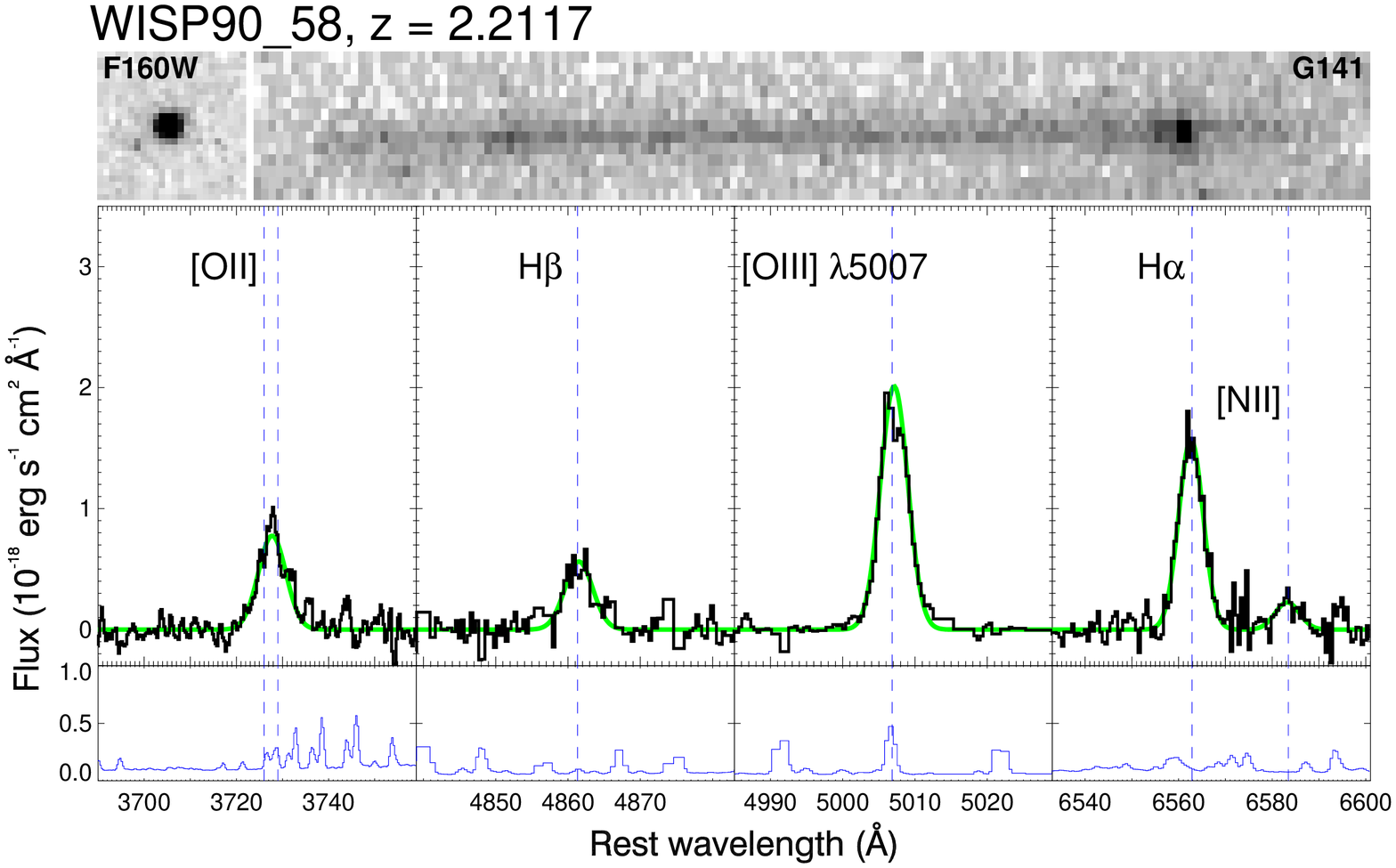} &
    \includegraphics[width=.47\textwidth]{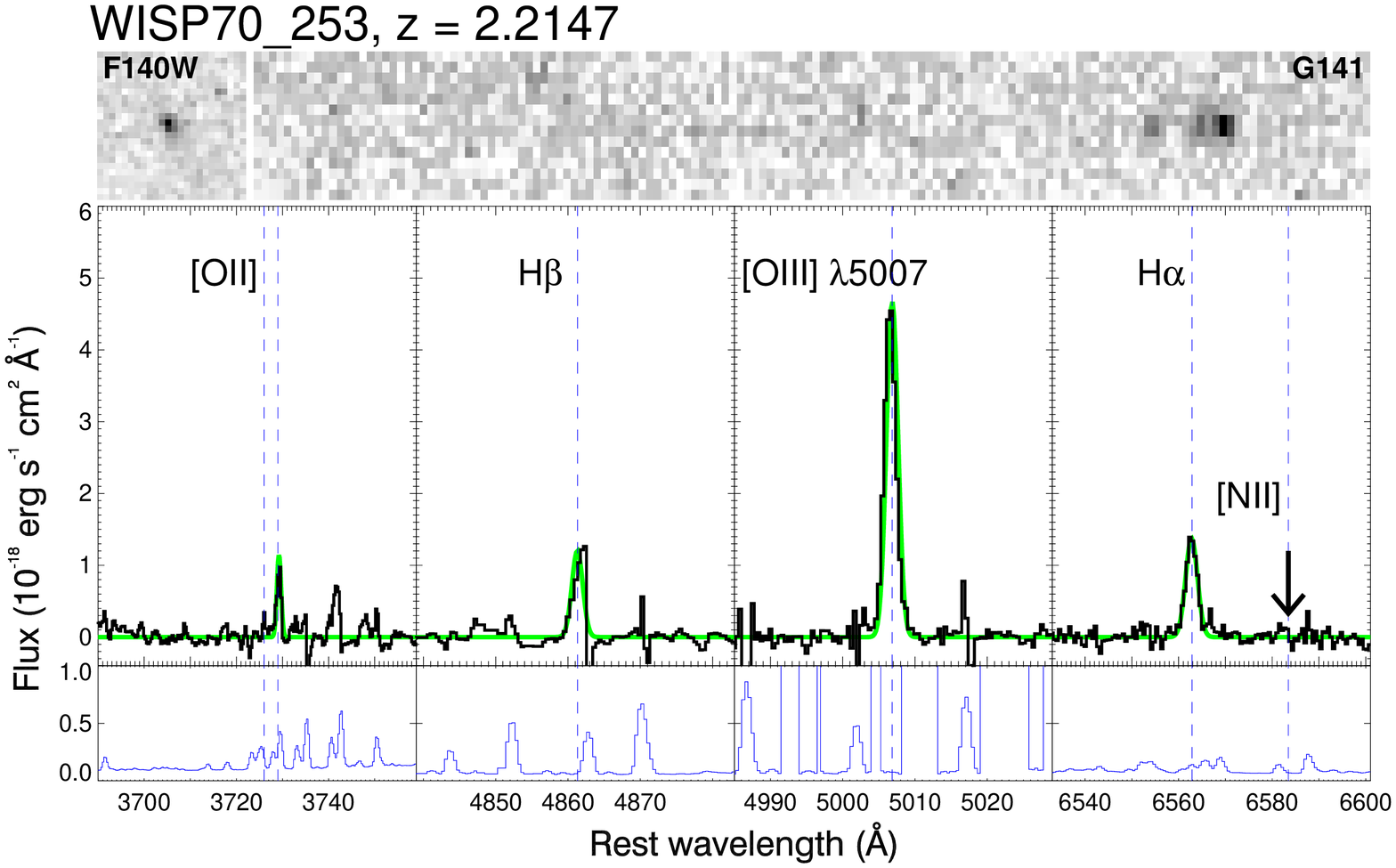} \\
  \end{tabular}
  \label{figure:overview1}
  \caption{\scriptsize{Summary of the $z\sim2.2$ sources selected from the WISP
    survey for follow-up spectroscopy with FIRE. For each object we
    show the WFC3 direct image and grism spectrum (with clear [\oiii]
    emission, the basis of the selection) on top and the 1D
    emission lines measured from the FIRE spectroscopy on the
    bottom (the 2D grism spectrum is not on the same wavelength scale as the 1D
    FIRE spectrum). The error spectra are shown in the lower panels. Fits to the emission lines are shown in
    green. Non-detections of [\nii]$\lambda6583$ are indicated with
    down arrows. When lines are not shown they typically were not
    measurable due to interference from OH sky emission lines.}}
\end{figure}

\begin{figure}[htb]
\centering
  \begin{tabular}{@{}cc@{}}
    \includegraphics[width=.47\textwidth]{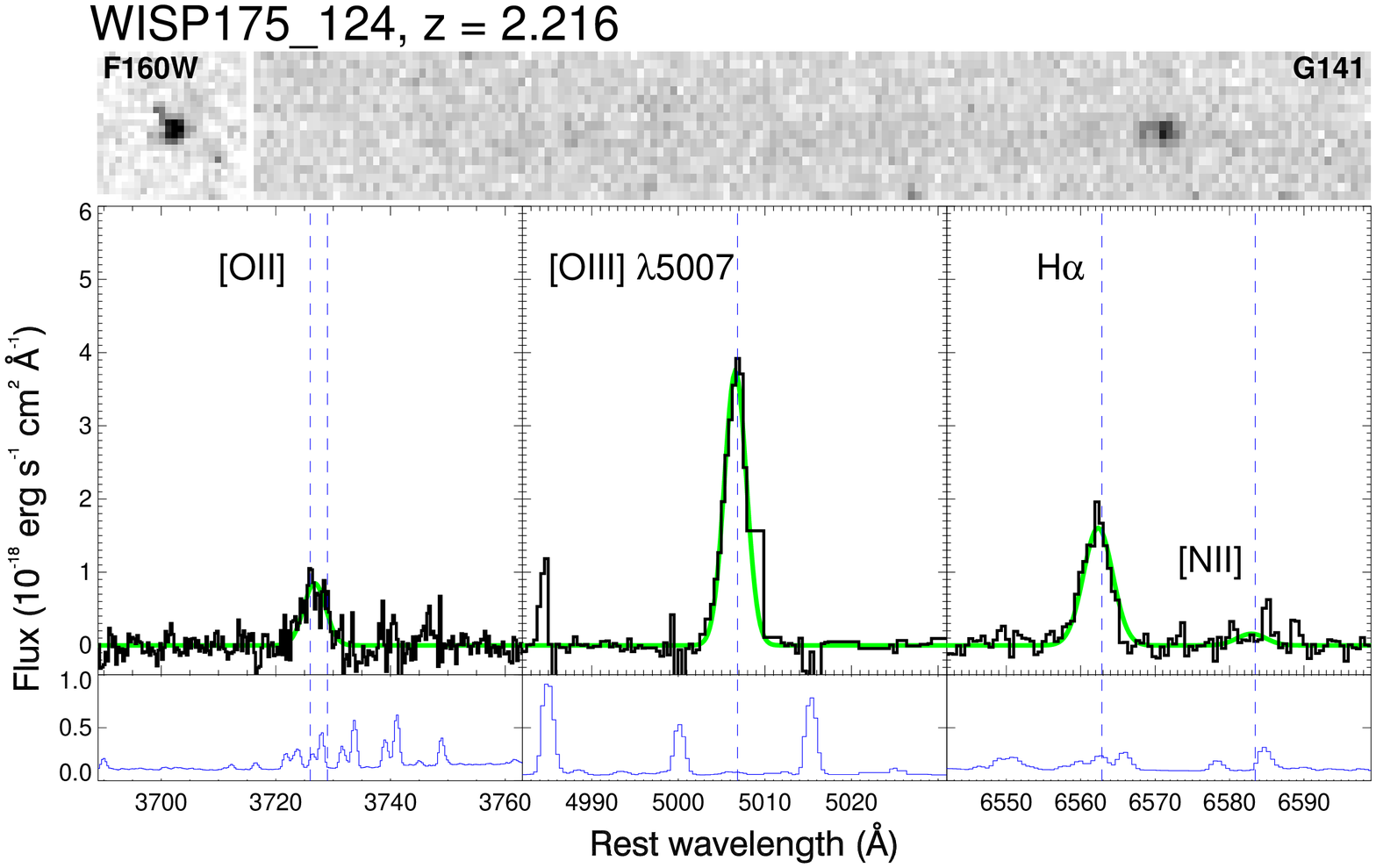} &
    \includegraphics[width=.47\textwidth]{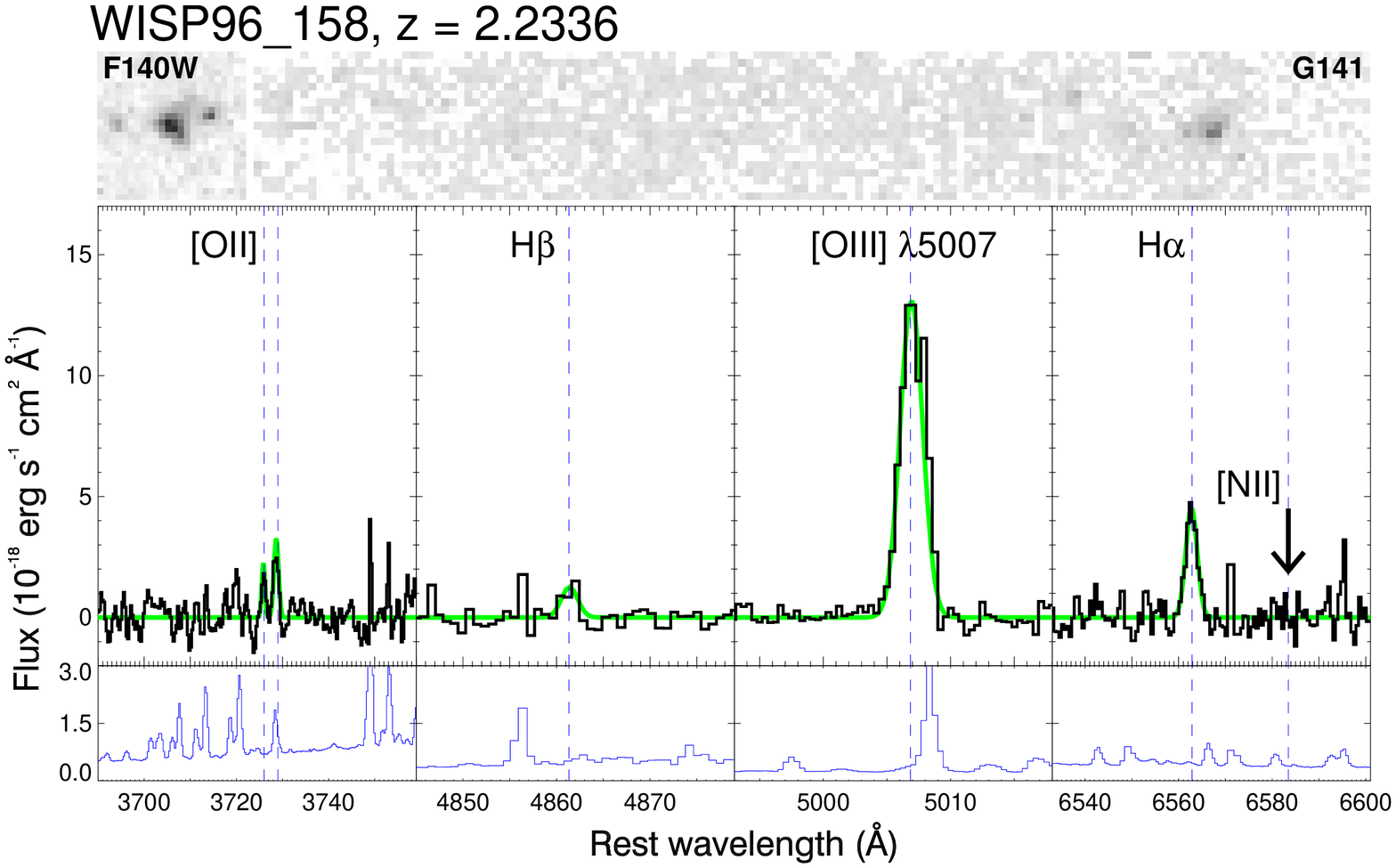} \\
    \includegraphics[width=.47\textwidth]{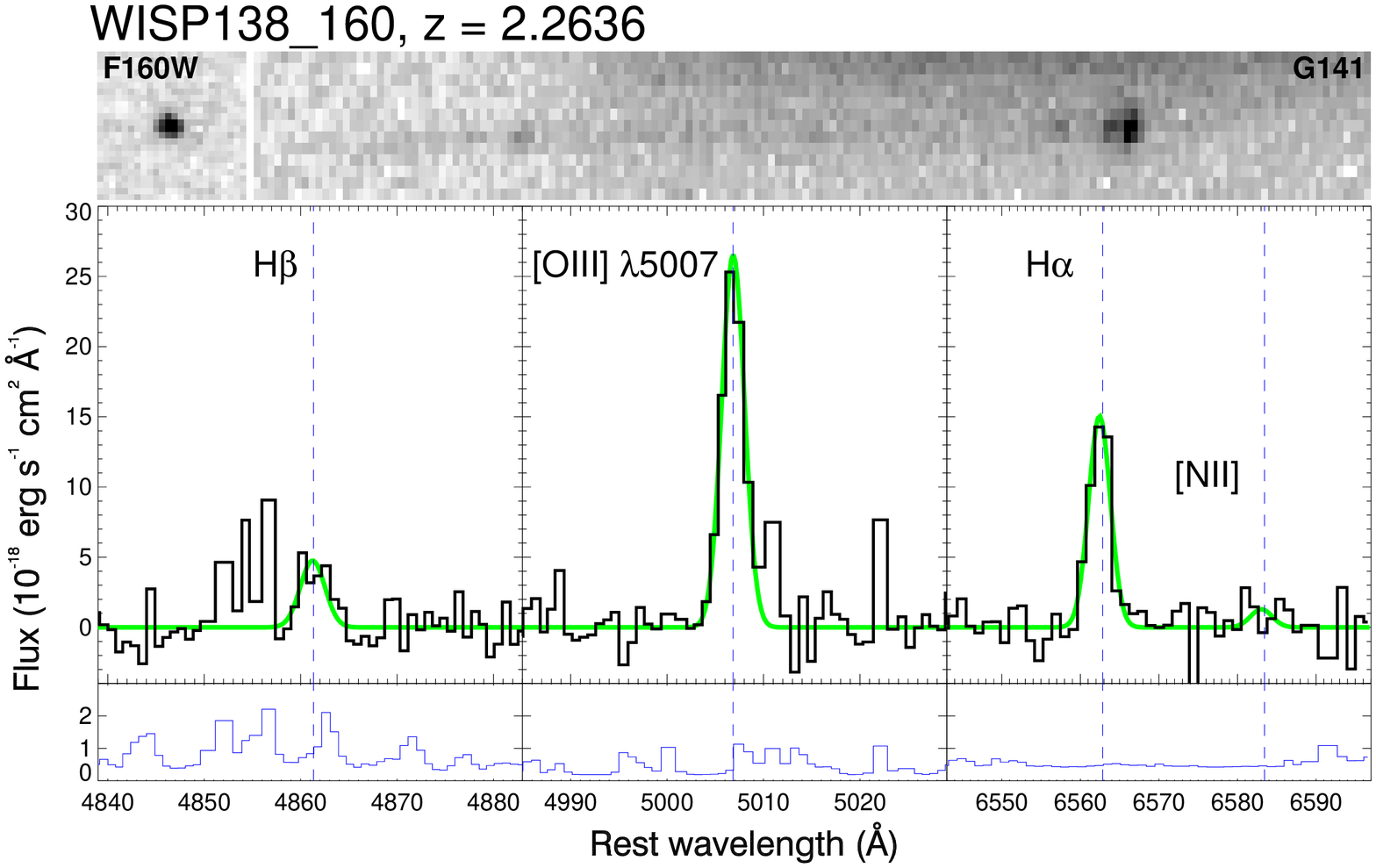} &
    \includegraphics[width=.47\textwidth]{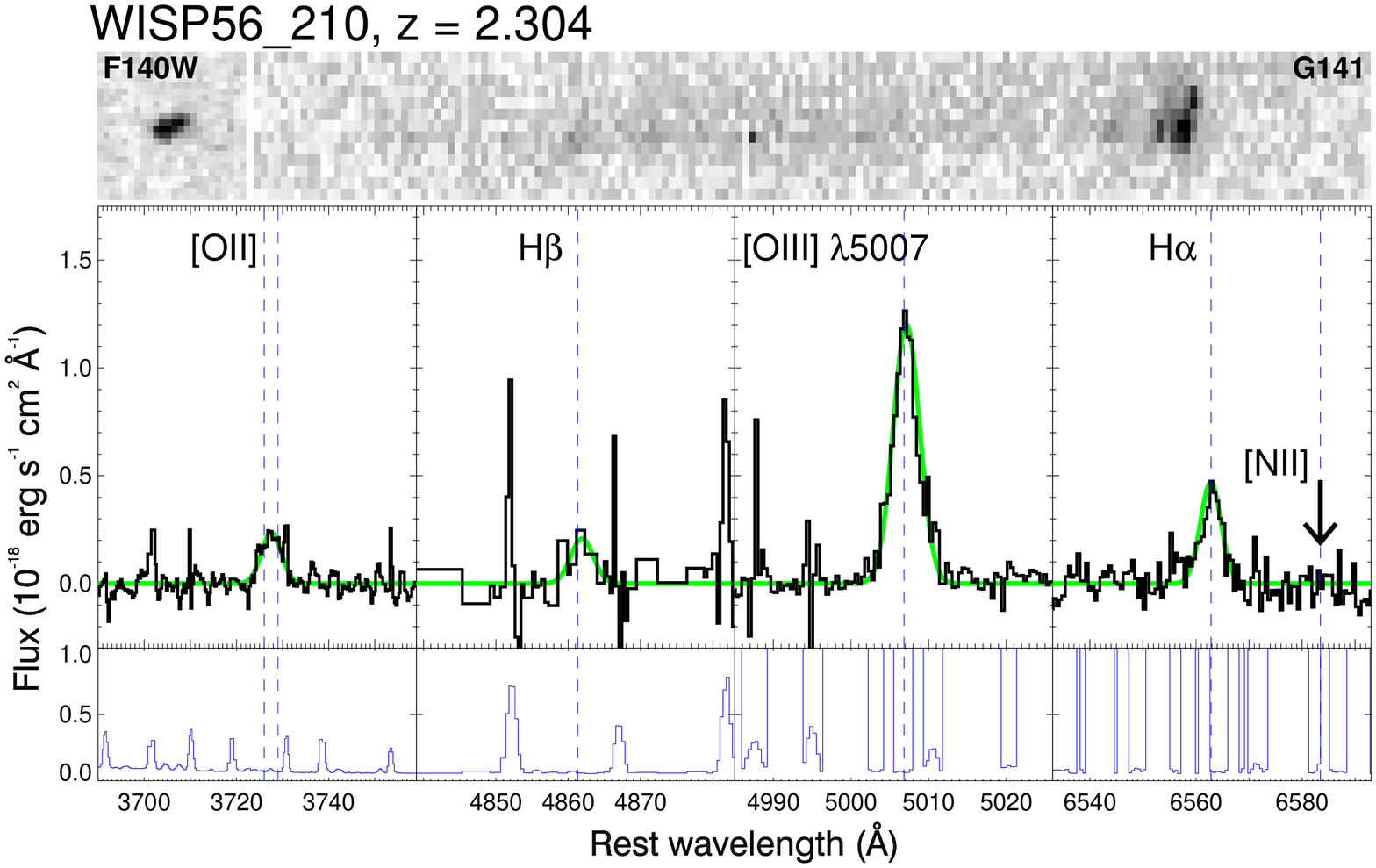} \\
    \multicolumn{2}{c}{\includegraphics[width=.49\textwidth]{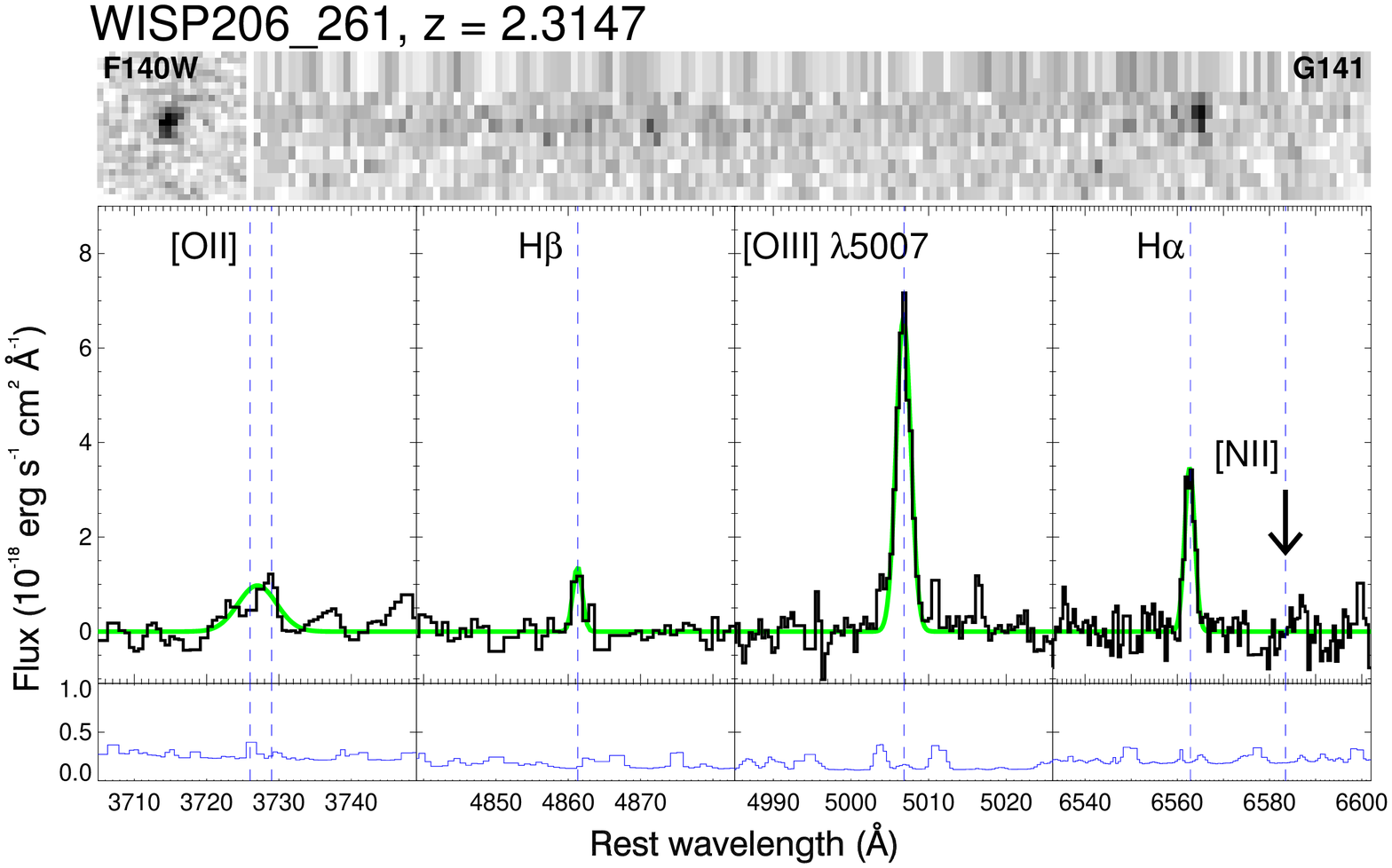}}
 \end{tabular}
 \label{figure:overview2}
  \caption{Same as for Figure~13.}
\end{figure}

\begin{figure}[htb]
\centering
  \begin{tabular}{@{}cc@{}}
    \includegraphics[width=.47\textwidth]{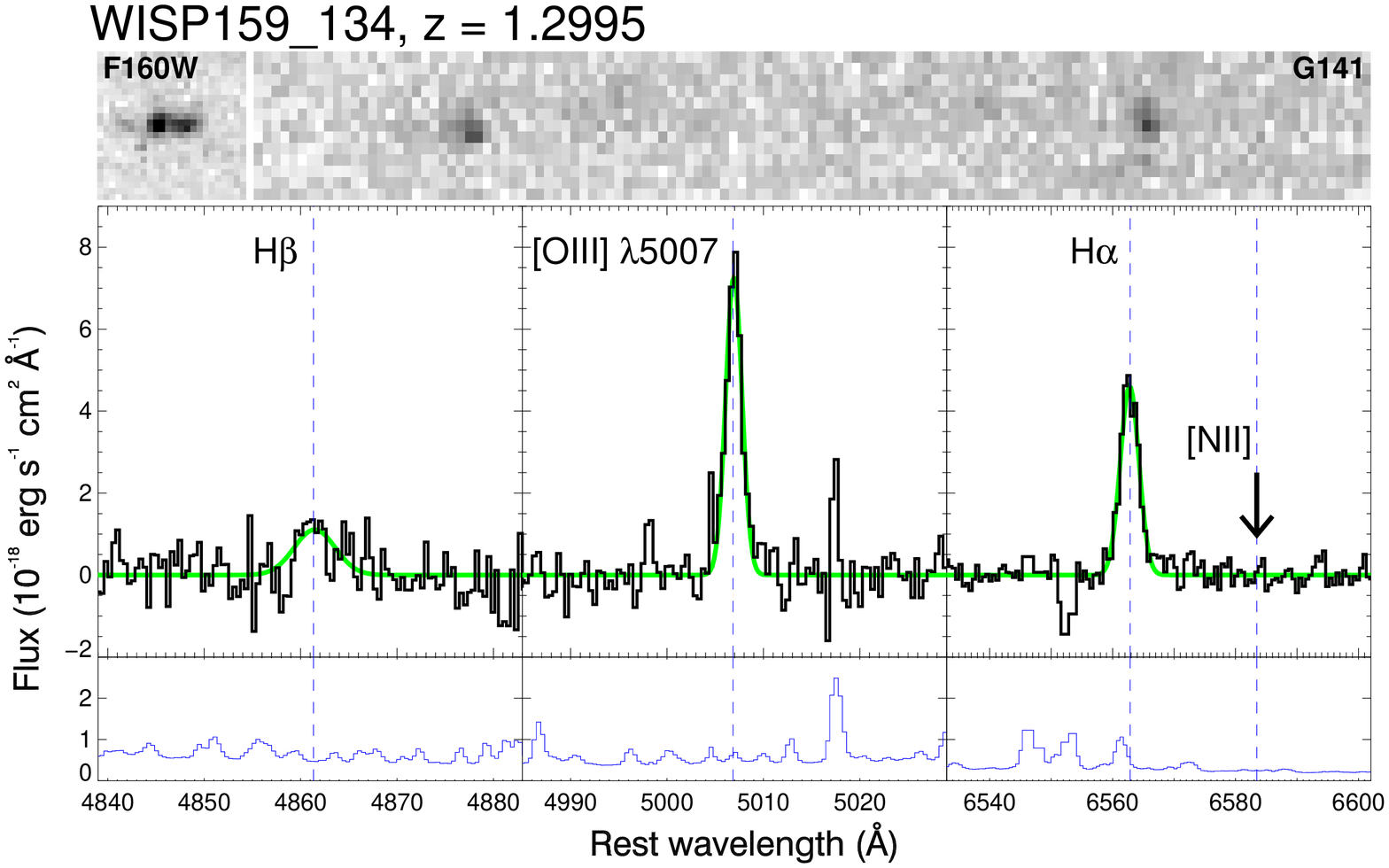} &
    \includegraphics[width=.47\textwidth]{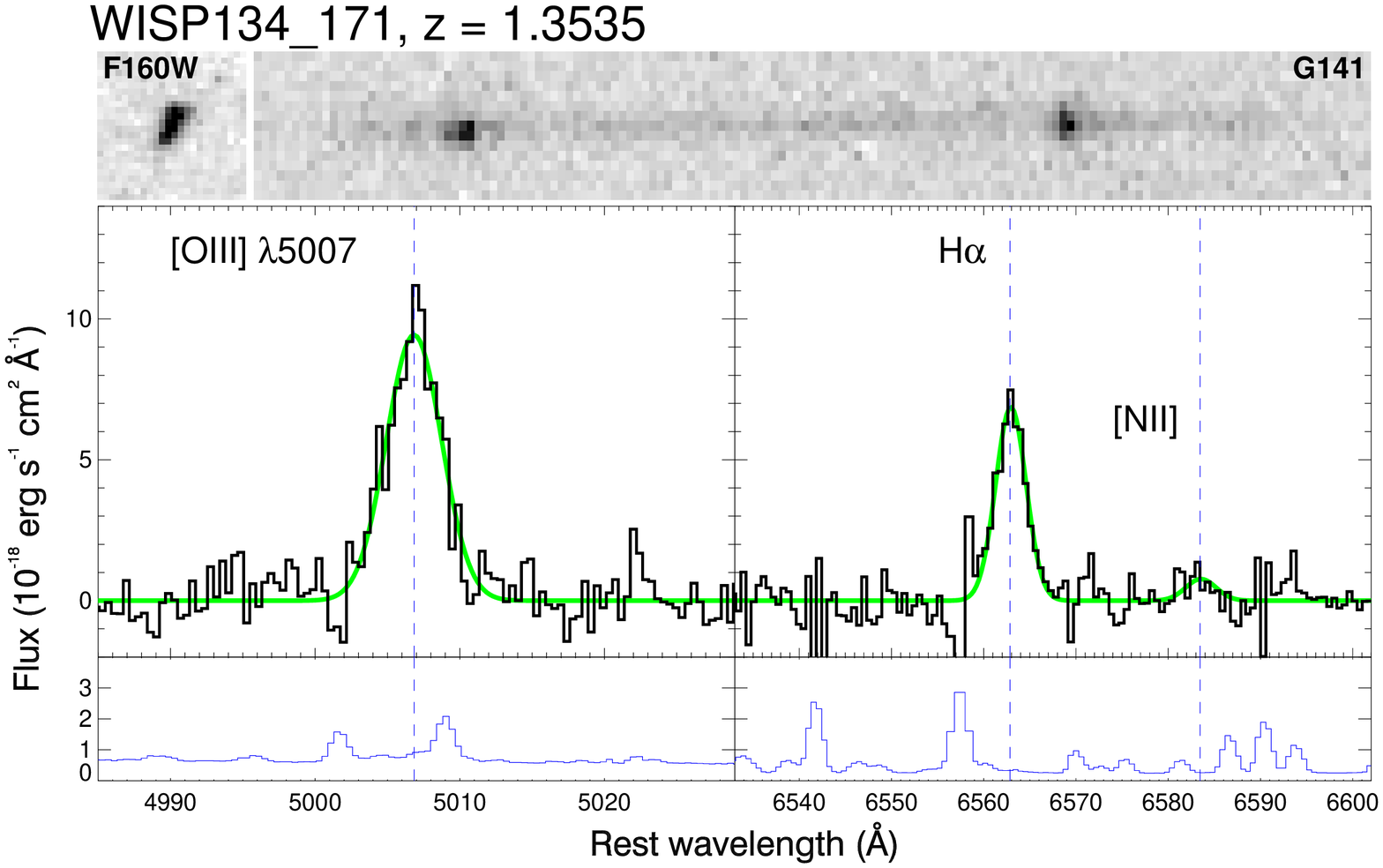} \\
    \includegraphics[width=.47\textwidth]{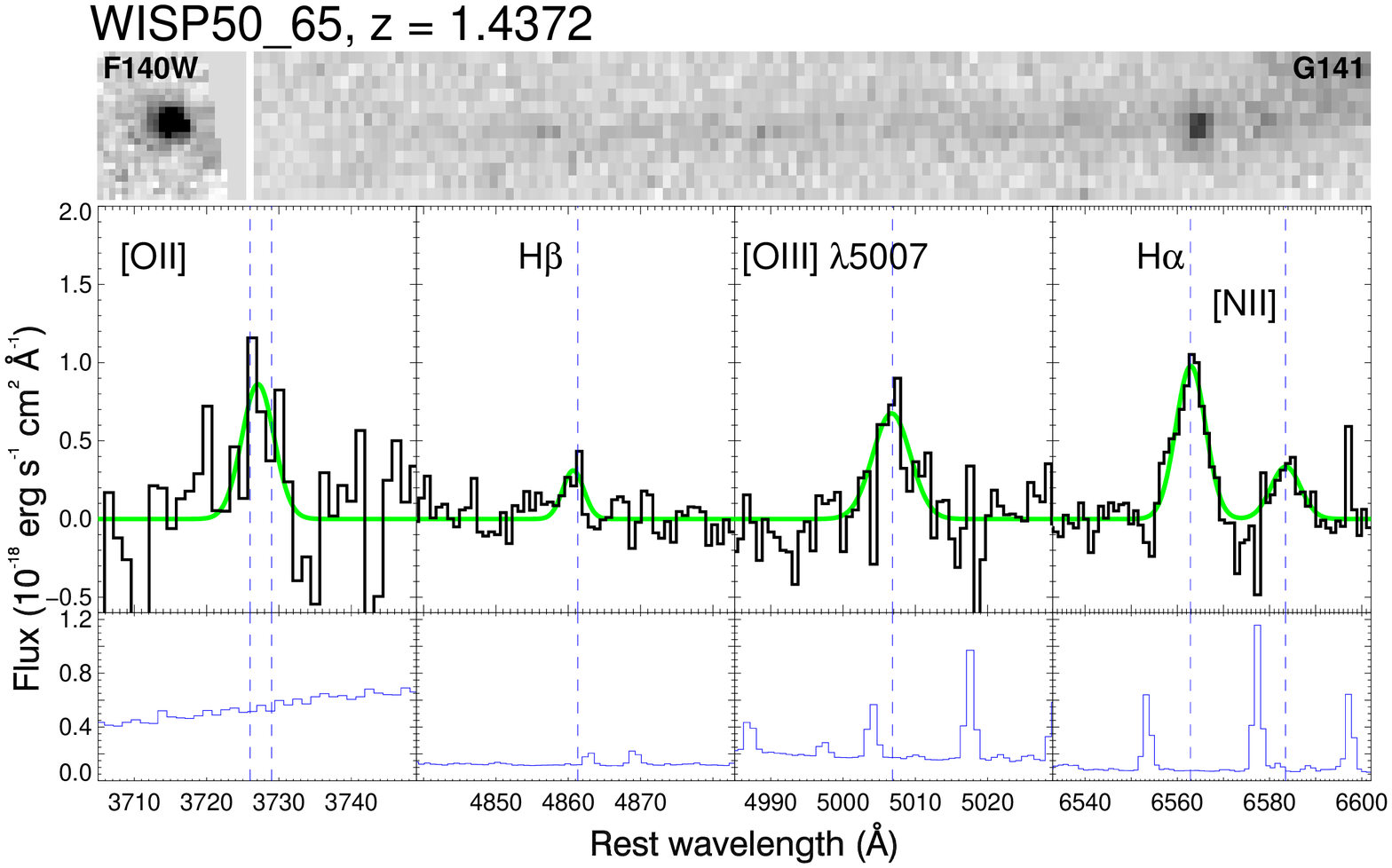} &
    \includegraphics[width=.47\textwidth]{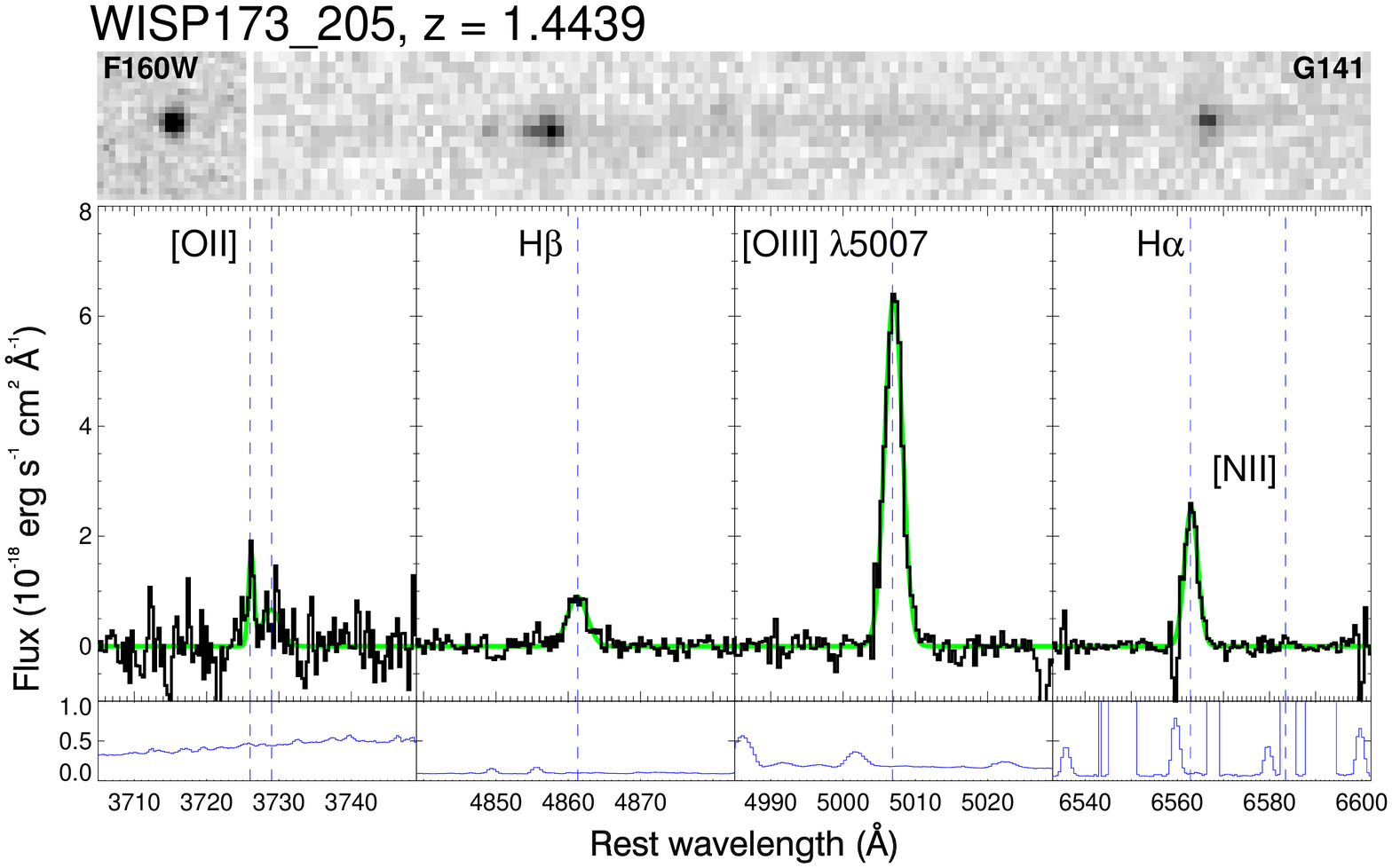} \\
    \includegraphics[width=.47\textwidth]{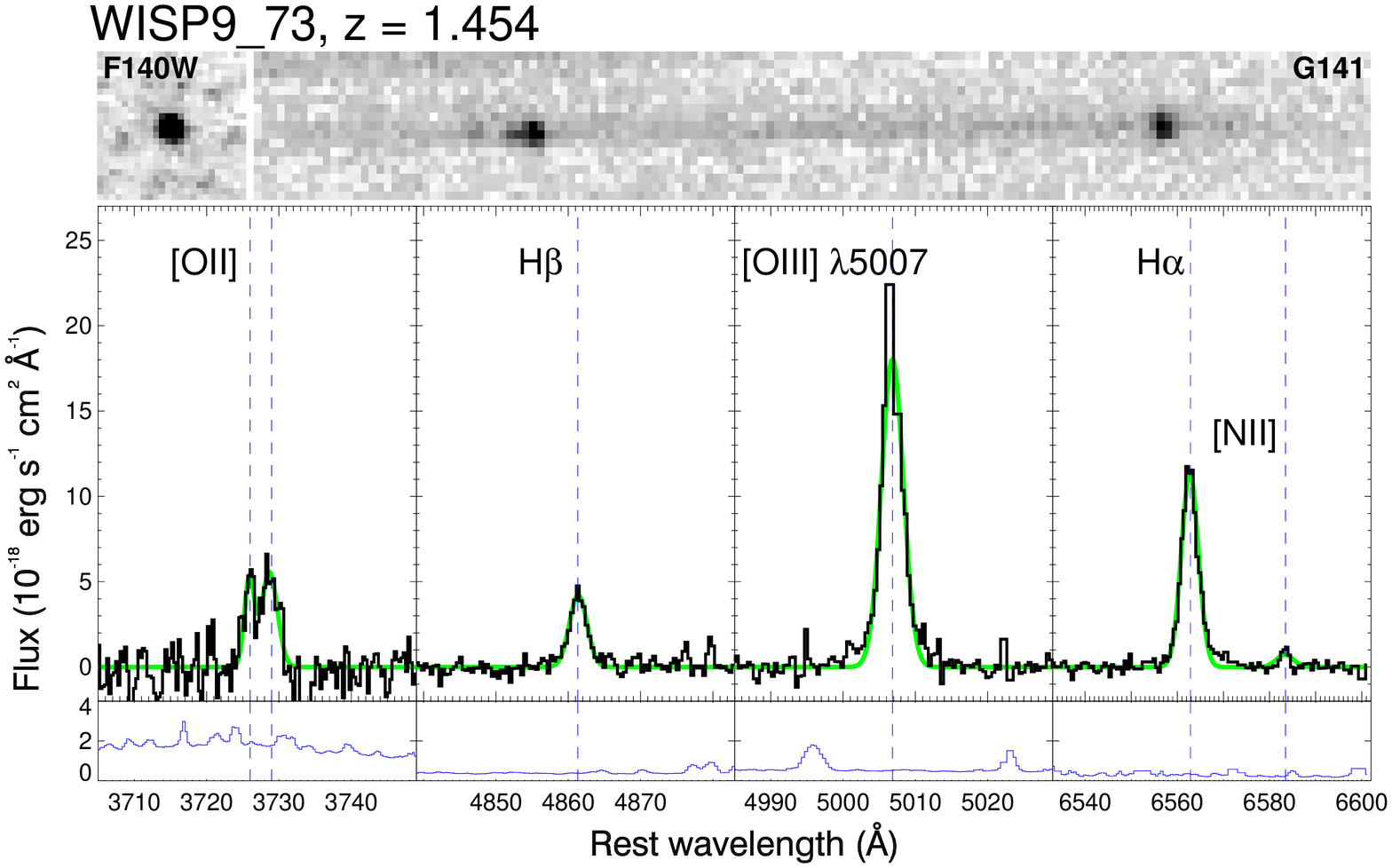} &
    \includegraphics[width=.47\textwidth]{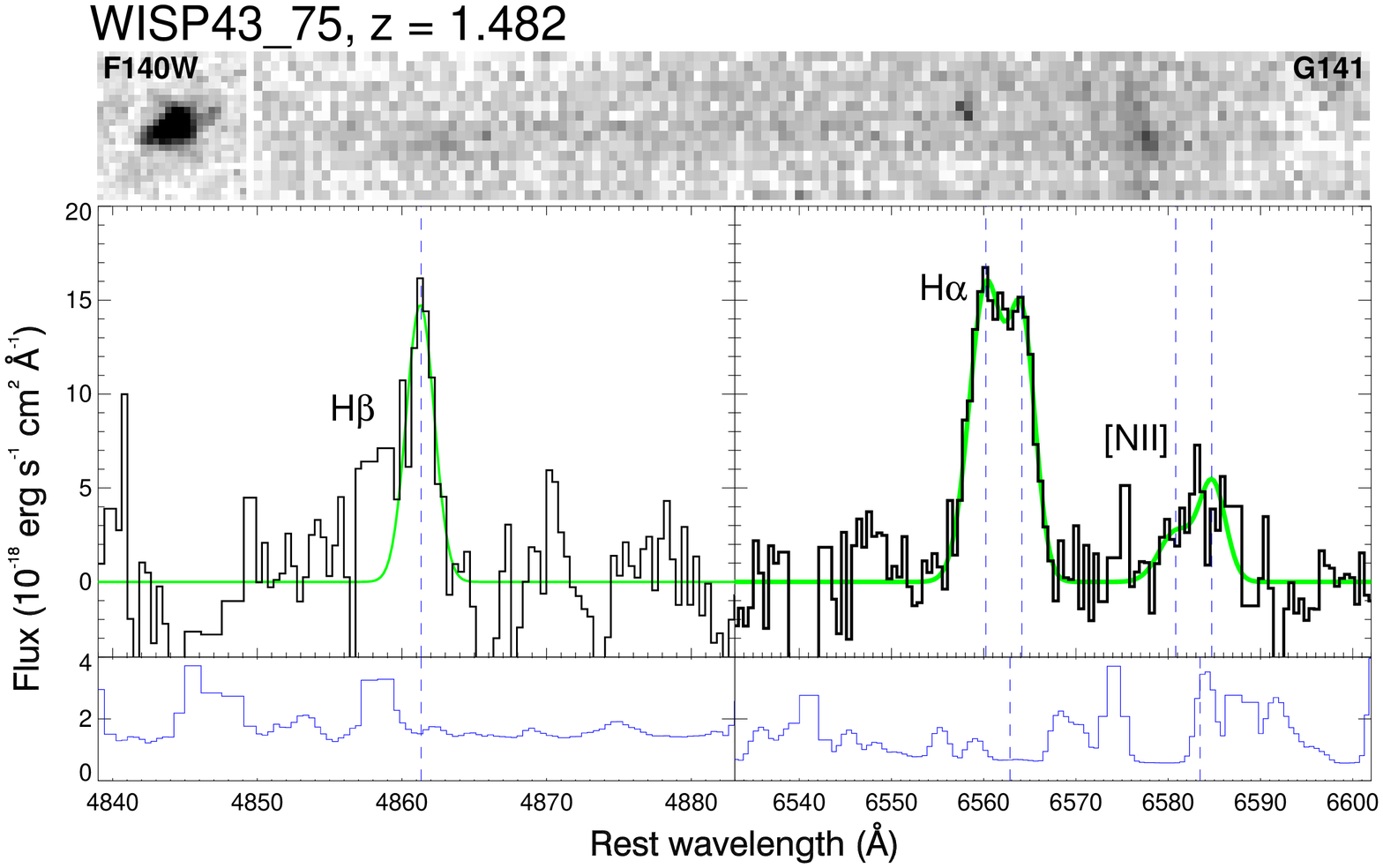} \\
    \includegraphics[width=.47\textwidth]{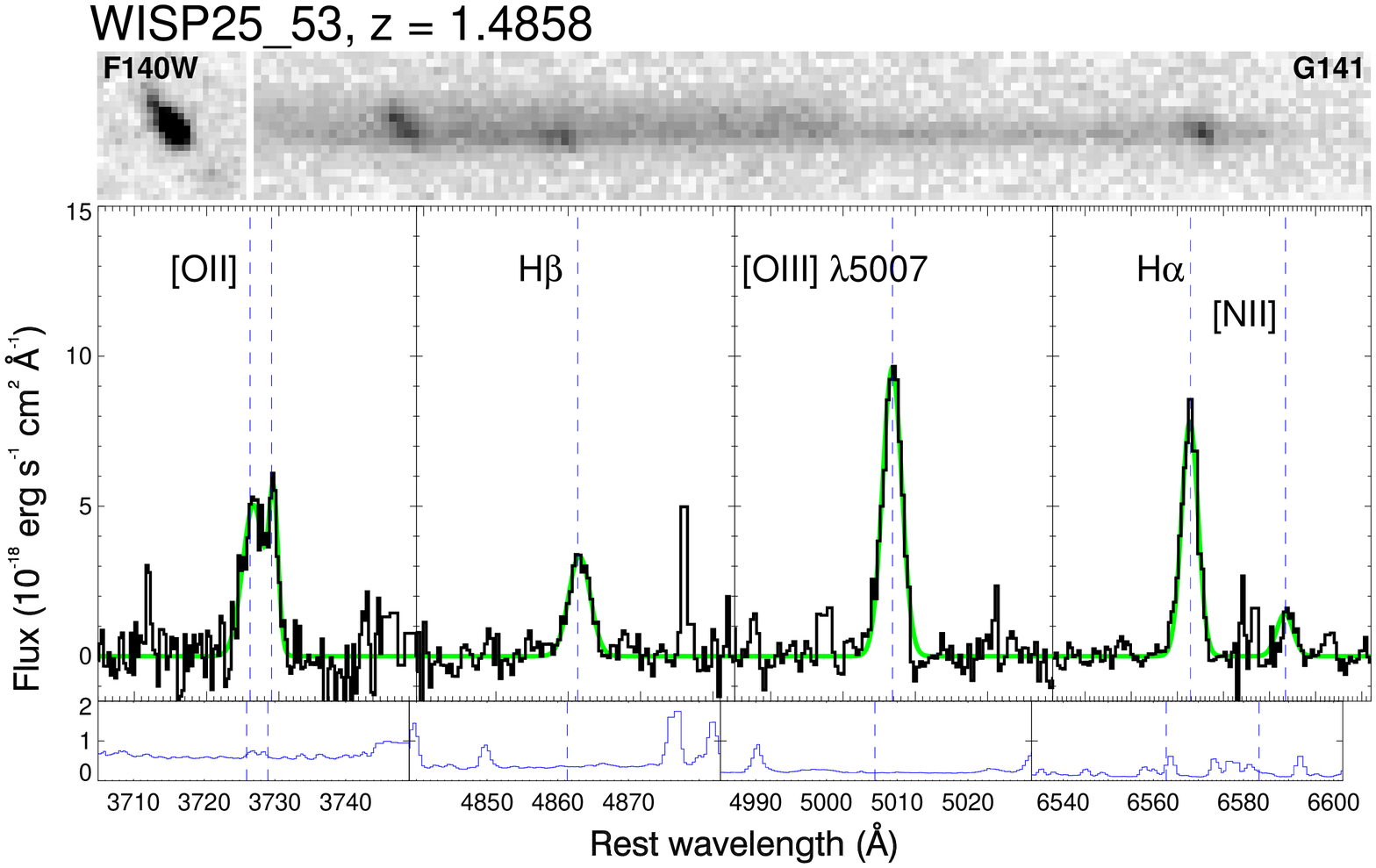} &
    \includegraphics[width=.47\textwidth]{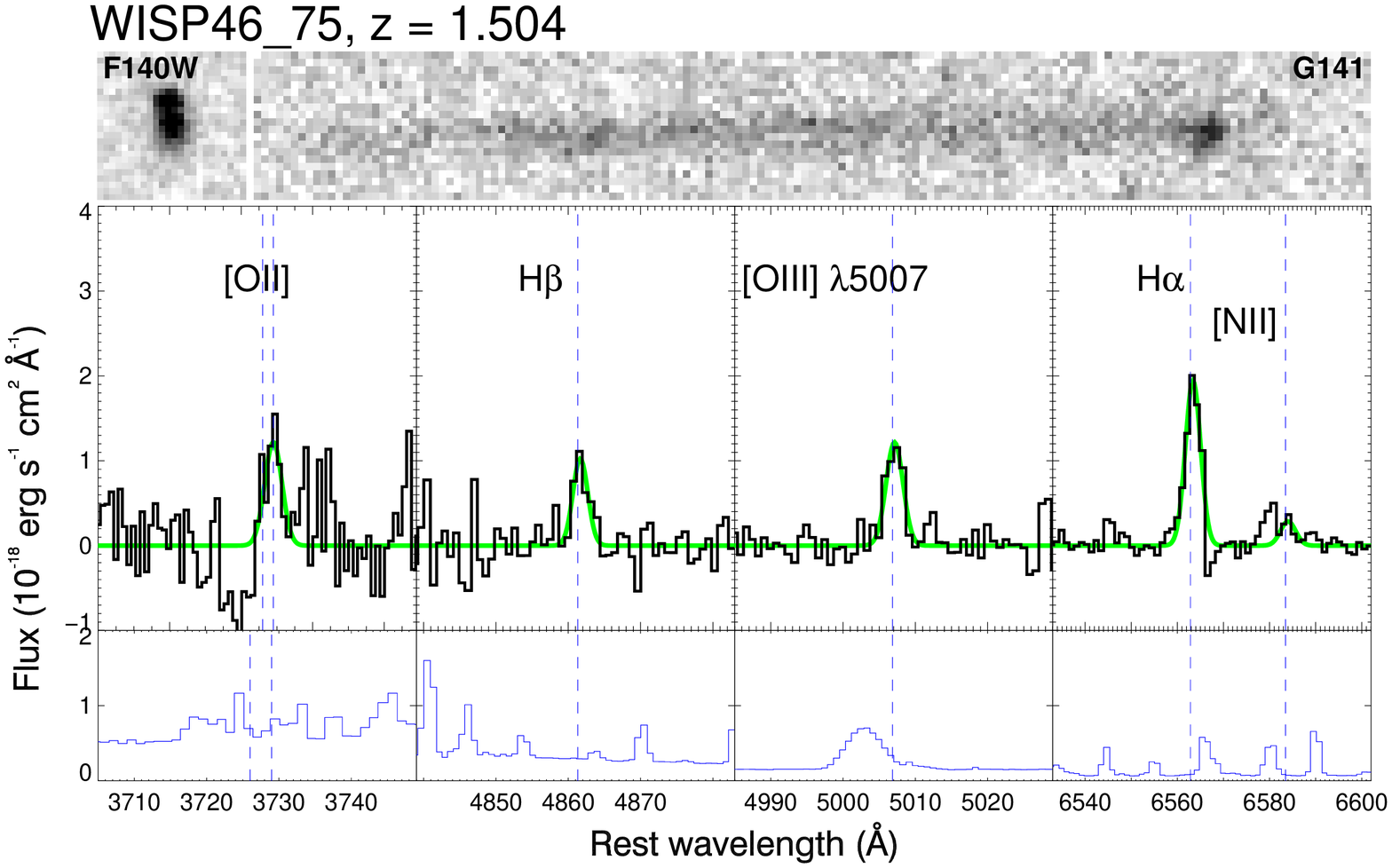} \\  
  \end{tabular}
   \label{figure:overview3}
  \caption{Summary of the $z\sim1.5$ sample, as in Figures~13 and 14. In the
    G141 grism images both [\oiii] and \ha\ are visible for most objects.}
\end{figure}

\begin{figure}[htb]
\centering
  \begin{tabular}{@{}cc@{}}
    \includegraphics[width=.47\textwidth]{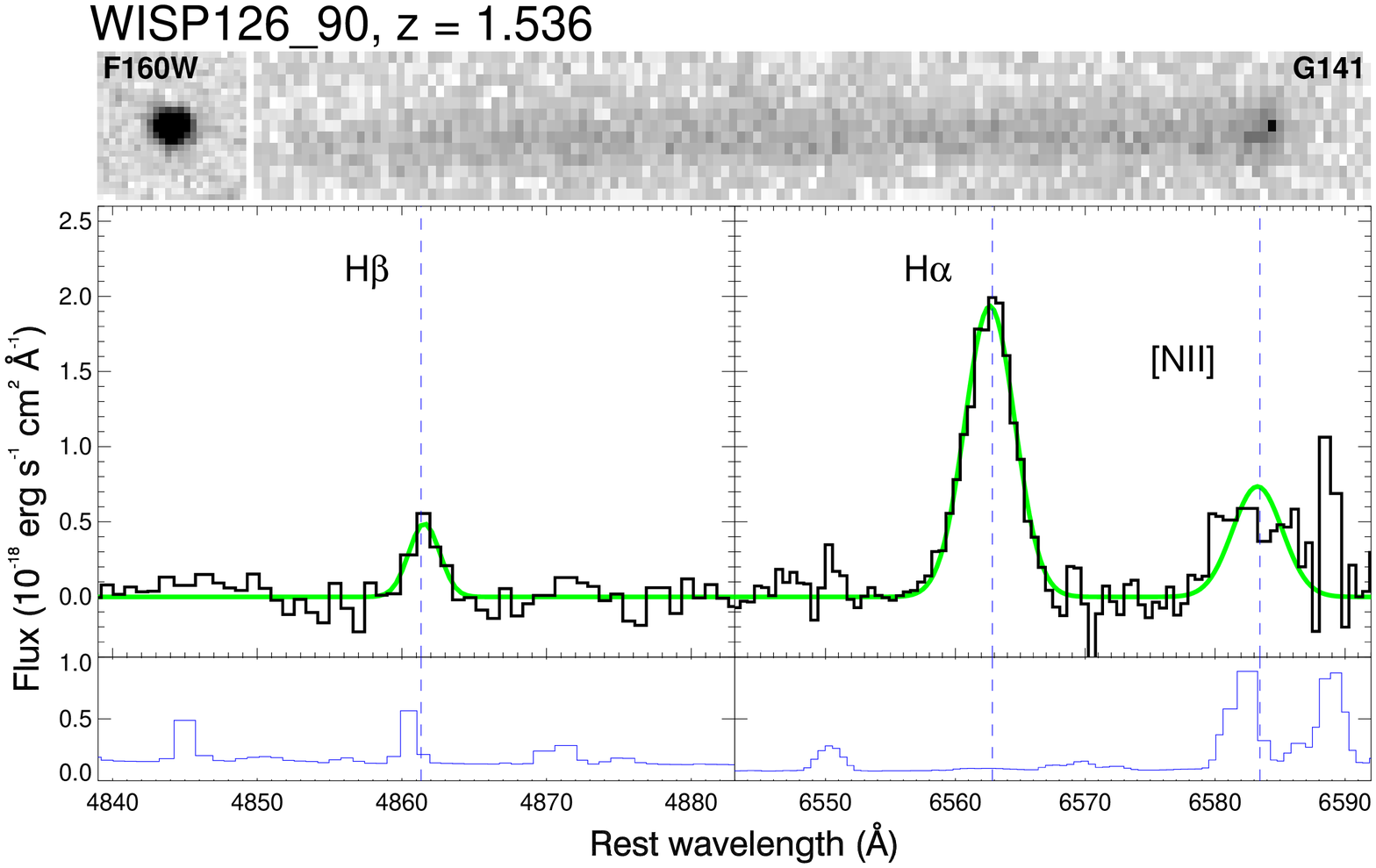} &
    \includegraphics[width=.47\textwidth]{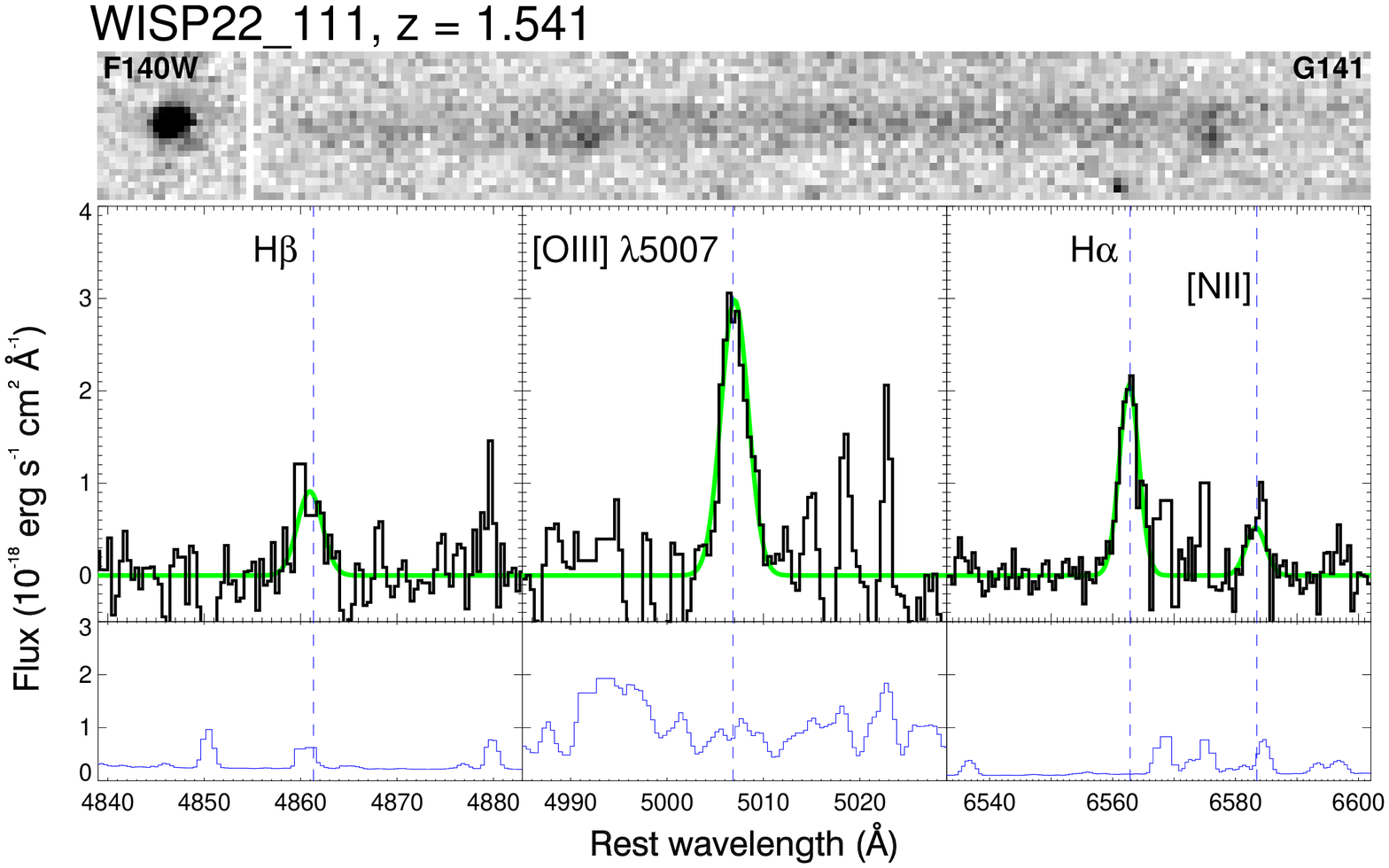} \\
    \includegraphics[width=.47\textwidth]{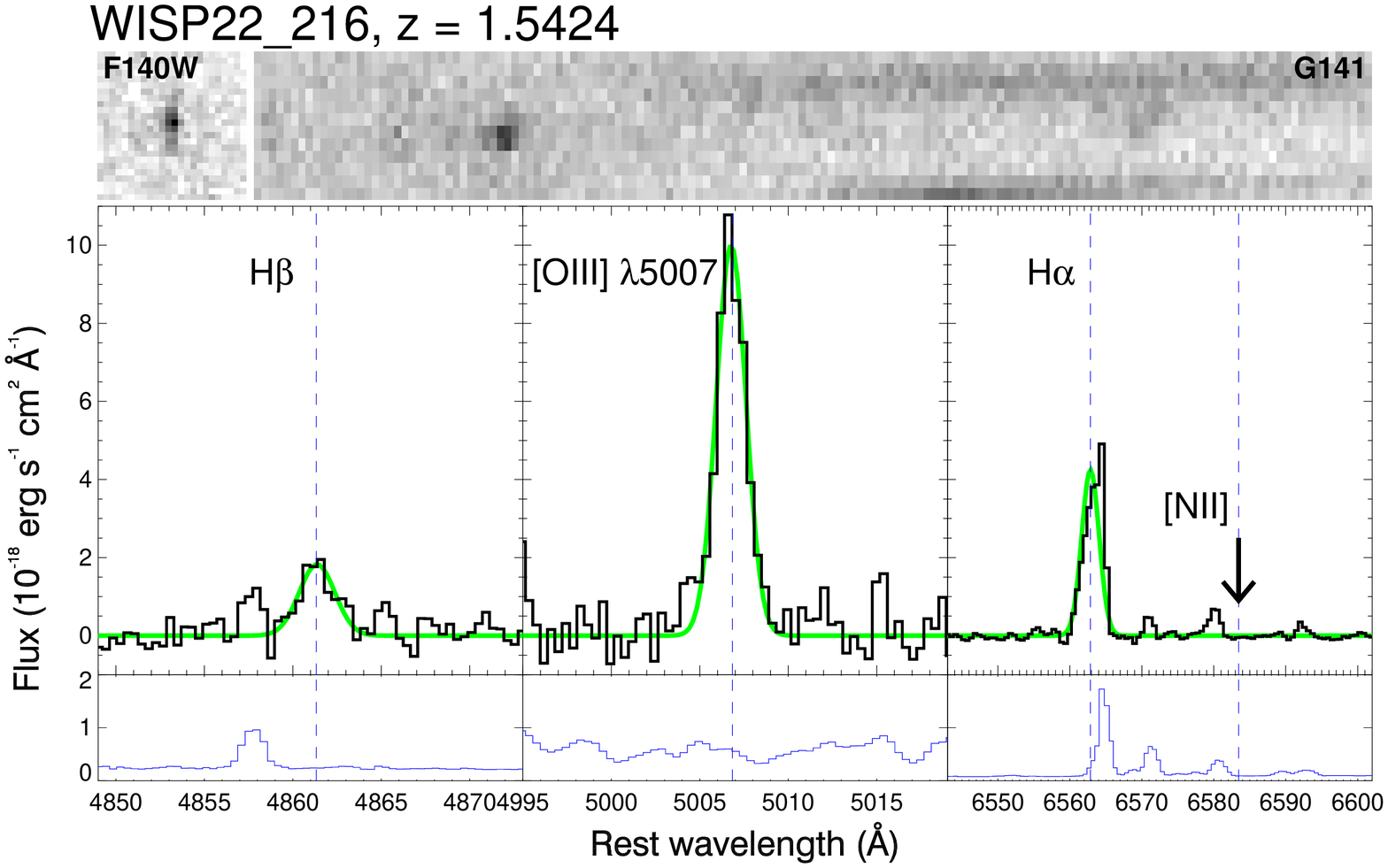} &
    \includegraphics[width=.47\textwidth]{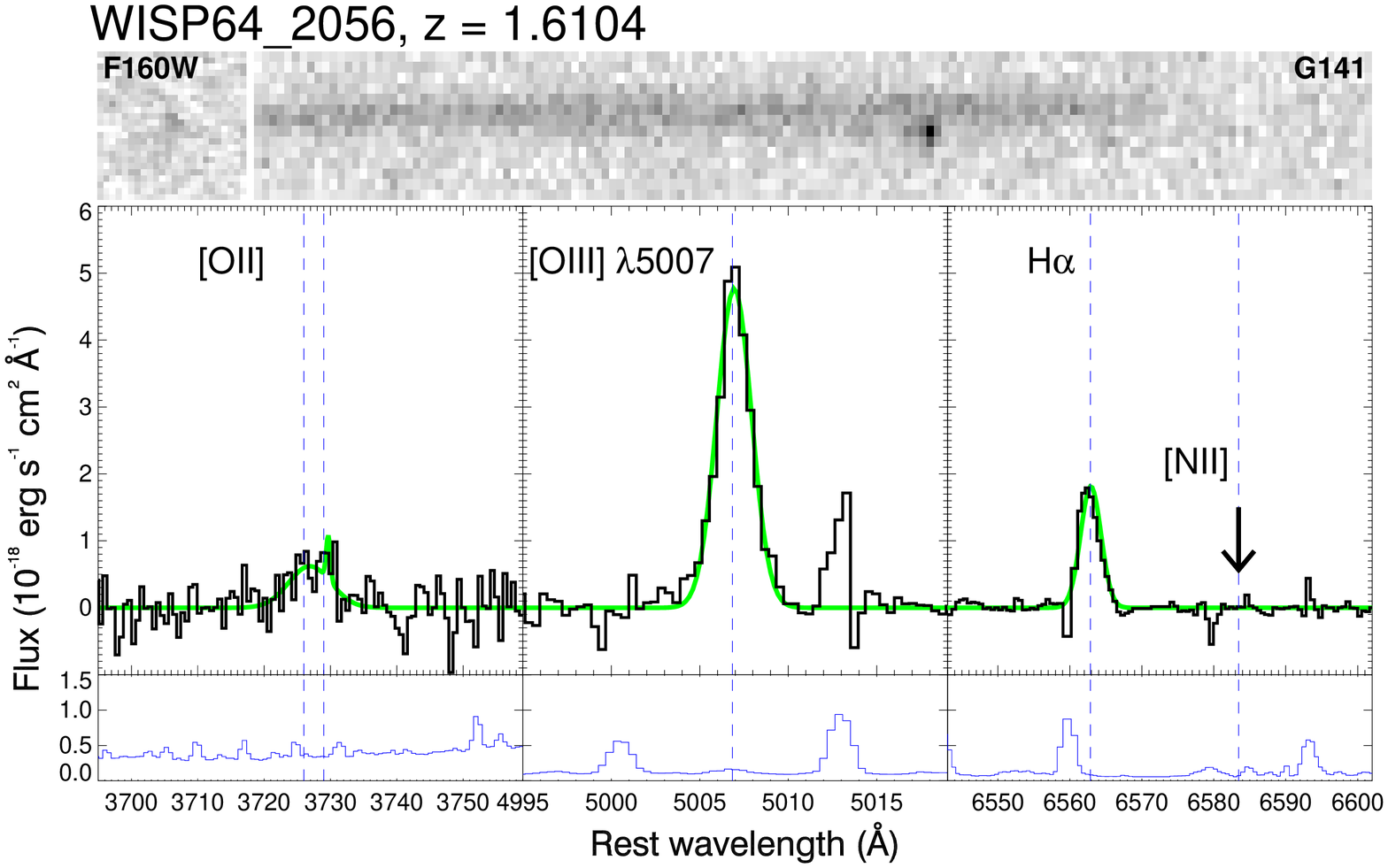} \\
    \multicolumn{2}{c}{\includegraphics[width=.49\textwidth]{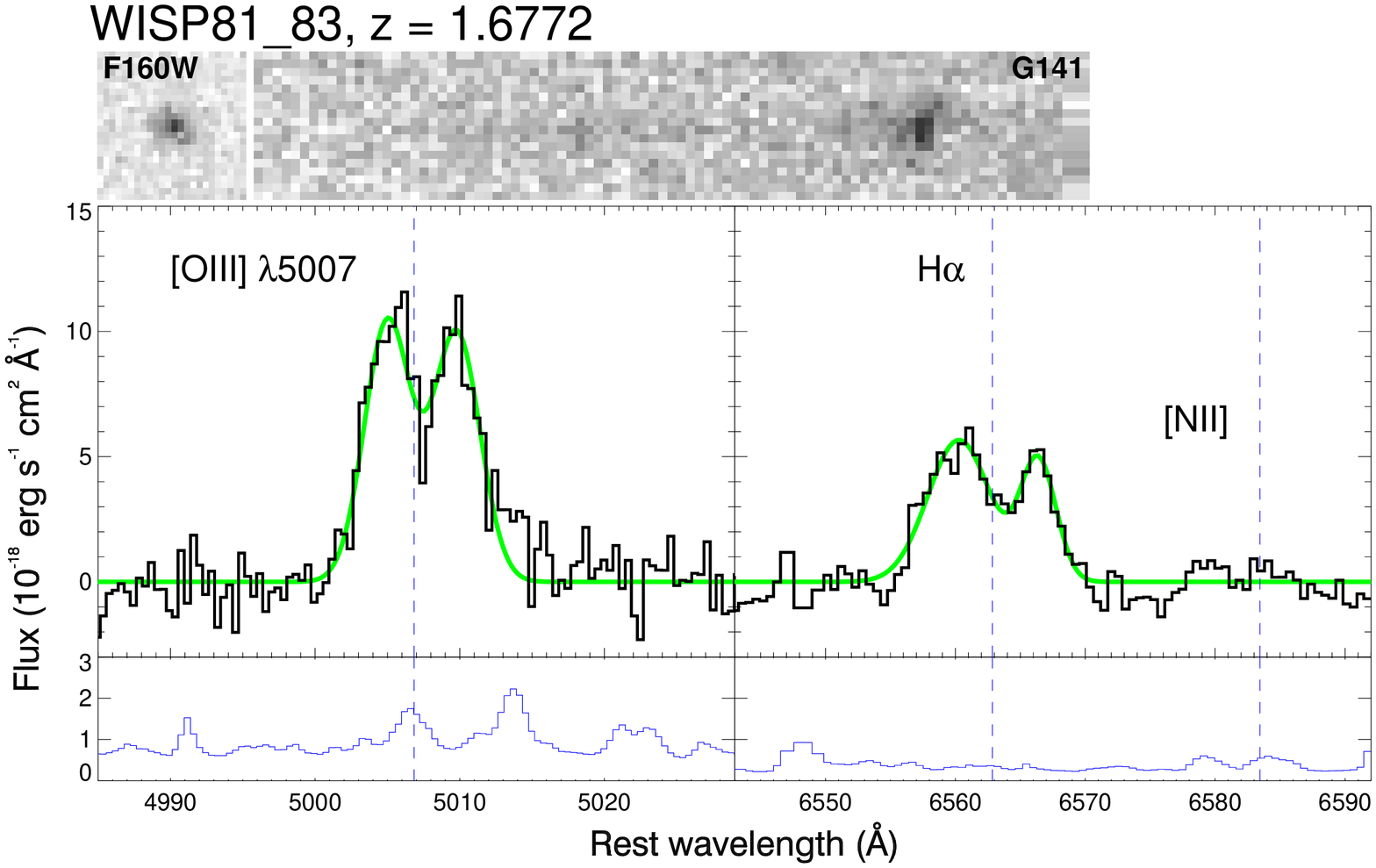}}
  \end{tabular}
 \label{figure:overview4}
  \caption{Same as for Figure~15.}
\end{figure}

\end{document}